\documentclass{pasj00}

\usepackage{graphicx}
\draft

\def\ep{E_{\rm peak}}

\begin{document}
\SetRunningHead{Author(s) in page-head}{Running Head}
\Received{2010/01/18}
\Accepted{2009/12/31}


\title{Spectral Cross-calibration of the Konus-Wind, \\the Suzaku/WAM, 
and the Swift/BAT Data \\using Gamma-Ray Bursts}

%

%
 \author{%
   Takanori \textsc{SAKAMOTO},\altaffilmark{1,2,3}
   Valentin \textsc{PAL$^{'}$SHIN},\altaffilmark{4} 
   Kazutaka \textsc{YAMAOKA},\altaffilmark{5}\\
   Masanori \textsc{OHNO},\altaffilmark{6}
   Goro \textsc{SATO},\altaffilmark{6}
   Rafail \textsc{APTEKAR}, \altaffilmark{4}
   Scott S. \textsc{BARTHELMY},\altaffilmark{3}\\
   Wayne H. \textsc{BAUMGARTNER},\altaffilmark{1,2,3}
   Jay R. \textsc{CUMMINGS},\altaffilmark{1,2,3} 
   Edward E. \textsc{FENIMORE},\altaffilmark{7}\\
   Dmitry \textsc{FREDERIKS}, \altaffilmark{4}
   Neil \textsc{GEHRELS},\altaffilmark{3}
   Sergey \textsc{GOLENETSKII}, \altaffilmark{4}\\
   Hans A. \textsc{KRIMM},\altaffilmark{1,8,3}
   Craig B. \textsc{MARKWARDT},\altaffilmark{1,9,3}
   Kaori \textsc{ONDA},\altaffilmark{12}\\
   David M. \textsc{PALMER},\altaffilmark{7}
   Ann M. \textsc{PARSONS},\altaffilmark{3}
   Michael \textsc{STAMATIKOS}, \altaffilmark{10,3}\\
   Satoshi \textsc{SUGITA}, \altaffilmark{13}
   Makoto \textsc{TASHIRO}, \altaffilmark{12}
   Jack \textsc{TUELLER},\altaffilmark{3}
   and 
   Tilan N. \textsc{UKWATTA},\altaffilmark{11,3}}
 \altaffiltext{1}{Center for Research and Exploration in Space Science and Technology 
(CRESST), \\NASA Goddard Space Flight Center, Greenbelt, MD 20771, U.S.A.}\email{Taka.Sakamoto@nasa.gov}
 \altaffiltext{2}{Joint Center for Astrophysics, University of Maryland, Baltimore County,\\
1000 Hilltop Circle, Baltimore, MD 21250, U.S.A.}
 \altaffiltext{3}{NASA Goddard Space Flight Center, Greenbelt, MD 20771, U.S.A.}
 \altaffiltext{4}{Ioffe Physico-Technical Institute, Laboratory for Experimental Astrophysics,\\
 26 Polytekhnicheskaya, St Petersburg 194021, Russian Federation}
 \altaffiltext{5}{Department of Physics and Mathematics, Aoyama Gakuin University, \\
5-10-1 Fuchinobe, Chuo-ku, Sagamihara, Kanagawa 252-5258}
 \altaffiltext{6}{Institute of Space and Astronautical Science (ISAS/JAXA), \\3-1-1 Yoshinodai, 
Sagamihara, Kanagawa 229-8510}
 \altaffiltext{7}{Los Alamos National Laboratory, P.O. Box 1663, Los Alamos, NM 87545, U.S.A.}
 \altaffiltext{8}{Universities Space Research Association, 10211 Wincopin Circle, Suite 500, 
	Columbia, MD 21044-3432, U.S.A.}
 \altaffiltext{9}{Department of Astronomy, University of Maryland, College Park, MD 20742, U.S.A.}
 \altaffiltext{10}{Oak Ridge Associated Universities, P.O. Box 117, Oak Ridge, TN 37831-0117, U.S.A.}
 \altaffiltext{11}{Center for Nuclear Studies, Department of Physics, George Washington University, 
	Washington, DC 20052, U.S.A.}
 \altaffiltext{12}{Department of Physics, Saitama University, 255 Shimo-Okubo, Sakura, Saitama 338-8570}
 \altaffiltext{13}{coTopia Science Institute, Nagoya University, Furo-cho, chikusa, Nagoya 464-8603}

\maketitle

\begin{abstract}
We report on the spectral cross-calibration results of the {\it Konus-Wind}, 
the {\it Suzaku}/WAM, and the {\it Swift}/BAT instruments using simultaneously 
observed gamma-ray bursts (GRBs).  This is the first attempt to use 
simultaneously observed GRBs as a spectral calibration source to understand 
systematic problems among the instruments.  
Based on these joint spectral fits, we find 
that 1) although a constant factor (a normalization factor) agrees within 20\% 
among the instruments, the BAT constant factor shows a systematically smaller value 
by 10-20\% compared to that of {\it Konus-Wind}, 2) there is a systematic trend 
that the low-energy photon 
index becomes steeper by 0.1-0.2 and $\ep$ becomes systematically higher by 10-20\% 
when including the BAT data in the joint fits, and 3) the high-energy 
photon index agrees within 0.2 among the instruments.  
Our results show that cross-calibration based on joint spectral 
analysis is an important step to understanding the instrumental effects which could 
be affecting the scientific results from the GRB prompt emission data.  
\end{abstract}

\section{Introduction}
Precise measurements of prompt emission spectra of Gamma-Ray Bursts 
(GRBs) is essential for understanding the physics of relativistic shocks.  
For instance, the peak energy in the observed prompt GRB $\nu$F$_{\nu}$ 
spectrum ($E_{\rm peak}$) is believed to correspond to the 
critical energy of synchrotron radiation from accelerated electrons with 
a minimum Lorentz factor (e.g. \cite{sari1998}).  The burst bolometric fluence, which 
requires an accurate measurement of the broad-band spectrum to calculate, reflects the total 
radiated energy in the prompt emission phase.  Moreover, the power-law photon index 
below $E_{\rm peak}$ can be used for testing whether the 
origin of the emission is indeed synchrotron (e.g. \cite{preece1998}).  
Similarly, the power-law photon index above $E_{\rm peak}$ should inform us 
about the power-law index, $p$, of electrons ($N(\gamma_{e}) \propto \gamma_{e}^{-p}$) 
in the framework of the synchrotron shock model (e.g. \cite{sari1998,kaneko2006}).  

Additionally, several empirical relationships based on $E_{\rm peak}$ have been proposed.  
They are A) the correlation between $E_{\rm peak}$ in the GRB rest frame 
($E^{\rm src}_{\rm peak}$) and the isotropic radiated energy ($E_{\rm iso}$), 
the so called the $E^{\rm src}_{\rm peak}$-$E_{\rm iso}$ (Amati) relation 
\citep{amati2002,amati2006}, B) the correlation between the $E^{\rm src}_{\rm peak}$
energy and the collimation-corrected energy ($E_{\gamma}$), the so called
$E^{\rm src}_{\rm peak}$-$E_{\gamma}$ (Ghirlanda) relation \citep{ghirlanda2004}, 
C) the correlation between $E^{\rm src}_{\rm peak}$, $E_{\rm iso}$, and the 
achromatic break time in the afterglow light curve (t$_{jet}$) \citep{liang2005}, 
D) the relationship between $E^{\rm src}_{\rm peak}$ and the isotropic peak luminosity 
($L^{\rm peak}_{\rm iso}$), the so called the $E^{\rm src}_{\rm peak}$-
$L^{\rm peak}_{\rm iso}$ (Yonetoku) relation \citep{yonetoku2004}, and E) 
the correlation between $L^{\rm peak}_{\rm iso}$, $E^{\rm src}_{\rm peak}$, and 
the time scale of the brightest 45 per cent of the background subtracted counts 
in the light curve of the prompt emission \citep{firmani2006,rossi2008}.  However, there is much 
discussion whether these empirical relationships reflect the fundamental physics 
of GRBs or are due to instrumental effects 
(e.g. \cite{cabrera2006,butler2007,sato2007,ghirlanda2007}).  The measurements of 
the broad-band GRB spectra by multiple GRB instruments is a necessary step 
to understand instrumental effect on these empirical relations.  

$Swift$ \citep{gehrels2004} is providing more details for understanding 
GRBs (e.g. \cite{zhang2007}).  However, because of the narrow 
energy band of the Burst Alert Telescope (BAT; 15-150 keV; \cite{barthelmy2005}) 
on board $Swift$, very limited information about the spectrum of the prompt emission 
is available from the BAT data alone.  The Wide-band All-sky Monitor (WAM; 50-5000 
keV; \cite{yamaoka2009}) which is the active shield of the Hard X-ray detector 
(HXD; \cite{takahashi2007,kokubun2007}) aboard $Suzaku$ \citep{mitsuda2007} 
has also been detecting $Swift$ GRBs.  The WAM has very large 
effective area from 300 keV to 5 MeV (400 cm$^{2}$ even at 1 MeV), where 
BAT has no sensitivity.  The $Konus-Wind$ instrument \citep{aptekar1995},  
which has been on-orbit since 1994 collecting spectral and temporal 
data from GRBs, has also been detecting $Swift$ GRBs and providing information about 
the spectral properties of $Swift$ GRBs using its broad-band energy coverage 
(10-10,000 keV).  

In this paper, we report on the cross-calibration effort of the energy response for 
{\it Konus-Wind} (hereafter KW), $Suzaku$/WAM (hereafter WAM), and $Swift$/BAT 
(hereafter BAT), using simultaneously observed GRBs.  Each instrument 
has its own pros and cons for the calibration of its energy response.  For instance, 
the BAT uses the Crab nebula as a standard source to calibrate its energy 
response using its imaging capability.  On the other hand, there is an uncertainty 
in the energy response for a source which has a hard spectrum, such as a GRB, due to 
the lack of a known steady bright hard X-ray source with a similar spectrum as GRBs.  
The $\gamma$-ray instruments without an imaging capability such as WAM 
require the Earth occultation technique to collect a special 
set of the data to calibrate their energy response using a steady source 
like the Crab.\footnote{Since the Wind spacecraft is at distances of 1-7 
light-seconds from Earth, 
it is not possible to use the Earth occultation techniques to collect the data of 
a steady source.}
%
However, their instruments have a much simpler energy response 
compared to that of an imaging detector such as the BAT.  This work allows us to compensate 
the weak points in the spectral calibration of each instrument by combining the data, 
and to understand the systematic uncertainty of the energy response when using GRBs as the 
spectral calibration sources.  The paper is organized in the following manner.  In \S 2, 
we describe the instrumentation of KW, WAM and BAT.  In \S 3, 
we summarize the GRB samples used in this cross-calibration work and 
the methods of analyzing the data of each instrument.  We show the results in \S 4.  
Our conclusions are summarized in \S 5.  The quoted errors in this work are at the 
90\% confidence level, unless otherwise stated.  

\newpage

\section{Instruments}
\subsection{Konus-Wind}
The KW is a gamma-ray spectrometer designed to study 
temporal and spectral characteristics of gamma-ray bursts, solar
flares, SGR bursts, and other transient phenomena in a wide energy
range from 10 keV to 10 MeV. It consists of two identical
omnidirectional NaI(Tl) detectors (S1 and S2) one of which points
toward the south ecliptic pole thereby observing the south ecliptic
hemisphere (S1), and the other observes the north ecliptic
hemisphere (S2). Each detector has an effective area of
$\sim$80--160 cm$^2$ depending on the photon energy and incident
angle. In interplanetary space far outside the Earth's
magnetosphere, the KW has the advantages over Earth-orbiting GRB 
monitors of continuous coverage, uninterrupted by Earth occultation,
and a steady background, undistorted by passages through the Earth's
trapped radiation. 

The instrument operates in two modes: waiting and triggered. In the 
triggered mode 64 spectra are measured in two partially overlapping
energy ranges with nominal bounds 10--750 keV (PHA1) and 0.2--10 MeV (PHA2).  
Each range has 63 channels. The first four spectra are obtained with a 
fixed accumulation time of 64 ms. For the subsequent 52 spectra, the
adaptation system determines the accumulation times which can vary
from 0.256 to 8.192 s. The last 8 spectra are obtained in 8.192 s
each. As a result the minimum duration of spectral measurements is
79.104 s, and the maximum, 491.776 s. Due to the degradation of the
photomultiplier tubes, the overall energy range has shifted to
$\sim$20 keV -- $14$~MeV at the present time. Further details can be 
found in \citet{aptekar1995}.  

\begin{figure}
\begin{center}
\includegraphics[scale=0.8,angle=0]{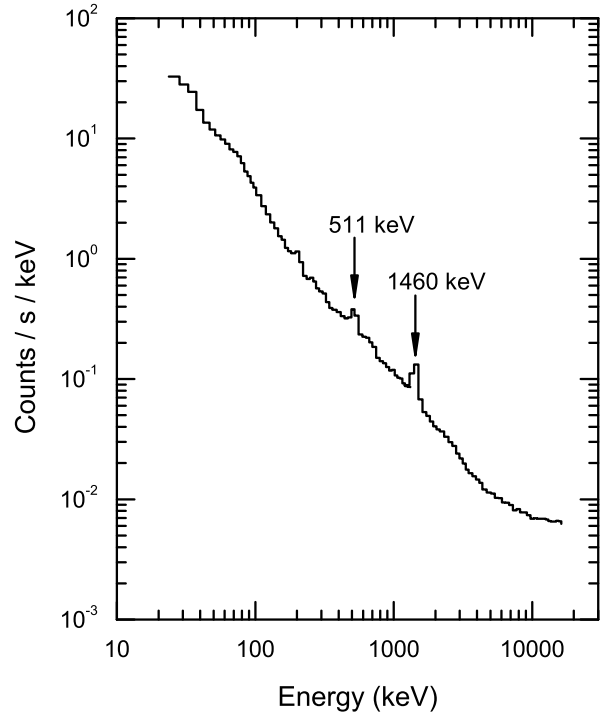}
\end{center}
\caption{The background spectrum of the Konus-Wind instrument.}\label{kw_bgd}
\end{figure}

The instrument calibration is described in \citet{terekhov1998}. The 
detector response matrix (DRM), which is a function only of the burst 
angle relative to the instrument axis, is computed using the GEANT4 
package \citep{agostinelli2003}. The last version of DRM contains 
responses for 225 photon energies between 5 keV and 16 MeV on a 
quasi-logarithmic scale for incident angles between 0$^{\circ}$ and 
90$^{\circ}$ with a step of 5$^{\circ}$.  
The energy scale is calibrated 
in-flight using 1460 keV line of $^{40}$K and the 511 keV annihilation line 
(see the background spectrum shown in Figure \ref{kw_bgd}).

The KW has been detecting GRBs in the triggered mode at a rate 
of $\sim$125 per year. It has detected 100 triggered GRBs
simultaneously with Swift-BAT with an average rate of 20 bursts per
year.  

\subsection{Suzaku/WAM}

The WAM is a large and thick BGO active shield for 
the {\it Suzaku} HXD.  The WAM consists of four perpendicular walls, 
references as WAM 0 to 3, each with a geometrical area of $\sim$ 800cm$^2$.  
The nominal energy range is 50-5000 keV, but depends on the gain of the 
photo-multipliers.  The field of view is about half of the sky.  
Since the {\it Suzaku} launch on July 10, 2005, the WAM has been working 
nominally and has been detecting GRBs at a rate 
of 140 per year including both triggered and un-triggered events.  
The coincident rate with the KW and the BAT 
triggers is $\sim$70 and $\sim$14 per year, respectively. 

There are two kinds of the WAM data for GRB analysis: transient (TRN) 
data and burst (BST) data. The BST data will be available only 
when the GRB trigger is produced by an on-board trigger system. 
The TRN data has 1 s time resolution and 55 energy channels for any time, 
while the BST data, for 64
seconds (from 8 s before to 56 s after a GRB trigger), 
has both 1/64 s time resolution and 4 energy channels,
and 1/2 s time resolution and 55 channels.
The WAM is inside the spacecraft, hence, incident gamma-rays
suffer from heavy absorption due to other satellite
materials.  Therefore, the WAM response is very complex, and heavily 
dependent on incident angles.  We have constructed a {\it Suzaku} mass 
model using GEANT4 \citep{agostinelli2003} 
and verified the Monte Carlo simulation results by comparison between 
the pre-launch calibration data and the experimental data taken during 
in-flight calibrations for certain directions. 

\begin{figure}
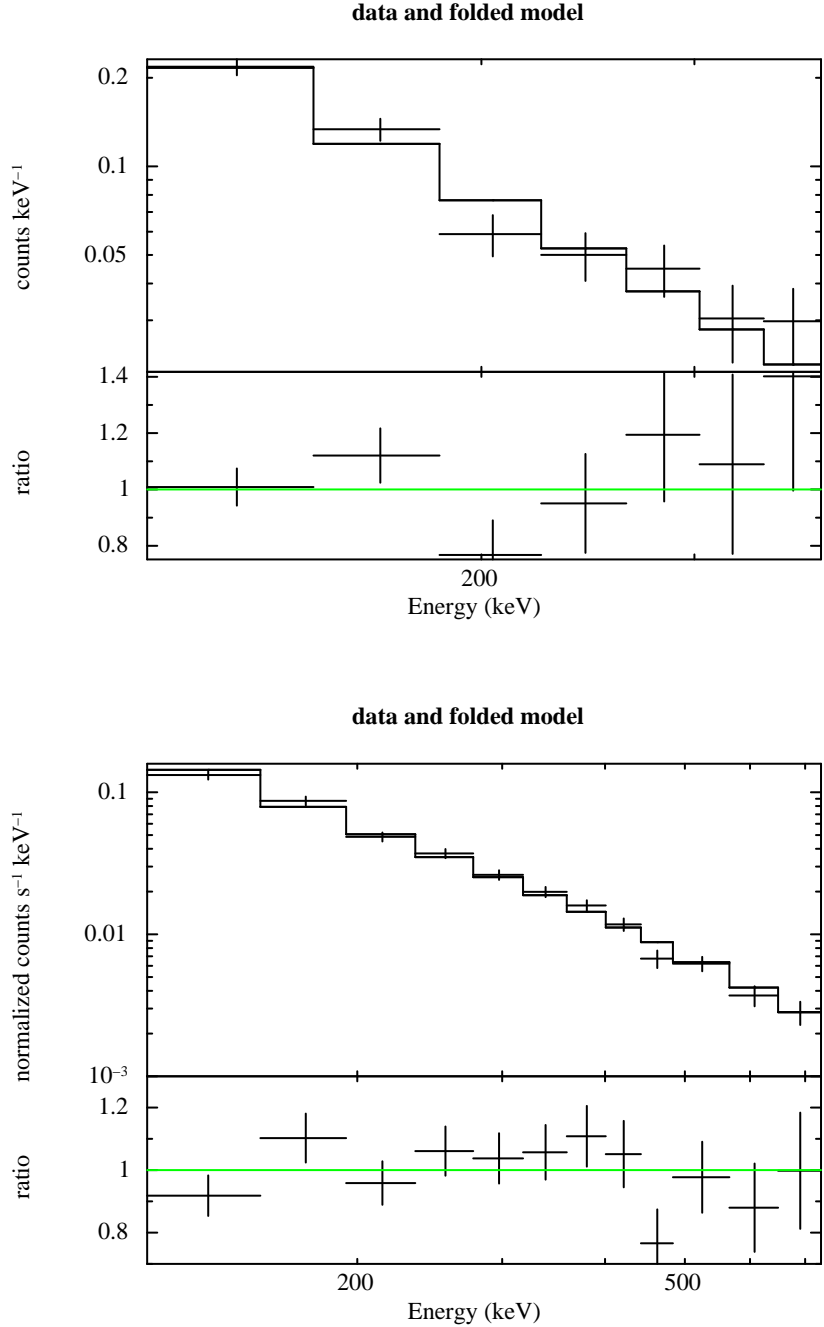

\begin{center}
\includegraphics[scale=0.42,angle=-90]{figure2a.eps}
\end{center}
\hspace{0.5cm}
\begin{center}
\includegraphics[scale=0.42,angle=-90]{figure2b.eps}
\end{center}
\caption{The WAM Crab spectra obtained by the earth occultation technique.  
Top panel shows the data 
integrated about one day. The Crab incident angle is about 20 degrees from the WAM 0 detector.  
Bottom panel is the three-year integration data. All of the data with incident angles smaller 
than 30 degrees from the normal direction of each detector are added.  
The energy response is averaged over each incident angle, weighting by the number of 
occultation.  The spectrum above 100 keV from one-day integration data is well fitted by a single power-law 
model with a photon index of $2.16 \pm 0.27$, and its energy flux in the 100-500 keV band 
is $9.9_{-1.6}^{+1.2} \times 10^{-9}$ erg cm$^{-2}$ s$^{-1}$ ($\chi^{2}$/d.o.f. = 6.96/5).  
For the three-year integration data, the spectrum is also well fit by a single power-law 
with a photon index of $2.23 \pm 0.09$, and its energy flux in the 100-500 keV band is 
$9.0_{-1.1}^{+0.9} \times 10^{-9}$ cm$^{-2}$ s$^{-1}$ by adding 7\% systematic errors for 
all PHA channels ($\chi^{2}$/d.o.f. = 11.79/10).  
}
\label{wam_crab}
\end{figure}

In the in-flight calibration, we calibrate the gain drift of the detectors  
after every SAA passage using the 511 keV annihilation line \citep{yamaoka2009}.  
For the energy response, the Crab spectrum obtained by the earth 
occultation technique is used to verify the energy response.  Figure \ref{wam_crab} shows 
the examples of the Crab spectra.  
We found that the WAM Crab spectrum is well fit by the single power-law model 
with a photon index of $\sim$2.1.  The energy flux in the 100-500 keV band is 
$\sim$$1 \times 10^{-8}$ erg cm$^{-2}$ s$^{-1}$.  This is roughly consistent 
with the value reported by other gamma-ray instruments (e.g. \cite{sizun2004}).  
%
However, in this Crab calibration, it is difficult to calibrate the energy 
response above several hundreds of keV for various incident directions.  
This is because the exposure time of the Crab data for 
a certain direction by this technique is strongly limited by the fact 
how long the satellite 
can keep the same attitude (typically a day).  Therefore, a GRB 
would be a suitable source to calibrate the energy response for various 
incident angles, and also, up to a higher energy range than the Crab.  You can find 
a detailed description of the Crab calibration and the earth occultation 
technique of the WAM data in \citet{kira2009}.  

\subsection{Swift/BAT}

The BAT is a highly sensitive (the effective area including the mask modulation 
is $\sim$ 1,400 cm$^{2}$ at 50 keV on-axis), large field of view (FOV) 
(2.2 sr for $>$ 10\% coded FOV), coded-aperture 
telescope that detects and localizes GRBs in real time.  The BAT 
detector plane is composed of 32,768 pieces of CdZnTe 
(CZT: $4 \times 4 \times 2$ mm), and the coded-aperture mask is 
composed of $\sim$ 52,000 lead tiles ($5 \times 5 \times 1$ mm) with 
a 1 m separation between mask and detector plane.  The energy range 
is 14--150 keV for imaging or mask-weighting, which is a technique 
to subtract the background based on the modulation resulting from 
the coded mask, with a non-coded response up to 350 keV.  Further 
detailed descriptions, references and the in-orbit calibration 
status of the BAT can be found in the BAT1 GRB catalog \citep{sakamoto2008}.  

The important update on the spectral calibration of the BAT since the BAT1 
GRB catalog has been published is the understanding of the problem in the 
energy response above 100 keV.  The BAT team investigated the ground calibration data more 
deeply and found that the mobility-lifetime products of electrons and holes ($\mu\tau$) 
which determine the characteristics of an individual CZT detector 
\citep{sato2005,suzuki2005} 
have to be 1.7 
larger than those originally determined.  After updating the $\mu\tau$ values, 
the BAT team confirmed that the adjustment which we were applying to reproduce the canonical 
Crab spectrum\footnote{The Crab canonical spectral parameters in the BAT energy range are $-2.15$ for 
the photon index and $2.11 \times 10^{-8}$ ergs cm$^{-2}$ s$^{-1}$ for the flux in the 
15-150 keV band.}
 above 100 keV is no longer needed.  Therefore, the BAT team reduces 
the systematic error above 100 keV to a flat 4\% (previously, the systematic error was 
4\% at 100 keV, and then increased to 12\% at 150 keV; see Figure 2 of 
\cite{sakamoto2008}).  The Crab data and the GRB data have been re-analyzed using 
the updated energy response and applying the new systematic error vector.  The spectral 
shape and the flux agree within the systematic uncertainties, which are mainly due to the off-axis 
response, to previous results 
(within $\sim$5\% in the photon index and within $\sim$10\% in the flux from the canonical Crab 
spectrum).  These changes in the energy response and the systematic error 
are available to the public (CALDB 20081026).  

Another important update is related to the gain change of the detectors.  
The BAT team has recognized shifts of peak energies in spectra from the 
on-board $^{241}$Am tagged source towards lower energies.  An analysis of four years 
of on-board $^{241}$Am spectra shows the shift is about 2.5 keV 
for the 59.5 keV peak.  The BAT team also noticed that the Crab data on 
January 21-23, 2009 shows a systematically steeper photon index of 0.05 and also a 
lower flux of about 5\% comparing to previous years.  Motivated by these results, 
the BAT team has developed new calibration files to store the information 
(coefficients to convert from PHA channel to energy) to correct this gain change 
as a function of time.  
After applying the new gain correction, the scatter of the 59.5 keV line energy is $\sim$0.1 keV 
over the four-year period.  Furthermore, no systematic trends in the photon index and the 
flux have been seen in the Crab data on 2009 by applying this new gain correction.  
The BAT team is planning to provide new calibration files every 6 months to correct 
for additional gain drifts as they may occur.\footnote{  
The calibration files will be 
available to the public in 2010 or early 2011.}

The results of the Crab spectral analysis based on these latest in-orbit calibrations are 
presented at \citet{sakamoto2010}.
We have used the updated energy response and the systematic error, and also the 
spectral file applying the new gain correction for this cross-calibration work.  

\section{Analysis}

We have selected 14 GRBs which were simultaneously detected by all three instruments, and also 
have enough statistics to perform spectral analysis.  We extracted 
36 spectra to do the joint analysis.  Table \ref{tab:grb_trigtime} 
shows the GRB samples in this work and the trigger time of each instrument in UTC in 
the form of $YYYY/MM/DD$ $hh:mm:ss.sss$ where $YYYY$ is year, $MM$ is month, $DD$ is day 
of month, $hh$ is hour, $mm$ is minute, and $ss.sss$ is second with three decimals.  
Table \ref{tab:list_grb} shows the GRB position in the sky from the BAT data \citep{sakamoto2008} 
and the incident angles of 
GRBs for each instrument in the unit of degrees.  
In Table \ref{tbl:fluence} and \ref{tbl:peakflux}, we summarize the energy fluence 
and the 1-s peak energy flux measured by KW, WAM and BAT for those samples.
Table \ref{tab:spec_interval} shows 
the spectral time intervals in the start time from the trigger time, t$_{\rm start}$ and 
the stop time from the trigger time, t$_{\rm stop}$ in the unit of seconds (the trigger time 
of each instrument is shown in Table \ref{tab:grb_trigtime}).  
Since the KW spectral data are binned by the flight software in a 
time variable manner, we adjusted the BAT and the WAM spectral intervals to that of the 
KW.  To determine the same time region with the KW spectrum, we calculated 
the arrival time difference of the GRB photons from the spacecraft to the Earth center 
(Time of flight; ToF) for each instrument.  The ToFs of each instrument are shown 
in the last three columns of Table \ref{tab:spec_interval} in the unit of seconds.

During the process of finding the spectral time interval, 
we tried to select at least one region for each burst 
where the $Swift$ spacecraft is not slewing.  Since the current BAT energy response generator, 
{\tt batdrmgen}, performs its calculations for a fixed single incident angle, it is necessary to 
do the joint spectral fits for a time interval when the BAT data are not affected by a spacecraft 
slew.  For the time intervals which include the spacecraft slew in the BAT data, 
we used the averaged energy response.  The procedure for creating the averaged BAT energy 
response is as followings:  We created the energy response for every 5 s period taking into 
account the position of the GRB in detector coordinates.  And then, we weighted these energy 
responses by the 5 s count rates and created the averaged energy response.  The 
spectral intervals which contain the spacecraft slew in the BAT data are indicated by the 
superscript ``S'' in the 2nd column of Table \ref{tab:spec_interval}.  

The joint spectral fits are performed by {\tt xspec} (ver. 11.3.2).  
The spectrum was fitted by a simple power-law (PL) model,
\begin{eqnarray}
f(E) = K_{100}^{\rm PL}\left(\frac{E}{100  \: {\rm keV}}\right)^{\alpha^{\rm PL}}
\label{eq:pl}
\end{eqnarray}
where $\alpha^{\rm PL}$ is the power-law photon index and $K_{100}^{\rm
PL}$ is the normalization at 100 keV in units of
photons cm$^{-2}$ s$^{-1}$ keV$^{-1}$, by a cutoff power-law (CPL) model,
\begin{eqnarray}
f(E) = K_{100}^{\rm CPL}\left(\frac{E}{100 \: {\rm keV}}\right)^{\alpha^{\rm CPL}}
\exp\left(\frac{-E\,(2+\alpha^{\rm CPL})}{\ep}\right)
\label{eq:cpl}
\end{eqnarray}
where $\alpha^{\rm CPL}$ is the power-law photon index, $\ep$ is the
peak energy in the $\nu$F$_{\nu}$ spectrum and $K_{100}^{\rm CPL}$ is the
normalization at 100 keV
in units of photons cm$^{-2}$ s$^{-1}$ keV$^{-1}$, and by the Band function 
\citep{band1993}, 
\begin{eqnarray}
f(E) = \left \{
\begin{array}{ll}
K_{1}\left(\frac{E}{100  \: {\rm keV}}\right)^{\alpha} \exp\left(\frac{-E\,(2+\alpha)}{\ep}\right)
&\quad E < \left(\frac{(\alpha - \beta) \ep}{(2+\alpha)}\right) \\
K_{1}\left(\frac{(\alpha-\beta)\ep}{(2+\alpha) 100  \: {\rm keV}}\right)^{\alpha-\beta} \left(\frac{E}{100  \: {\rm keV}}\right)^{\beta}  
&\quad E \geq \left(\frac{(\alpha - \beta) \ep}{(2+\alpha)}\right)
\end{array}
\right.
\label{eq:band}
\end{eqnarray}
where $\alpha$ is the low-energy photon index, $\beta$ is the high-energy photon index, 
$\ep$ is the peak energy in the $\nu$F$_{\nu}$ spectrum and $K_{1}$ is the
normalization at 100 keV in units of photons cm$^{-2}$ s$^{-1}$ keV$^{-1}$.  We multiply 
the model by the constant factor to understand the calibration uncertainties among the 
instruments.  The constant factor for the KW data is fixed 
to 1, and the constant factors of the BAT and the WAM are kept as free parameters.  However, we 
fix the BAT constant factor to 1, and keep the WAM constant factor as a free parameter 
in the case of the BAT and WAM joint fit.  
Other spectral parameters such as photon indices, $\ep$ and a normalization are the same 
for all instruments.  If a GRB has been observed by two WAM detectors, 
we used both data in the joint fit analysis.  We call WAM$_{SP1}$ and WAM$_{SP2}$ for each 
detector throughout the paper.  The actual detector ID number of WAM$_{SP1}$ and 
WAM$_{SP2}$ are summarized in Table \ref{tab:list_grb}.  

\subsection{KW data analysis}
The KW data are processed using standard KW analysis tools, 
which convert the spectral data collected on-board from the 
internal format to PHA-files in FITS format suitable for analysis in
{\tt xspec}.  
The dead time correction has been applied.  
For the given incident angle, the DRM
is calculated by linear interpolation of the DRMs for the
nearest angles. Since the KW background is very stable, we 
use an average spectrum over a $\sim$100--300 s long interval after
the burst as the background spectrum.

\subsection{WAM data analysis}

The {\it Suzaku} WAM data were analyzed using HEADAS version 6.6.2.   
Spectral accumulations were carried out by {\tt hxdmkwamspec} and {\tt hxdmkbstspec} 
for the TRN data and BST data, respectively. The deadtime was corrected.  
For two bright bursts: GRB 061007 and GRB 080328, we added 5 \%
systematic errors to the energy channels lower than 400 keV, taking into account the 
ADC digitization errors.
The instrumental background varies with time, even during GRB time
intervals, hence we used a model background which interpolates the best-fit 
4th-order polynomial function of the data in each 500 s before and after 
the time intervals.  The energy response was calculated by the latest response generator
(version 1.9) based on the incident angles shown in Table 
\ref{tab:list_grb}. It cannot reproduce the low energy spectrum 
for energies less than 120 keV, corresponding to the 3 lowest energy channels.  
Thus, the fitting energy range was limited above $\sim$120 keV.  

\subsection{BAT data analysis}
We used HEAsoft 6.5 and the latest CALDB for the BAT data processing.
The BAT event-by-event data, which have a time resolution of 98 $\mu$sec \citep{palmer2005}, 
are used in the analysis.  
Since all of the GRBs in our sample have not been exceeding 
$2.6 \times 10^{6}$ s$^{-1}$ in the raw count-rate, the deadtime effect is 
negligible in the 
BAT data (see Palmer et al. 2005 for the details about the deadtime effect in 
the BAT data).
The spectral file (PHA file) is created 
with {\tt batbinevt} 
specifying the time interval (tstart and tstop options).  
The tool {\tt batphasyserr} is used to apply the systematic error into the PHA files.  We run 
{\tt batdrmgen} to create the 
energy response file.  We exclude the BAT data of the Reg2 and 
the Reg12 of GRB 051008 and the whole intervals of GRB 060124 from the cross-calibration 
work because the event-by-event data are not available for these intervals.  
Except for these intervals, the event-by-event 
data are available to create the spectrum.  Therefore, it is possible 
to choose exactly the same time interval selected by other missions for the BAT spectrum 
using event-by-event data.  

\section{Results}

\begin{figure}
\begin{center}
\includegraphics[scale=0.35,angle=90]{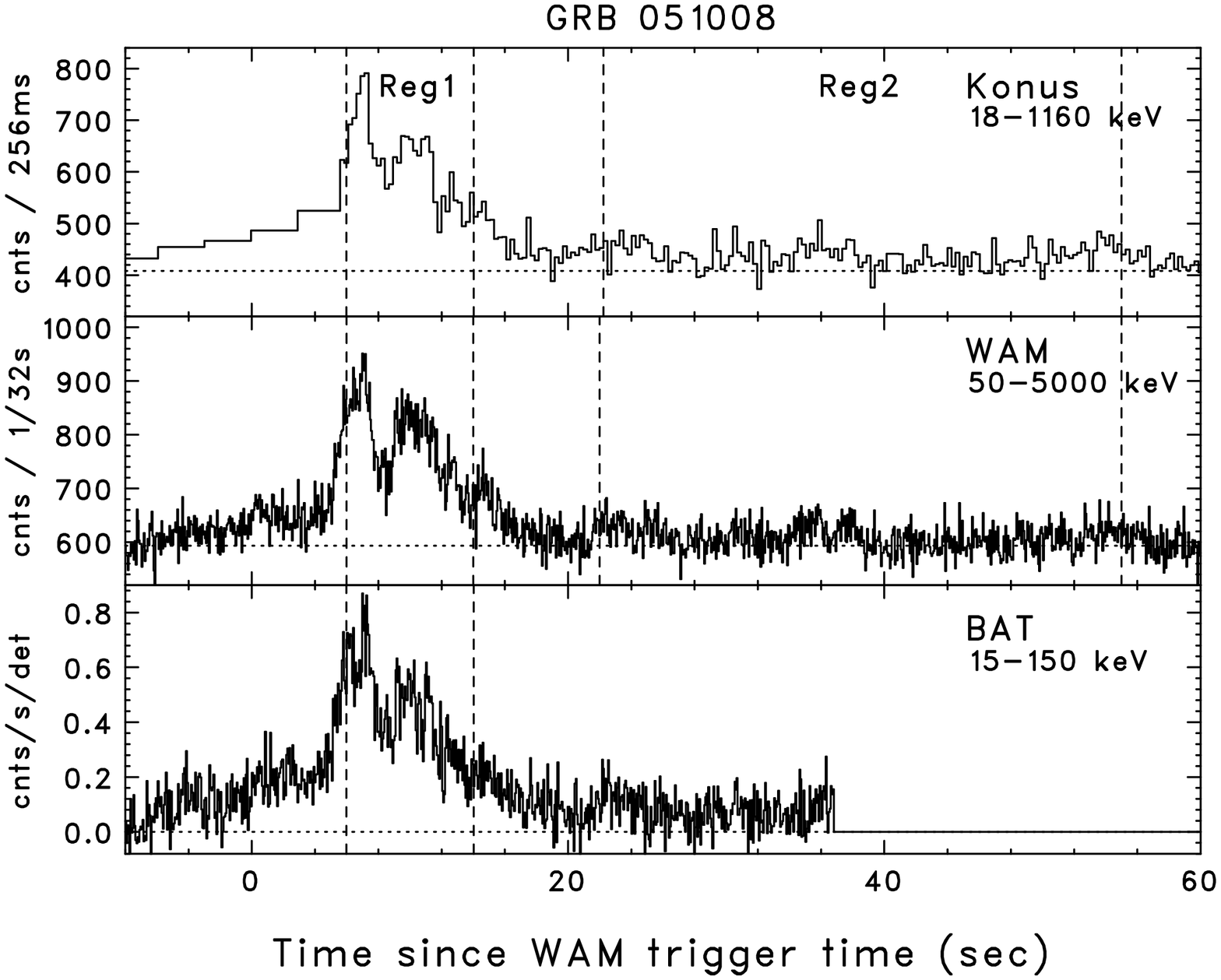}
\hspace{5mm}
\includegraphics[scale=0.35,angle=90]{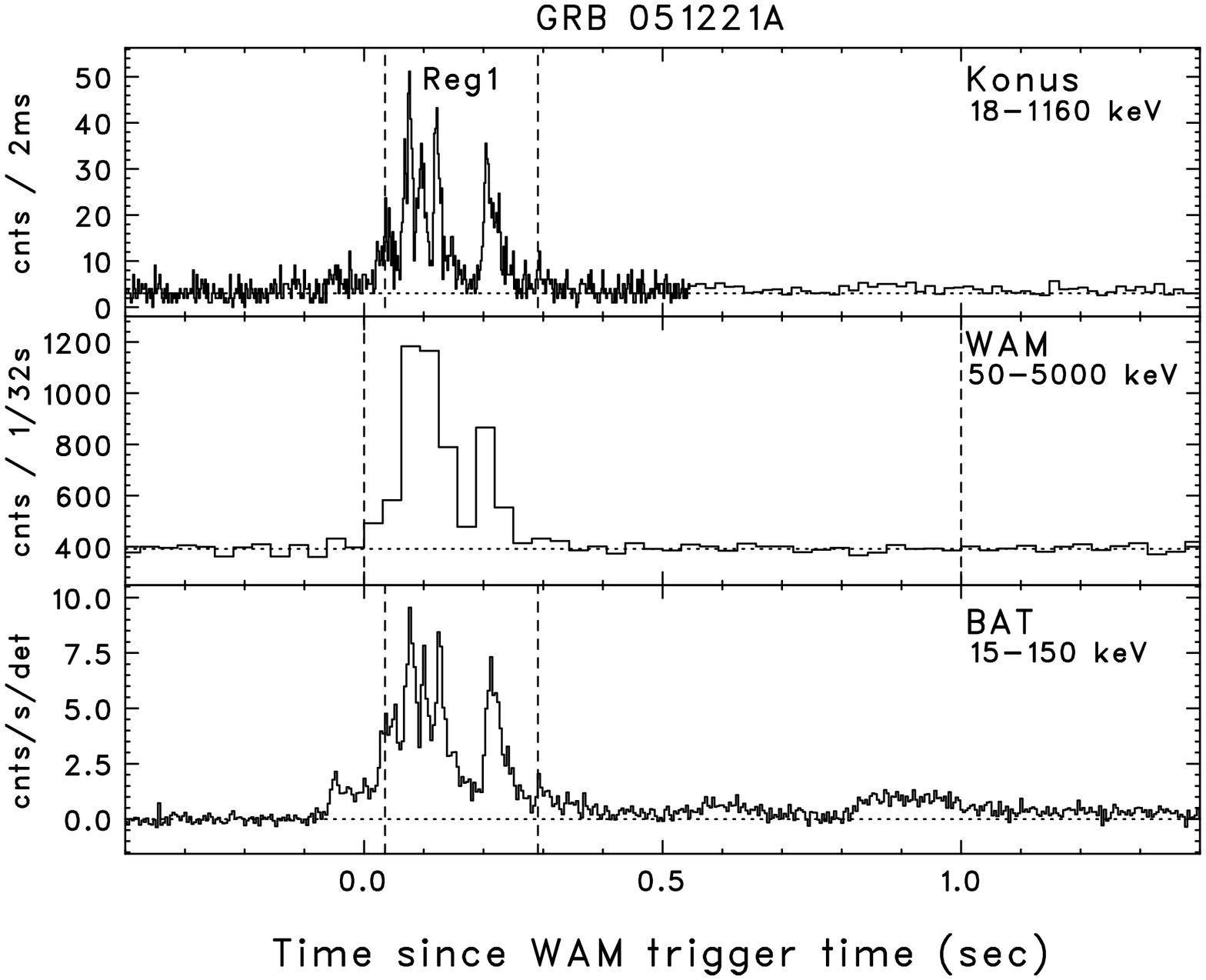}
\end{center}
\begin{center}
\includegraphics[scale=0.35,angle=90]{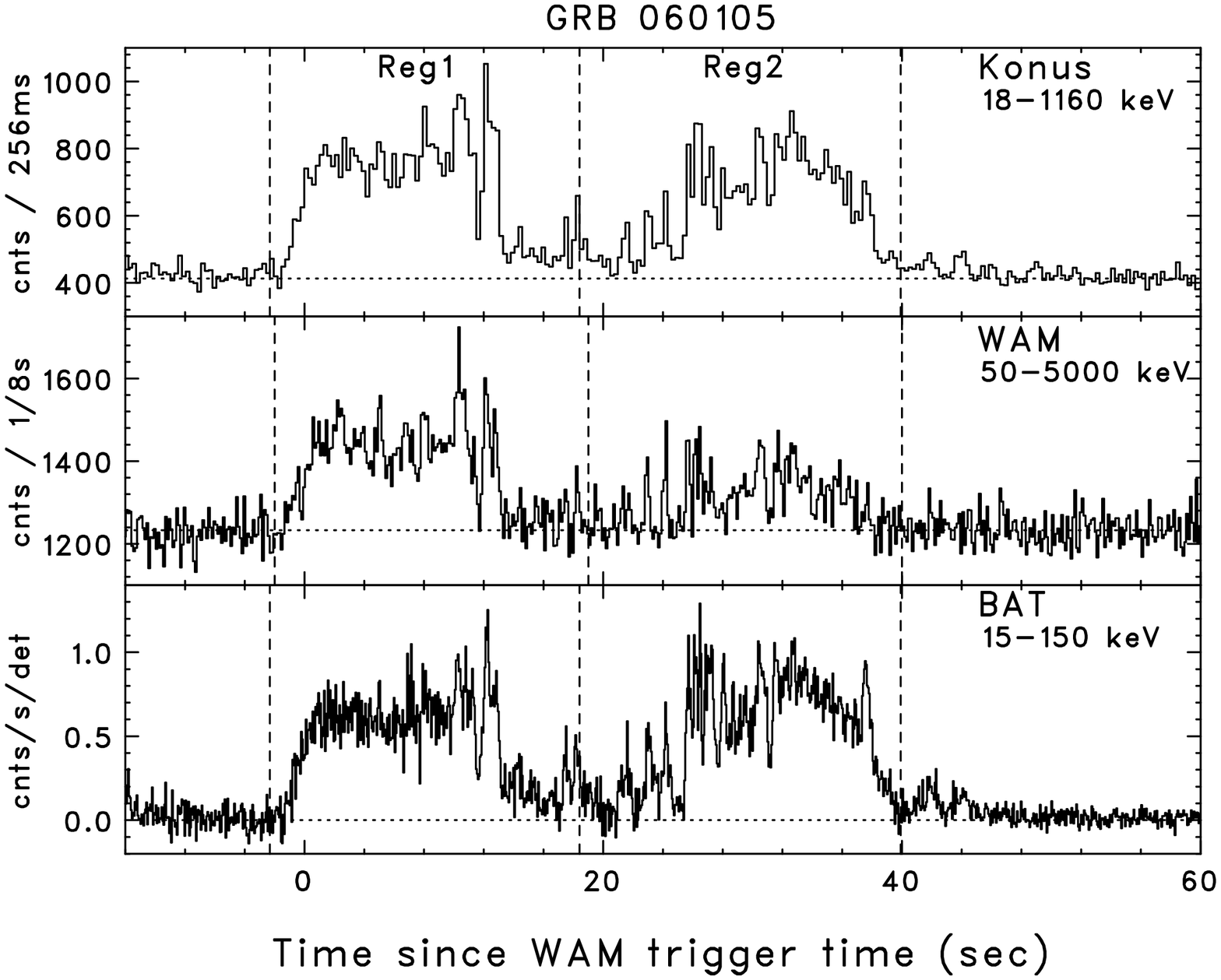}
\hspace{5mm}
\includegraphics[scale=0.35,angle=90]{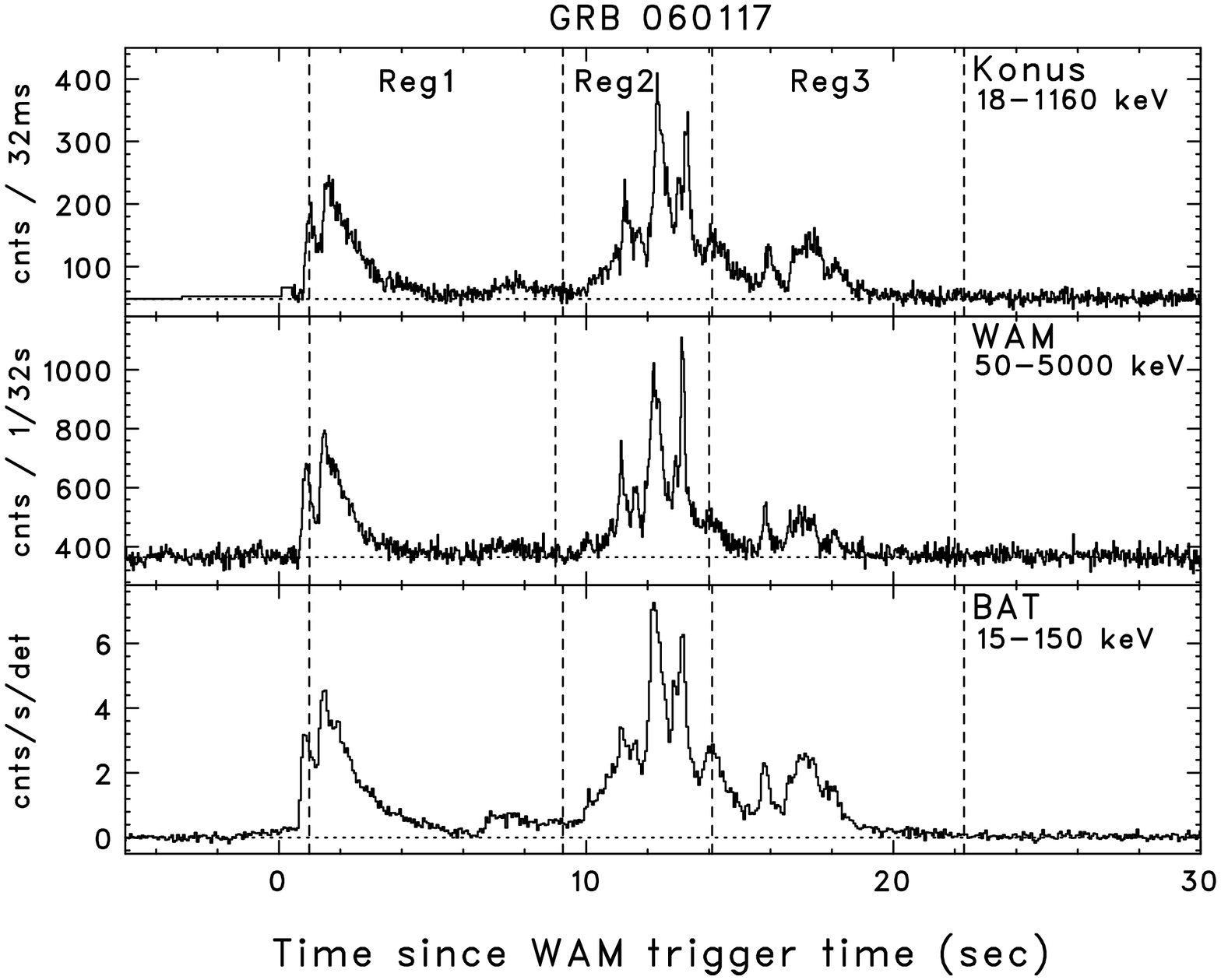}
\end{center}
\begin{center}
\includegraphics[scale=0.35,angle=90]{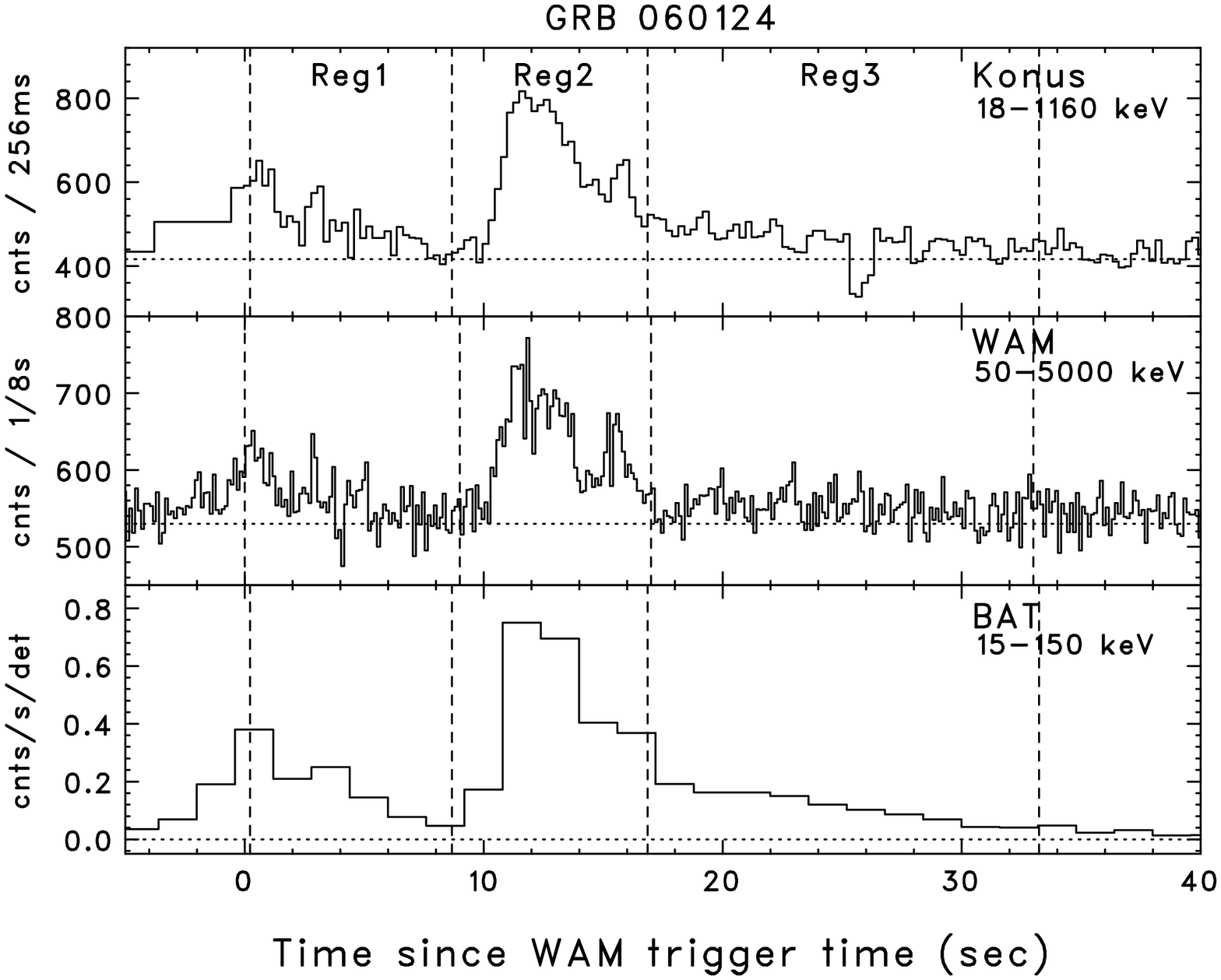}
\hspace{5mm}
\includegraphics[scale=0.35,angle=90]{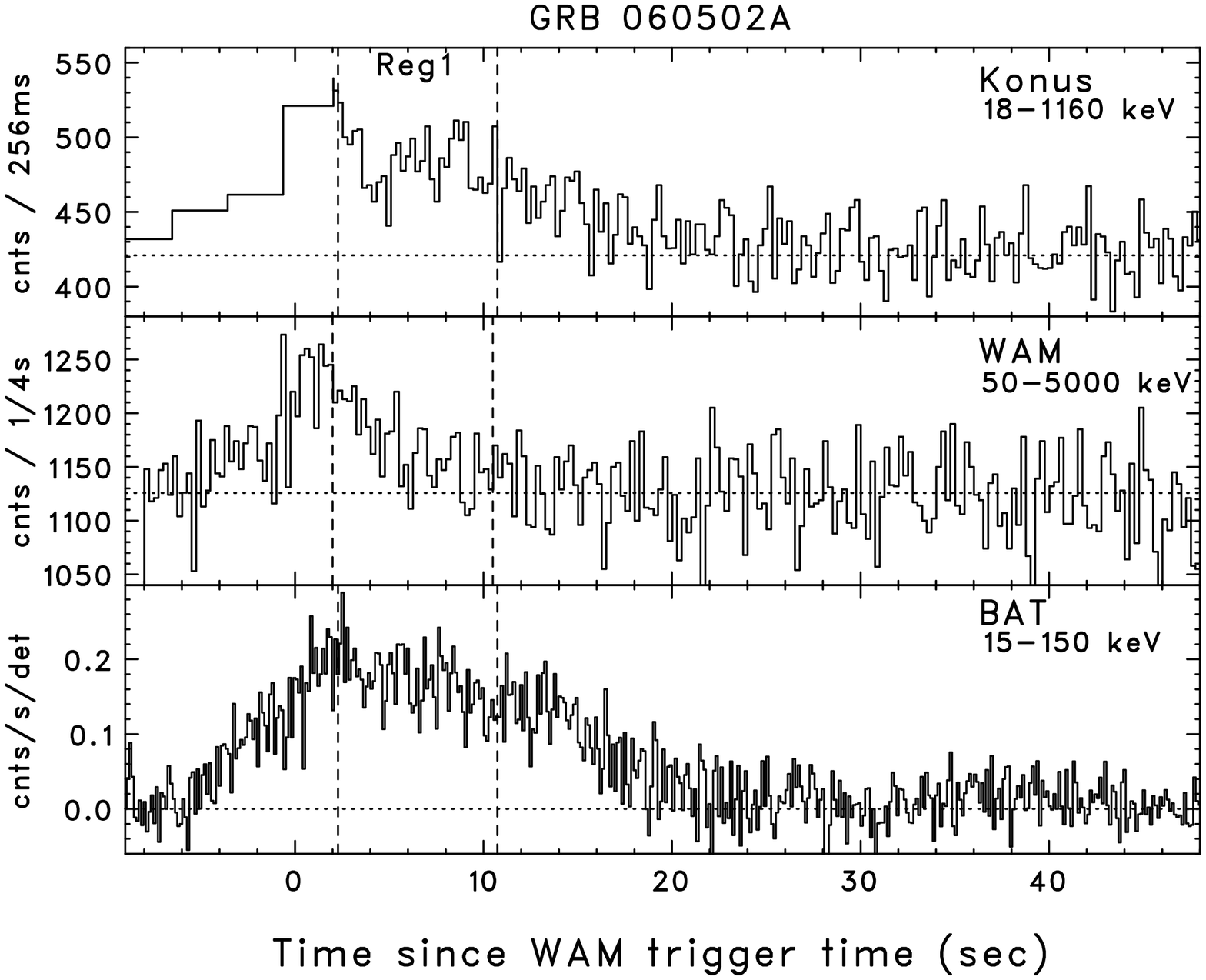}
\end{center}
\caption{Light curves of the KW (top), the WAM (middle), 
and the BAT (bottom) of our samples.  Note that the KW light curves 
before the KW trigger times were recorded in the 
waiting mode with 2.944 s time resolution.}\label{fig:LC1}
\end{figure}

\begin{figure}
\begin{center}
\includegraphics[scale=0.35,angle=90]{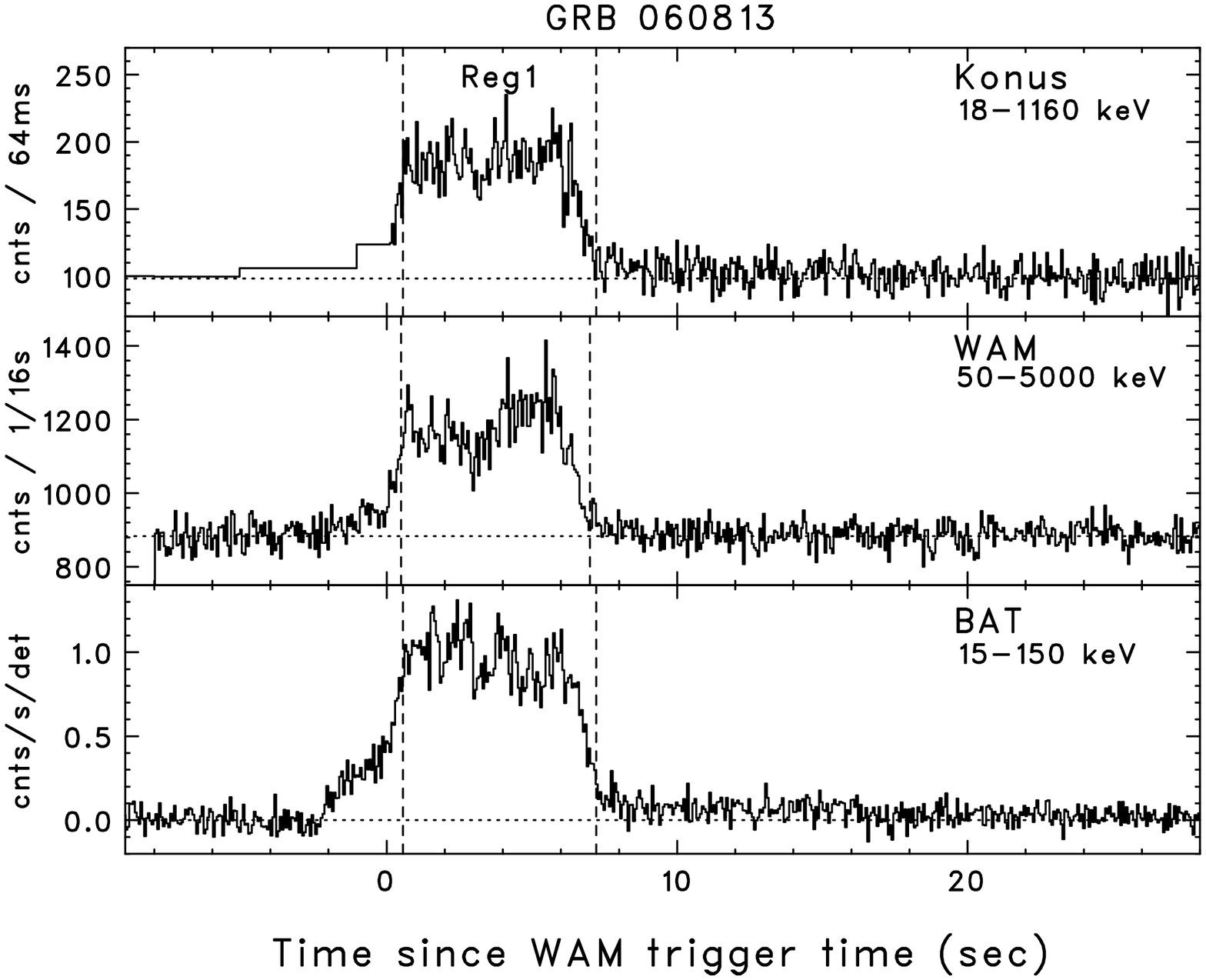}
\hspace{5mm}
\includegraphics[scale=0.35,angle=90]{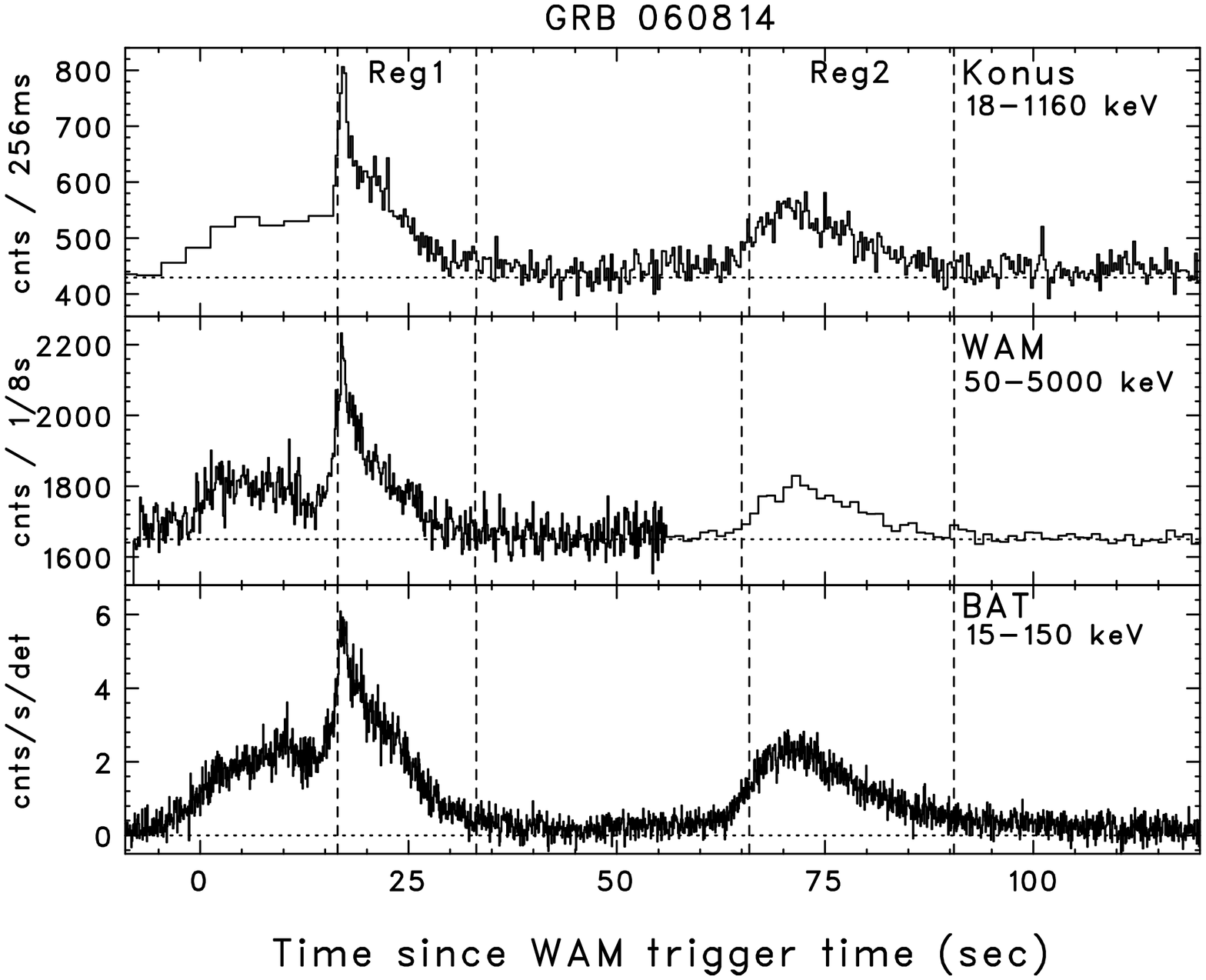}
\end{center}
\begin{center}
\includegraphics[scale=0.35,angle=90]{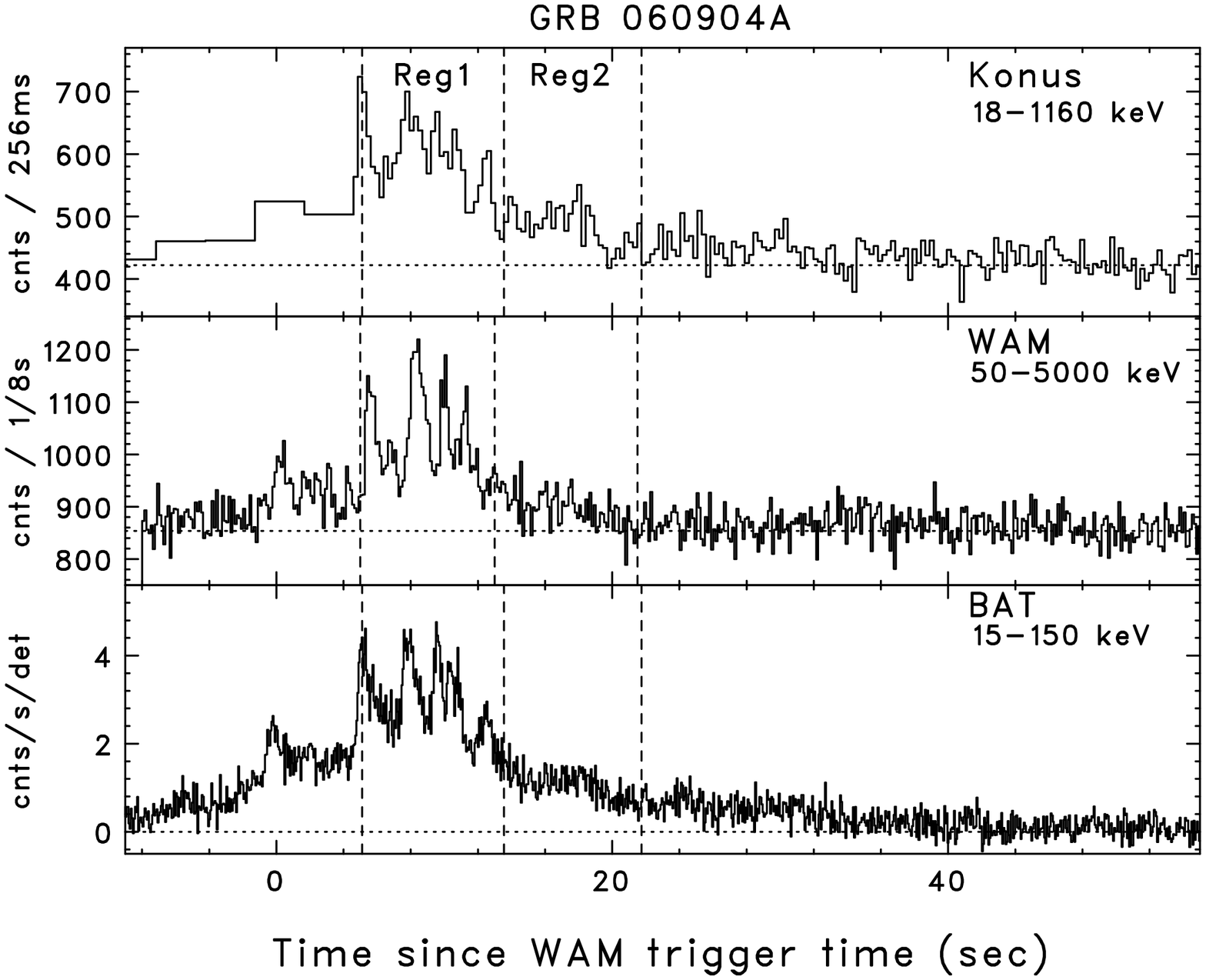}
\hspace{5mm}
\includegraphics[scale=0.35,angle=90]{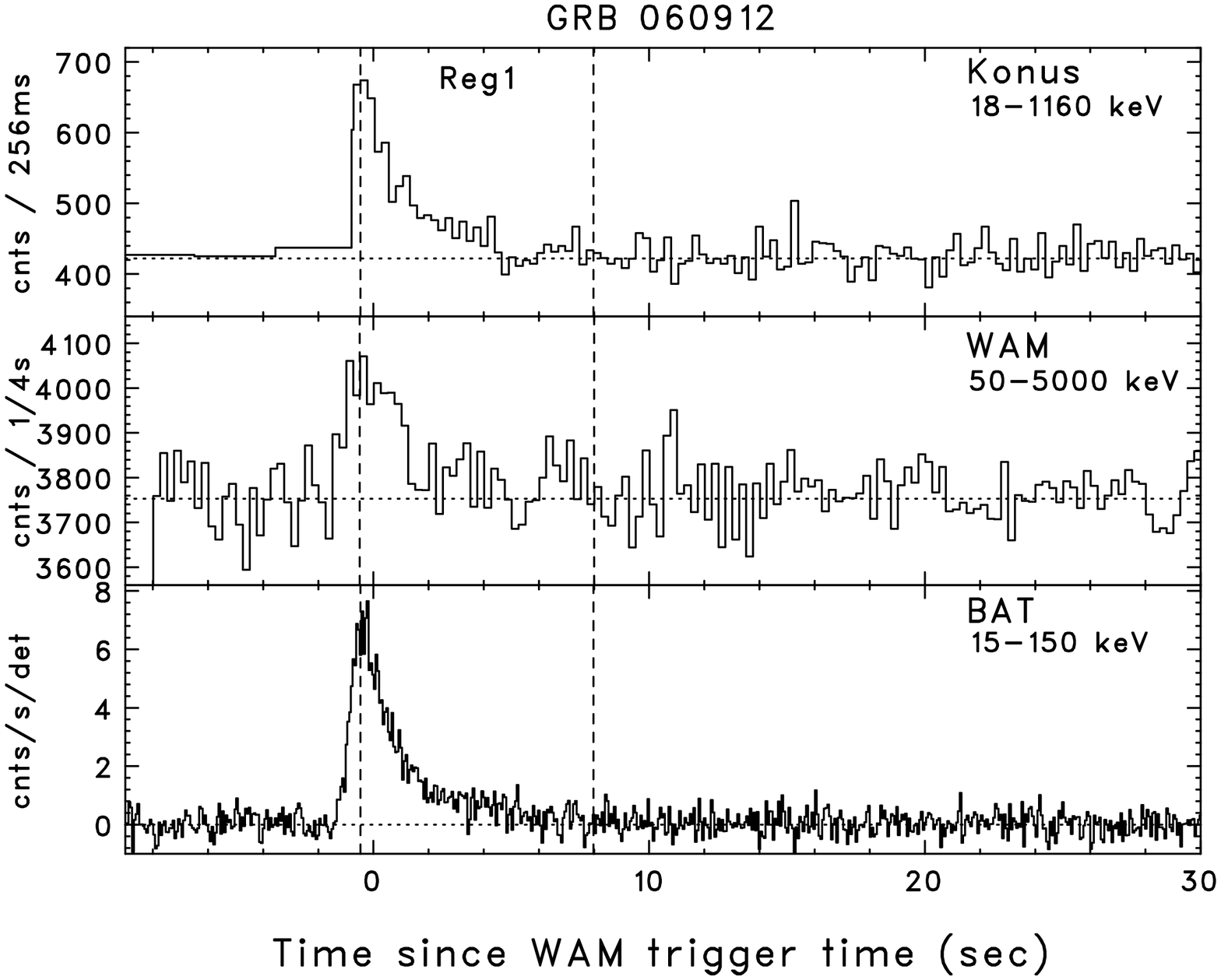}
\end{center}
\begin{center}
\includegraphics[scale=0.35,angle=90]{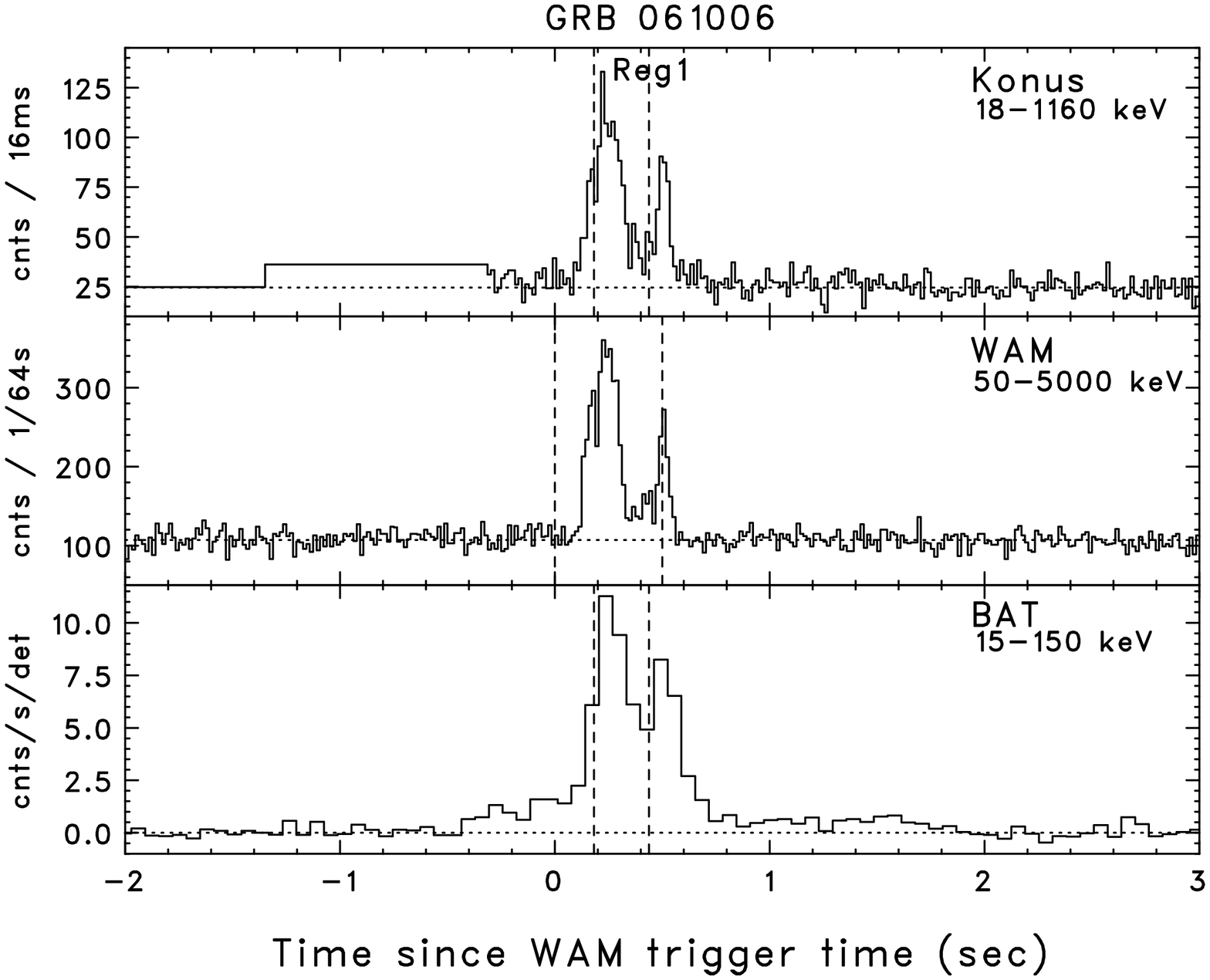}
\hspace{5mm}
\includegraphics[scale=0.35,angle=90]{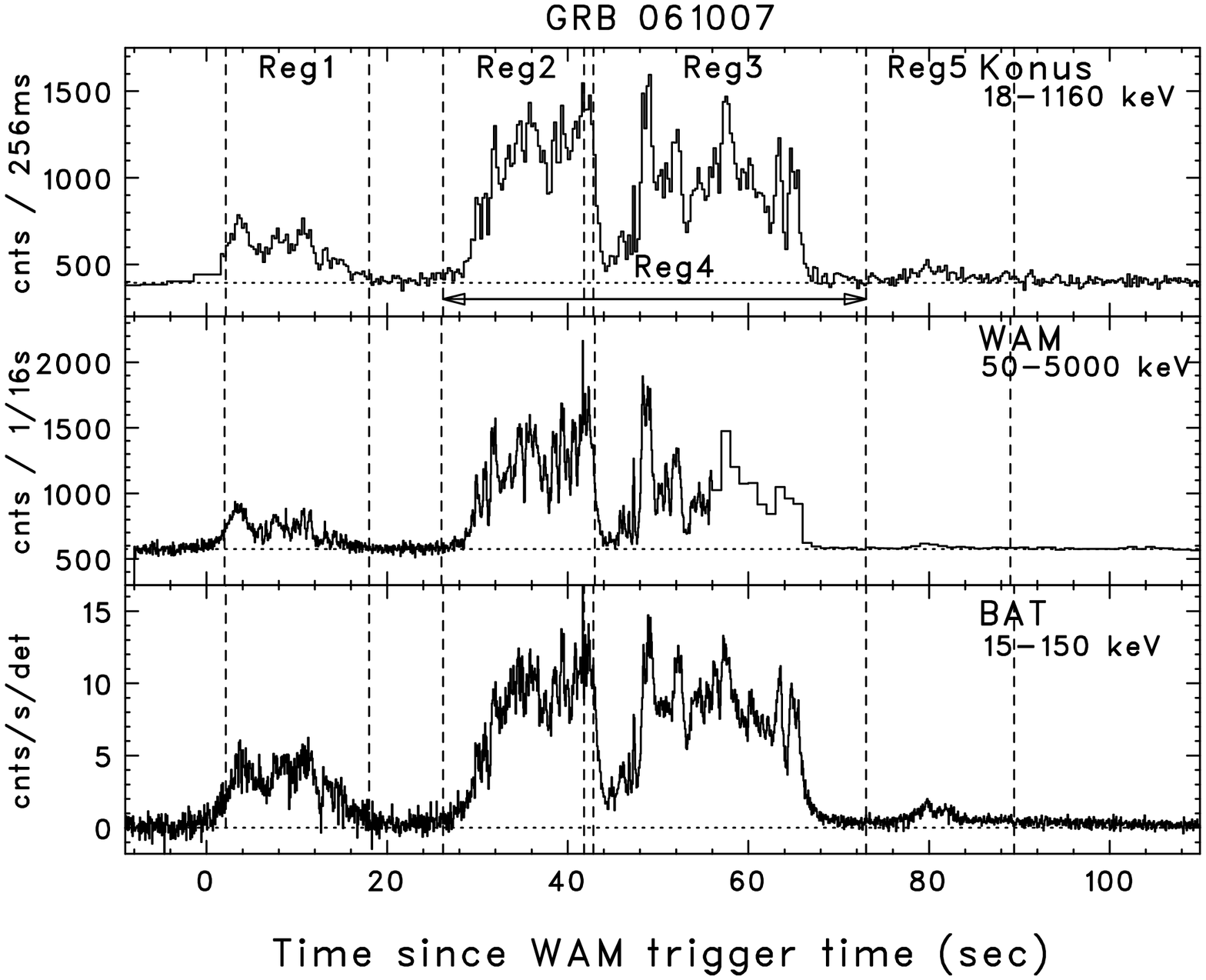}
\end{center}
\caption{Light curves (continued)}\label{fig:LC2}
\end{figure}

\begin{figure}
\begin{center}
\includegraphics[scale=0.35,angle=90]{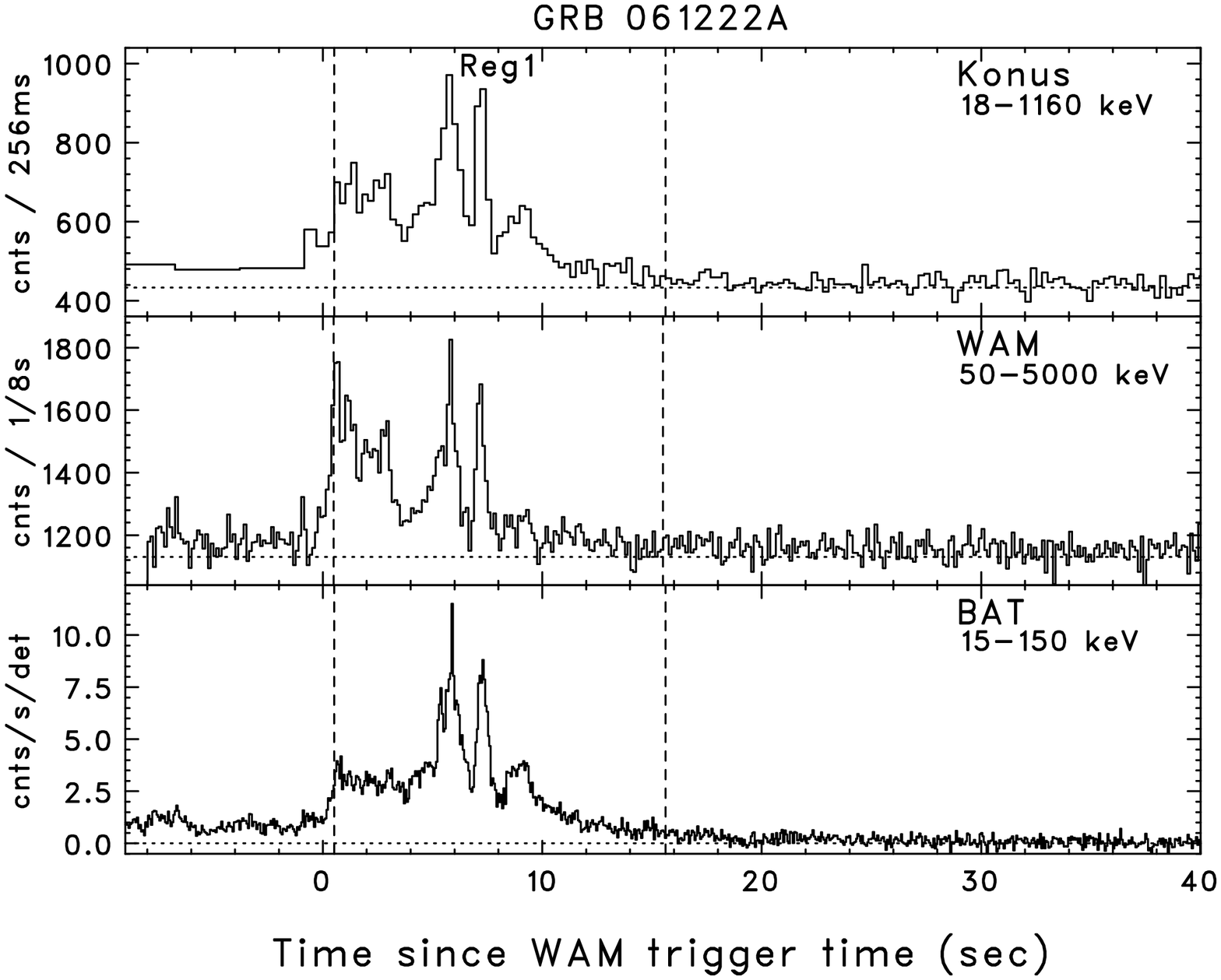}
\hspace{5mm}
\includegraphics[scale=0.35,angle=90]{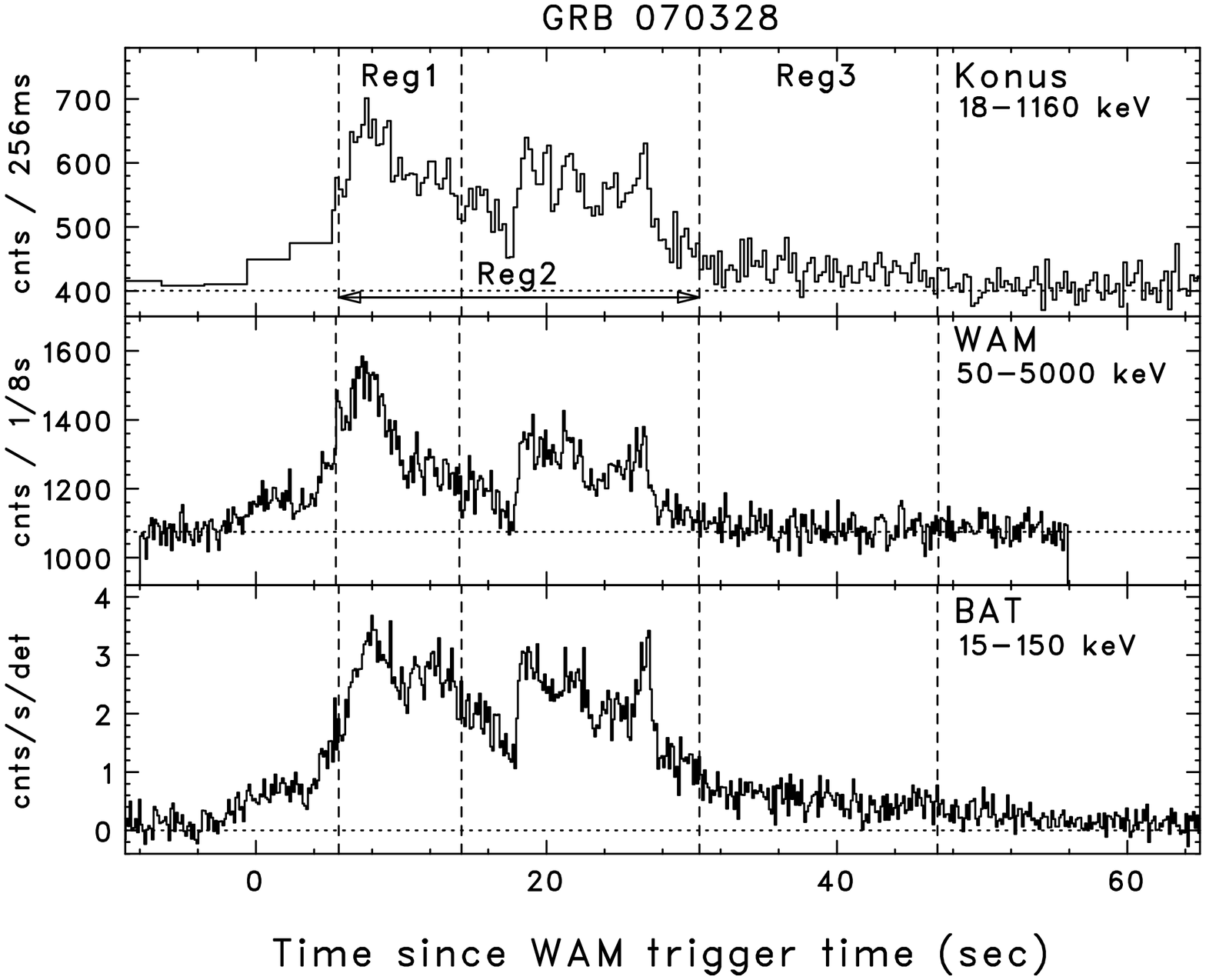}
\end{center}
\caption{Light curves (continued)}\label{fig:LC3}
\end{figure}

\begin{figure}
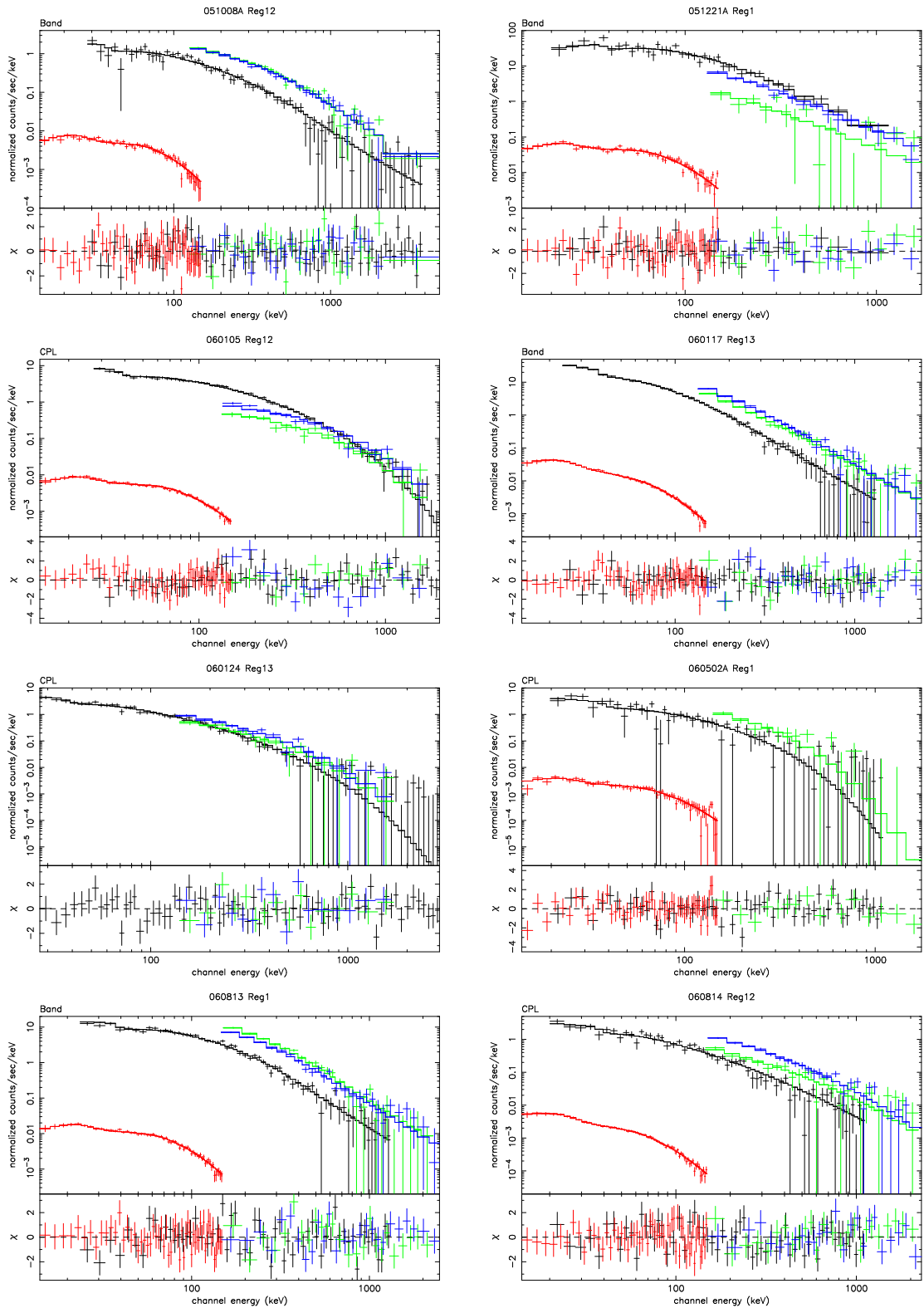

\begin{center}
\includegraphics[scale=0.3,angle=-90]{figure6a.eps}
\hspace{5mm}
\includegraphics[scale=0.3,angle=-90]{figure6b.eps}
\end{center}
\begin{center}
\includegraphics[scale=0.3,angle=-90]{figure6c.eps}
\hspace{5mm}
\includegraphics[scale=0.3,angle=-90]{figure6d.eps}
\end{center}
\begin{center}
\includegraphics[scale=0.3,angle=-90]{figure6e.eps}
\hspace{5mm}
\includegraphics[scale=0.3,angle=-90]{figure6f.eps}
\end{center}
\begin{center}
\includegraphics[scale=0.3,angle=-90]{figure6g.eps}
\hspace{5mm}
\includegraphics[scale=0.3,angle=-90]{figure6h.eps}
\end{center}
\caption{The joint fit spectra of the KW (black), the WAM (blue 
and green) and the BAT (red).  See text for details (\S 4).}\label{fig:SP1}
\end{figure}

Figures \ref{fig:LC1}-\ref{fig:LC3} show the light curves of the KW, the WAM and the BAT 
instruments.  The vertical lines on the figures correspond to the intervals from 
which we extract the spectra to perform the joint fits.  Figure \ref{fig:SP1}-\ref{fig:SP2} 
show examples of the KW (black), the WAM$_{SP1}$ (green), the WAM$_{SP2}$ (blue) and the BAT (red) 
observed spectra along with the best fit model as solid lines and the residuals from the best fit model 
in the bottom panel.  
The top label of each figure describes the GRB name and the interval of the fit.  The best fit model 
is shown at the left top of the figure.  
As shown in Figure \ref{fig:chi2}, the distribution of the reduced $\chi^{2}$ of the KW fit and 
all the joint fits based on the best fit model is centered around 1.  
The systematically smaller reduced $\chi^{2}$ seen in the inclusion of the BAT data is 
due to apply a large systematic error in the BAT spectral data.  
The linear fit to the data (a linear coefficient and an offset) between the best fit parameter 
obtained by the joint fit and the KW fit (results of Fig \ref{fig:kw_bat_cpl}, \ref{fig:kw_bat_band}, 
\ref{fig:kw_wam_cpl}, \ref{fig:kw_wam_band}, \ref{fig:wam_bat_cpl}, \ref{fig:wam_bat_band}, 
\ref{fig:kw_wam_bat_cpl}, and \ref{fig:kw_wam_bat_band})
are shown in Table \ref{tab:fit_param_cpl} for a CPL and 
in Table \ref{tab:fit_param_band} for the Band function.  
The spectral parameters and the fluxes of each spectral interval 
and instruments are presented in Table \ref{tbl:051008_reg1_spec}-\ref{tbl:070328_reg13_flux}.  
We report the instrument which fixes the constant factor to 1 for calculating the flux 
in the last column of the flux tables.  We are not reporting the spectral parameters of the 
model which does not constrain 
the parameters (e.g. parameters of the Band function fit based on the BAT data alone.)

\subsection{KW and BAT joint fit}

Figure \ref{fig:kw_bat_cpl} and \ref{fig:kw_bat_band} summarize the relationship 
between the best fit spectral 
parameters from the joint KW-BAT fit and the fit made by the KW data alone, 
and between the constant factor of BAT and the spectral parameters.  The results 
based on a CPL model 
and the Band function are shown in Figure \ref{fig:kw_bat_cpl} and in Figure 
\ref{fig:kw_bat_band} respectively.  
The BAT spectra which are not affected by the spacecraft slew are shown in blue in the 
figures.  As seen in the figures, there is no systematic trend between the spectral data 
affected by the spacecraft slew and not affected by the spacecraft slew.  Therefore, we 
concluded that there is no systematic problem in using the BAT spectra data during 
the slew using our weighted energy response.  

Most of the photon indices based on the joint fits agree with the KW best fits values 
within the uncertainties.  However, 
we do see a systematic trend that the joint fit photon indices are systematically 
steeper (smaller values) than in the KW fit.  
This systematic difference in the indices is 0.1 steeper for a CPL fit and 0.2 
steeper for the Band function fit on average as seen in Table \ref{tab:fit_param_cpl} and 
\ref{tab:fit_param_band}.  
Overall $\ep$ agrees between the joint fit and the KW fit.  
Although it is not well constrained, the high energy photon index $\beta$ of the Band function 
fit agrees between the joint fits and the KW fit.  
The most noticeable systematic difference in the joint fits is the systematically smaller 
constant factor for the BAT data.  The BAT constant factor is $\sim$12\% 
smaller than that of the KW.  
As can be seen in the bottom panels of Figure \ref{fig:kw_bat_cpl} 
and \ref{fig:kw_bat_band}, there is no systematic trend between the 12\% smaller BAT 
constant factor and the best fit photon indices, $\ep$, or the best fit spectral model.  
Since this trend has been seen for the most of the combination of joint fits 
using the BAT data as we describe in the following sections, 
we believe that this trend of the constant factor in the BAT data is 
intrinsic to the BAT energy response.  

To investigate the consistency in the spectral parameters especially $\alpha$ and 
$\ep$ derived from the BAT data alone
we select 
the fitting results which show the difference in chi-squared between a PL 
and a CPL fit to be greater than 6 ($\Delta \chi^{2} = \chi^{2}_{\rm PL} - \chi^{2}_{\rm CPL} > 6$) 
in the BAT fit.  This is the same $\Delta \chi^{2}$ criterion used 
in the BAT1 catalog (\cite{sakamoto2008}).  
As demonstrated in \citet{sakamoto2009}, the BAT spectrum alone can determine 
$\ep$ if the observed spectrum has a sufficiently high signal-to-noise ratio and also $\ep$ is 
located inside the BAT energy range.
The left and right panels of Figure \ref{fig:bat_comp_alpha_ep} 
show $\alpha$ and $\ep$ derived by the KW-BAT joint fit versus the BAT fit 
in a CPL model.  We do see a similar trend with the KW-BAT and the KW fit of 
$\sim$0.1-0.2 steeper $\alpha$ and $\sim$10\% higher $\ep$ on average between the KW-BAT and 
the BAT fits.  However, most of the $\alpha$ and $\ep$ values derived from the BAT data alone 
are consistent within the errors with the KW-BAT joint fit values.  We note that there is 
one spectrum, GRB 051008 Reg1, which show $\ep$ $\sim$800 keV in the KW-BAT joint fit (same as in 
the KW fit and in the KW-WAM-BAT fit), but $\ep$ 
of the spectrum is $\sim$200 keV in the BAT fit.  Since $\ep$ of this spectrum is around 
$\sim$800 keV in the different combination of joint fits, we believe that $\ep$ of these spectra 
are greater than 200 keV.  Therefore, we should use caution if $\ep$ derived by 
the BAT data alone is near the upper boundary of the effective BAT energy range of $\sim$150 keV.  
We also note that the values derived with the BAT spectrum alone 
are consistent with those derived with BAT-KW joint fit, only when the fit is done 
with a CPL model and $\ep$ can be determined.

\subsection{KW and WAM joint fit}

Figure \ref{fig:kw_wam_cpl} and \ref{fig:kw_wam_band} summarize the joint fit 
results of the KW and the WAM data.  
The parameters $\alpha$ and $\ep$ in both the CPL fit and the Band fit are consistent 
between the joint fit and the KW fit.  The difference in $\alpha$ is well 
less than 0.1 on average for both a CPL fit and the Band fit.  Furthermore, according to 
a linear fit of parameters between the KW-WAM fit and the KW fit, 
the difference is relatively small:  6\% in $\ep$ 
for both the CPL fit and the Band fit on average.  
The high energy photon index $\beta$ of the Band function is also consistent.  
The systematically small constant factor $\sim$0.2 of GRB 051221A is due to the four 
times larger time interval in the WAM data (see Table \ref{tab:spec_interval}).  
Also note that the WAM$_{SP2}$ constant factor (WAM detector ID 3) of GRB 060105 
is in the range of 2-3.  This is due to the large uncertainty of the energy response at the 
extreme off-axis incident angle\footnote{A small incident 
angle (on-axis) means that the source is coming from an off-axis angle for 
the WAM because the detector normal is at $\theta=90^{\circ}$.}
 ($\theta=5.7^{\circ}$).  

\subsection{BAT and WAM joint fit}

Figure \ref{fig:wam_bat_cpl} and \ref{fig:wam_bat_band} summarize the joint fit 
results of the BAT and the WAM data.  
The general trend in $\alpha$ and $\ep$, which we mentioned in section 4.1 
(the KW- BAT joint fit) is clearly seen in this joint fit.  The joint 
fit $\alpha$ are 0.2-0.3 steeper on average in the joint fit than in the KW fit for both 
the CPL model and the Band function.  
Moreover, the joint fit $\ep$ is systematically higher by 11\% for 
the CPL fit and 14\% for the Band fit on average based on a linear fit of parameters.  
The high energy photon index $\beta$ of the Band function fit agrees between the 
joint fits and the KW fit 
if we take into account the large uncertainties in the parameter.  

\subsection{KW, WAM and BAT joint fit}

Figure \ref{fig:kw_wam_bat_cpl} and \ref{fig:kw_wam_bat_band} summarize the joint 
fit results of all three instruments.  The systematic differences in $\alpha$ 
and $\ep$ which we see in the KW-BAT and the BAT-WAM joint fits are also apparent 
in this joint fit.  
$\alpha$ obtained by the joint fit is 0.1 and 0.2 steeper on average than the KW fit 
for the CPL fit and the Band fit respectively.  
There are eight spectral regions which the 90\% confidence level of $\alpha$ 
do not overlap between the KW and the KW-WAM-BAT joint fit for the CPL 
(Table 11, 18, 24, 30, 34, 35, 36 and 38) and the Band fit (Table 24, 33, 34, 35, 36, 38, 41 and 43).  
The largest un-overlapping confidence level in $\alpha$ between the KW and the KW-WAM-BAT joint fit 
is seen in GRB 051221A Reg1 of 0.04 for the CPL fit, and GRB 070328 Reg2 and Reg13 of 0.06 for 
the Band fit.  On averaged, the un-overlapping confidence level is 0.03 for the CPL fit and 0.04 
for the Band fit. 
According to a linear fit of parameters, $\ep$ based on the joint fit is 9\% and 12\% higher 
on average than the KW fit for the CPL model  
and the Band function respectively.  
There are ten and six spectral regions which the 90\% confidence level of $\ep$ 
do not overlap between the KW and the KW-WAM-BAT joint fit for the CPL 
(Table 26, 28, 30, 33, 34, 35, 36, 38, 41 and 43) 
and the Band fit (Table 34, 35, 36, 38, 41 and 43) 
respectively.  The largest un-overlapping confidence level in $\ep$ between the KW and the KW-WAM-BAT joint fit is 
seen in GRB 070328 Reg13 of 51 keV for the CPL fit, and GRB 070328 Reg2 of 151 keV for 
the Band fit.  Also, GRB 070328 Reg2 shows the relatively large gap of 41 keV in the CPL fit, 
and, similarly, GRB 070328 Reg13 has 142 keV gap in the Band fit.  Excluding those four outliers, 
the averaged un-overlapping confidence level is 15 keV for the CPL fit and 13 keV for the Band fit.
The constant factor of the BAT data is systematically smaller by 10\% in both 
a CPL fit and the Band fit.  
The WAM constant factor is smaller by 5\% for a CPL fit and 2\% for the Band fit 
excluding the outliers of GRB 051221A and GRB 060105 (see section 4.2).  

\section{Discussion and Conclusion}

We report the results of our attempt to use the simultaneously 
observed GRBs to cross-calibrate the energy response of the KW, the WAM and 
the BAT instruments.  Since every GRBs has a different brightness and spectral 
shape, it is difficult to use the observed GRB spectrum as 
``a standard candle'' to calibrate the instruments.  However, we can reveal the 
systematic problems among the instruments by comparing the best fit parameters 
derived for each instrument and from the joint analysis of the data.  

\begin{figure}
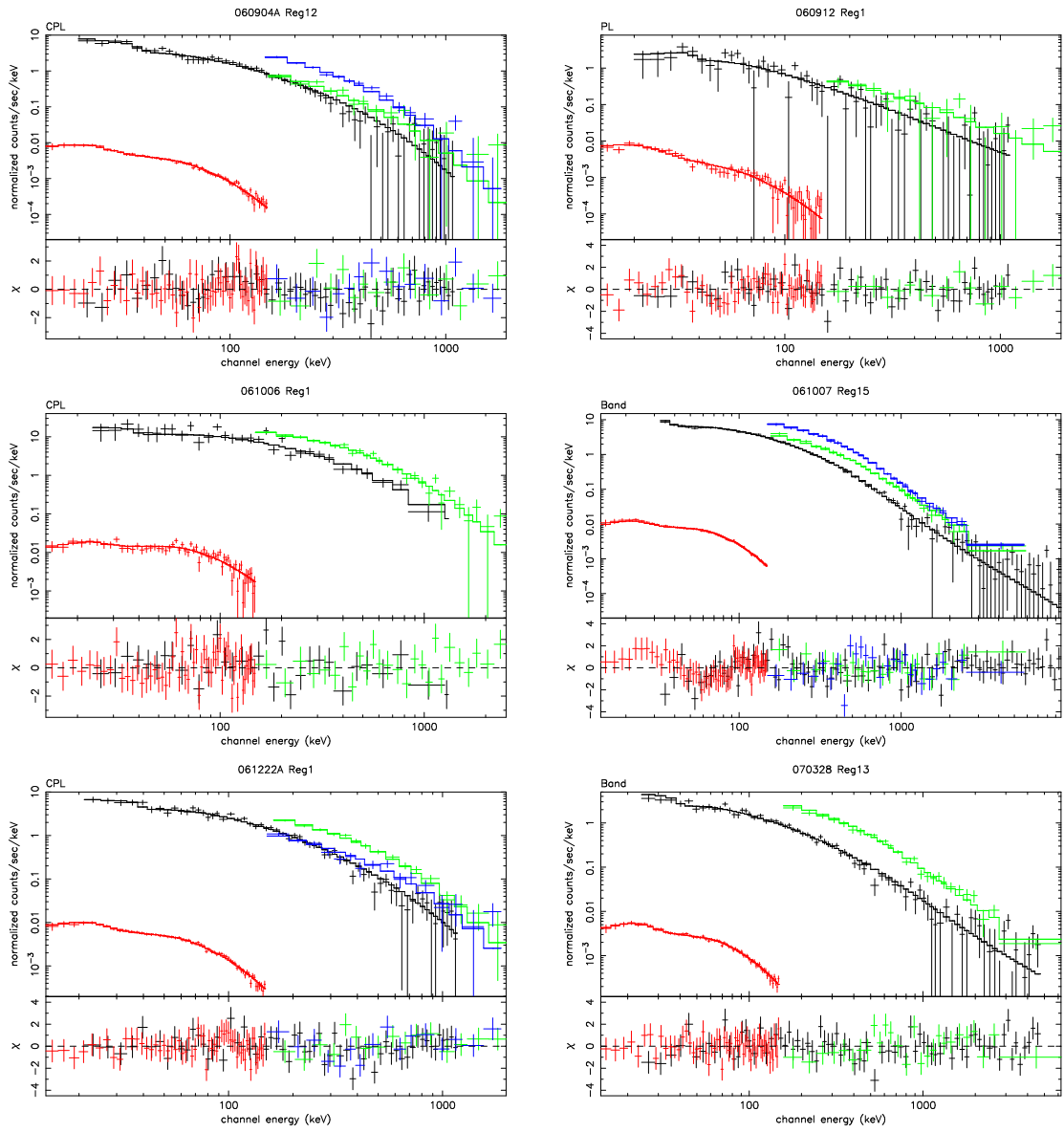

\begin{center}
\includegraphics[scale=0.3,angle=-90]{figure7a.eps}
\hspace{5mm}
\includegraphics[scale=0.3,angle=-90]{figure7b.eps}
\end{center}
\begin{center}
\includegraphics[scale=0.3,angle=-90]{figure7c.eps}
\hspace{5mm}
\includegraphics[scale=0.3,angle=-90]{figure7d.eps}
\end{center}
\begin{center}
\includegraphics[scale=0.3,angle=-90]{figure7e.eps}
\hspace{5mm}
\includegraphics[scale=0.3,angle=-90]{figure7f.eps}
\end{center}
\caption{The joint fit spectra (continued)}\label{fig:SP2}
\end{figure}

\begin{figure}
\begin{center}
\includegraphics[scale=0.8,angle=0]{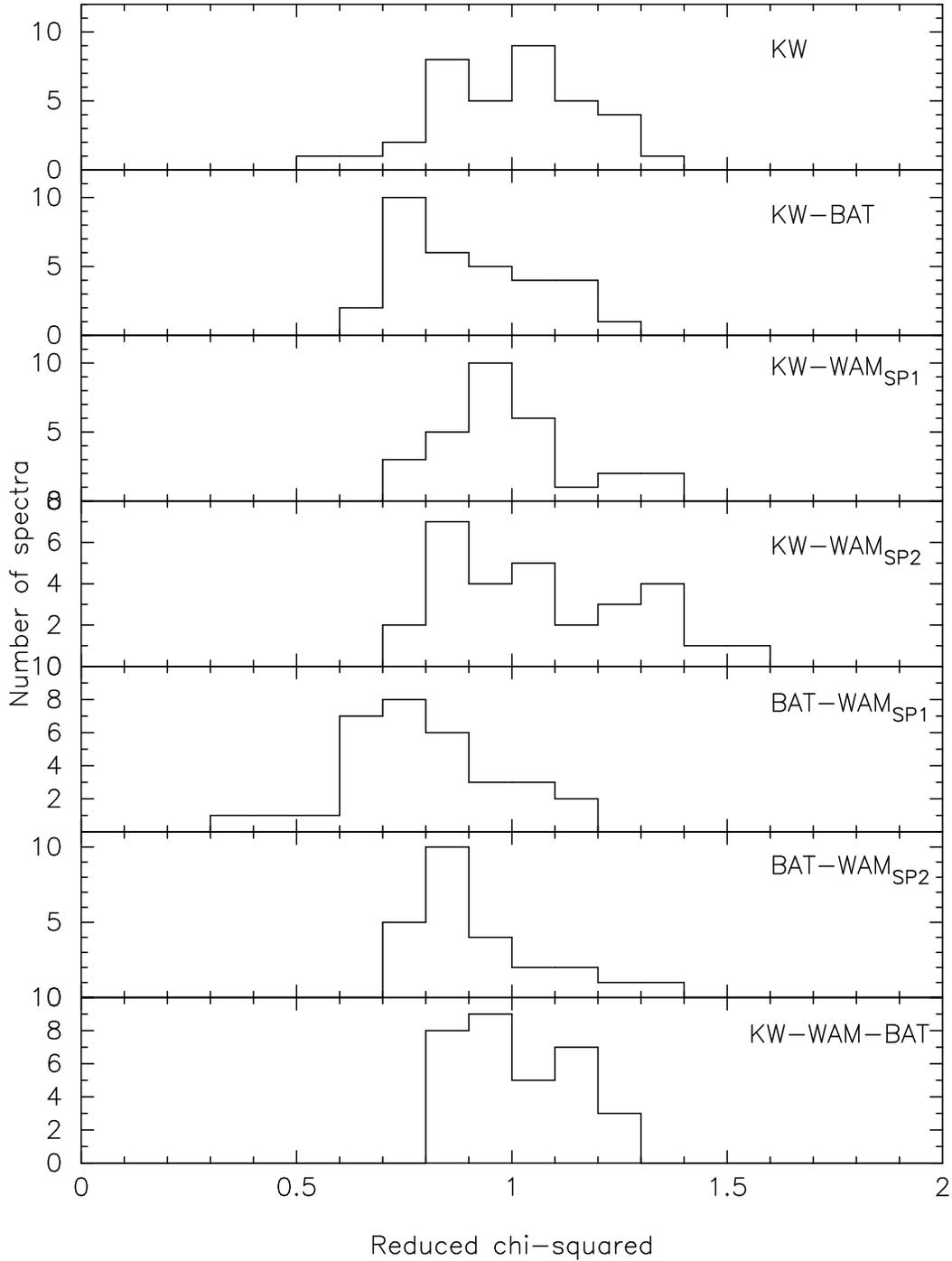}
\end{center}
\caption{Histograms of the reduced $\chi^{2}$ based on the best fit spectral model 
for the KW and joint fits as indicated on the figures.  
The best fit spectral model is determined based on the difference 
in $\chi^{2}$ between a CPL and the Band function fit.  If the value of $\Delta\chi^{2}$ 
between a CPL and the Band function fit is greater than 6 ($\Delta\chi^{2} \equiv 
\chi^{2}_{\rm CPL} - \chi^{2}_{\rm Band} > 6$), we determined that the Band function 
is a better representative spectral model for the data.  Otherwise, the reduced $\chi^{2}$ for a CPL fit 
is used.}
\label{fig:chi2}  
\end{figure}

First, there is a systematic trend in the joint fits that the low-energy photon 
index $\alpha$ becomes steeper by 0.2 in the Band function 
compared to that of the KW fit.  The lowering of 
the low-energy photon index is well represented by the examples of GRB 061007 
spectral fit results (e.g. Figure \ref{fig:SP2}).  
There are systematically positive 
residuals below 50 keV and also negative residuals above 50 keV in the 
BAT data from the best fit model of the KW fit to the Band function.  Along with 
the joint fit results, this suggests that the BAT data prefers a steeper 
power-law slope than that derived from the KW data.  
Note that the WAM data are less sensitive for deriving $\alpha$ since the lower 
energy boundary of its spectrum is $\sim$120 keV.  

Second, 
on average,
$\ep$ derived from the joint fits are systematically higher than those from 
the KW fit: 9\% higher for the joint KW-BAT fit, 6\% higher for the joint KW-WAM fit, 
14\% higher for the joint BAT-WAM fit, and 12\% higher for the joint KW-WAM-BAT fit based on 
the Band function.  
%
We note the relatively large inconsistency in the 90\% confidence level of $\ep$ 
for the spectra of GRB 070328 (Reg2 and Reg13) between the KW fit and the KW-WAM-BAT joint fit.  
The un-overlapping confidence level of $\ep$ is 40-50 keV for the CPL fit and 140-150 keV for 
the Band fit for those spectra.  
The low-energy photon index $\alpha$ and $\ep$ are strongly 
coupled.  $\ep$ will be higher if $\alpha$ goes steeper (left panel of Figure 
\ref{fig:alpha_ep}).  To investigate whether a systematically higher $\ep$ is due 
to a steeper $\alpha$, we check the differences in $\ep$ and $\alpha$ derived from 
the joint BAT-WAM fit (which has the largest inconsistency in $\alpha$) and the 
joint KW-BAT fit (which has the smallest inconsistencies in $\ep$ and $\alpha$ combining 
the BAT data).  
The result is shown in the right panel of Figure \ref{fig:alpha_ep}.  There is a very 
small difference in $\ep$ if $\alpha$ derived from the joint BAT-WAM fit is flatter 
than the joint KW-BAT fit ($\alpha$ (BAT-WAM) - $\alpha$ (KW-BAT) $>$ 0, right side of the figure).  
However the $\ep$ based on the joint BAT-WAM fit 
is systematically higher than that of the joint KW-BAT fit if the $\alpha$ based on the 
joint BAT-WAM fit is steeper than the joint KW-BAT fit 
($\alpha$ (BAT-WAM) - $\alpha$ (KW-BAT) $<$ 0, left side of the figure).  
This demonstrates 
that a systematically higher $\ep$ in the joint fit is mainly as a result of a 
steeper $\alpha$.  Therefore, we believe that the fundamental cause of this problem 
is due to a steeper $\alpha$ which comes when we include the BAT data as we discussed 
above.  

\begin{figure}
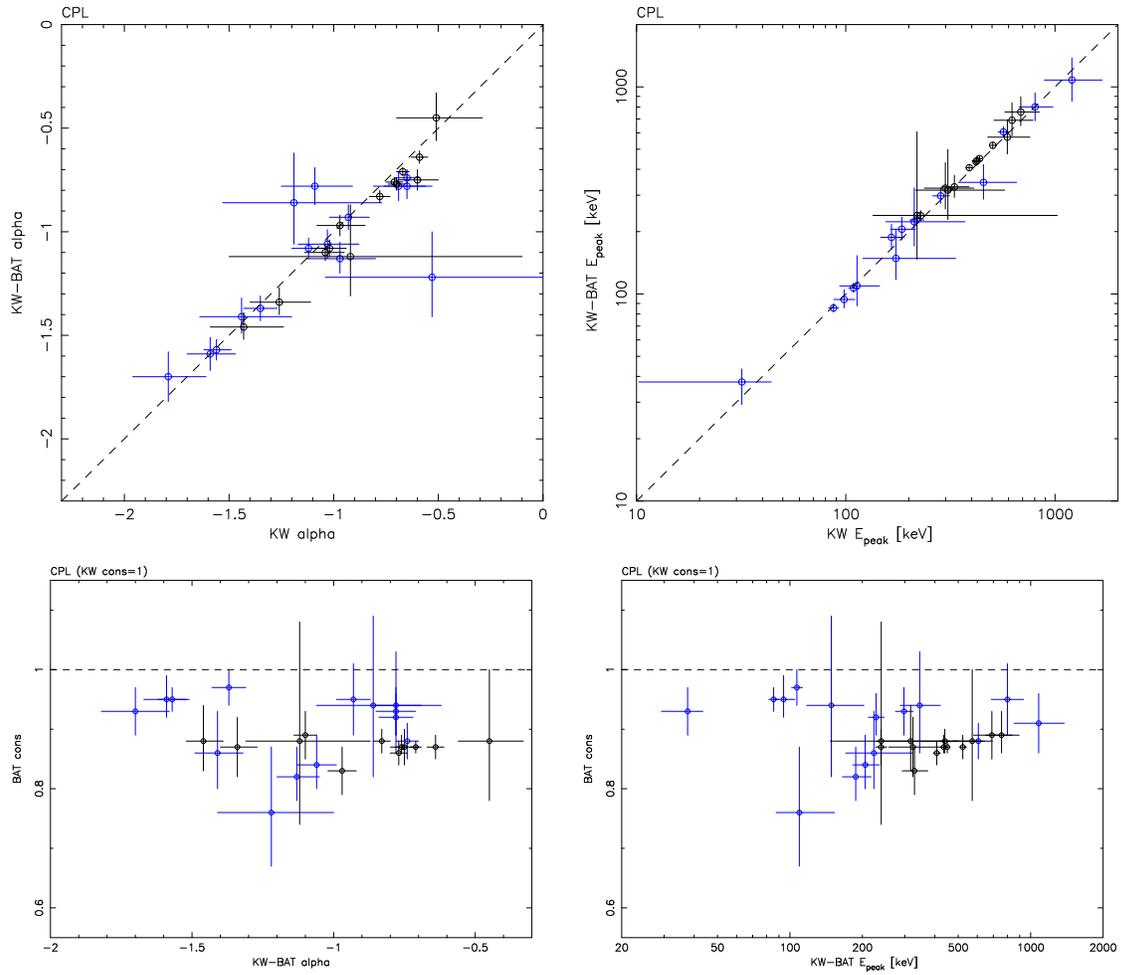

\begin{center}
\includegraphics[scale=0.4,angle=-90]{figure9a.eps}
\hspace{2mm}
\includegraphics[scale=0.4,angle=-90]{figure9b.eps}
\end{center}
\begin{center}
\includegraphics[scale=0.3,angle=-90]{figure9c.eps}
\hspace{4mm}
\includegraphics[scale=0.3,angle=-90]{figure9d.eps}
\end{center}
\caption{Correlation between the KW and BAT joint fit spectral parameters, 
the low-energy photon index $\alpha$ (left-top) and $\ep$ (right-top), and those 
derived from the KW fit based on a CPL model.  
The BAT constant factor as a function of $\alpha$ (left-bottom) and $\ep$ 
(right-bottom) based on a CPL model.  The blue data points are not affected by 
the $Swift$ spacecraft slew.  The KW constant factor is fixed to 1.}
\label{fig:kw_bat_cpl}
\end{figure}

\begin{figure}
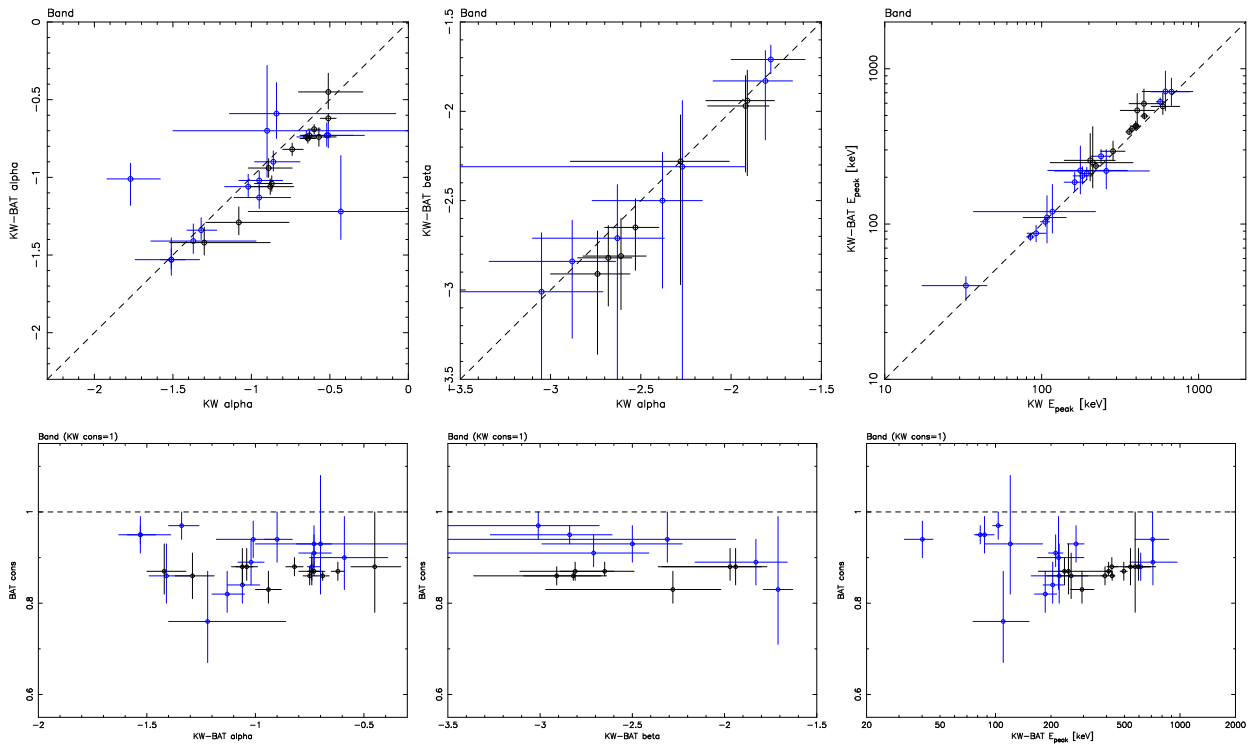

\begin{center}
\includegraphics[scale=0.3,angle=-90]{figure10a.eps}
\includegraphics[scale=0.3,angle=-90]{figure10b.eps}
\includegraphics[scale=0.3,angle=-90]{figure10c.eps}
\end{center}
\begin{center}
\includegraphics[scale=0.23,angle=-90]{figure10d.eps}
\includegraphics[scale=0.23,angle=-90]{figure10e.eps}
\includegraphics[scale=0.23,angle=-90]{figure10f.eps}
\end{center}
\caption{Correlation between the KW and BAT joint fit spectral parameters, 
the low-energy photon index $\alpha$ (left-top), the high-energy photon index $\beta$ 
(middle-top), and $\ep$ (right-top), and those derived from the KW fit based 
on the Band function.  
The BAT constant factor as a function of $\alpha$ (left-bottom), 
$\beta$ (middle-bottom) and $\ep$ (right-bottom) based on the Band function.  
The blue data points are not affected by the $Swift$ spacecraft slew.  The KW constant 
factor is fixed to 1.}
\label{fig:kw_bat_band}
\end{figure}

\begin{figure}
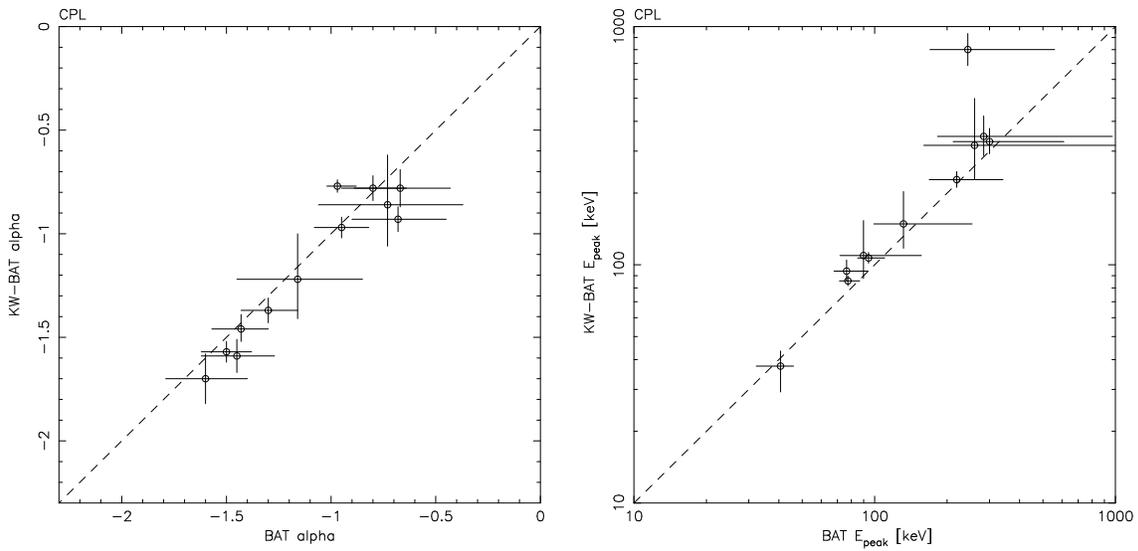

\begin{center}
\includegraphics[scale=0.4,angle=-90]{figure11a.eps}
\hspace{2mm}
\includegraphics[scale=0.4,angle=-90]{figure11b.eps}
\end{center}
\caption{Correlation between the KW and the BAT joint fit spectral parameters, the low 
energy photon index $\alpha$ (left) and $\ep$ (right), and those derived from the 
BAT fit based on a CPL model.}
\label{fig:bat_comp_alpha_ep}
\end{figure}

\begin{figure}
\begin{center}
\includegraphics[scale=0.4,angle=-90]{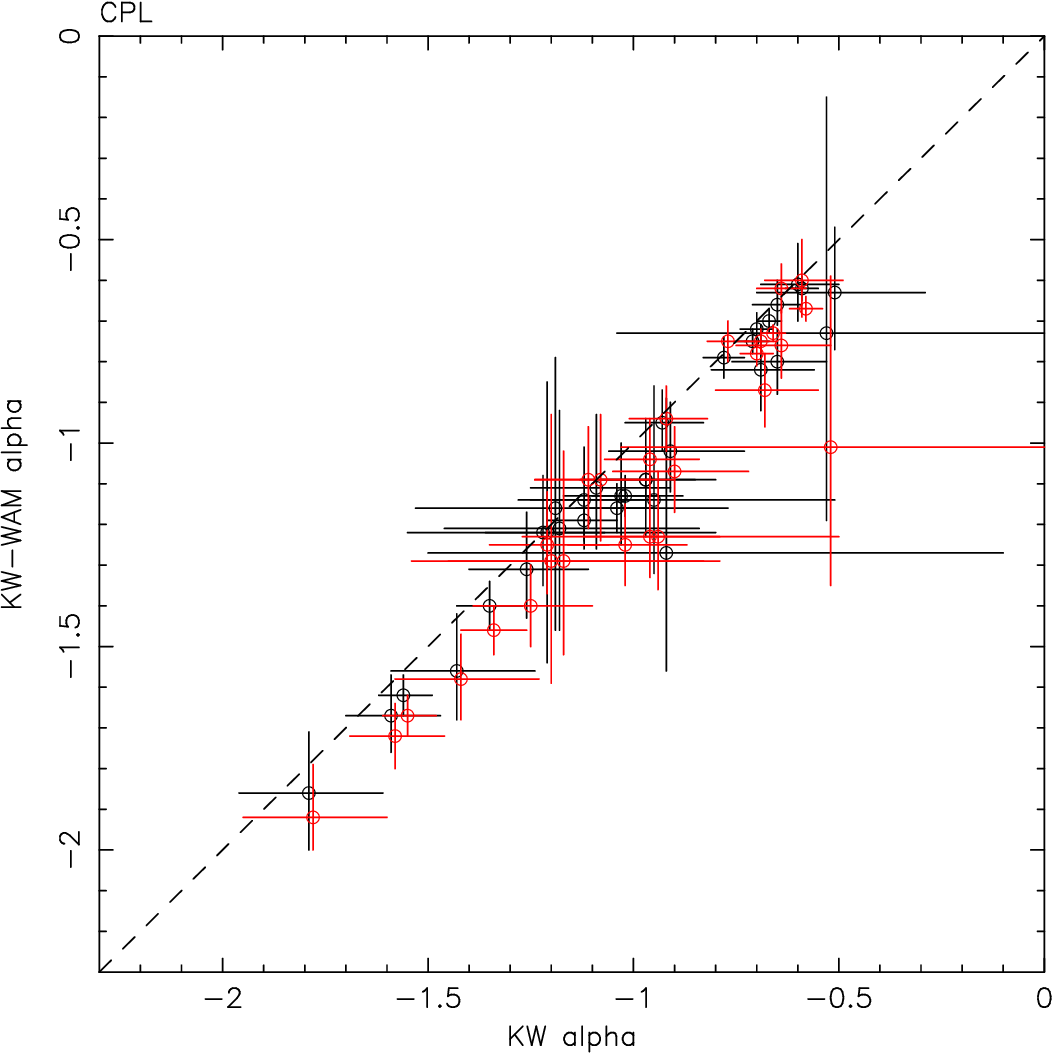}
\hspace{2mm}
\includegraphics[scale=0.4,angle=-90]{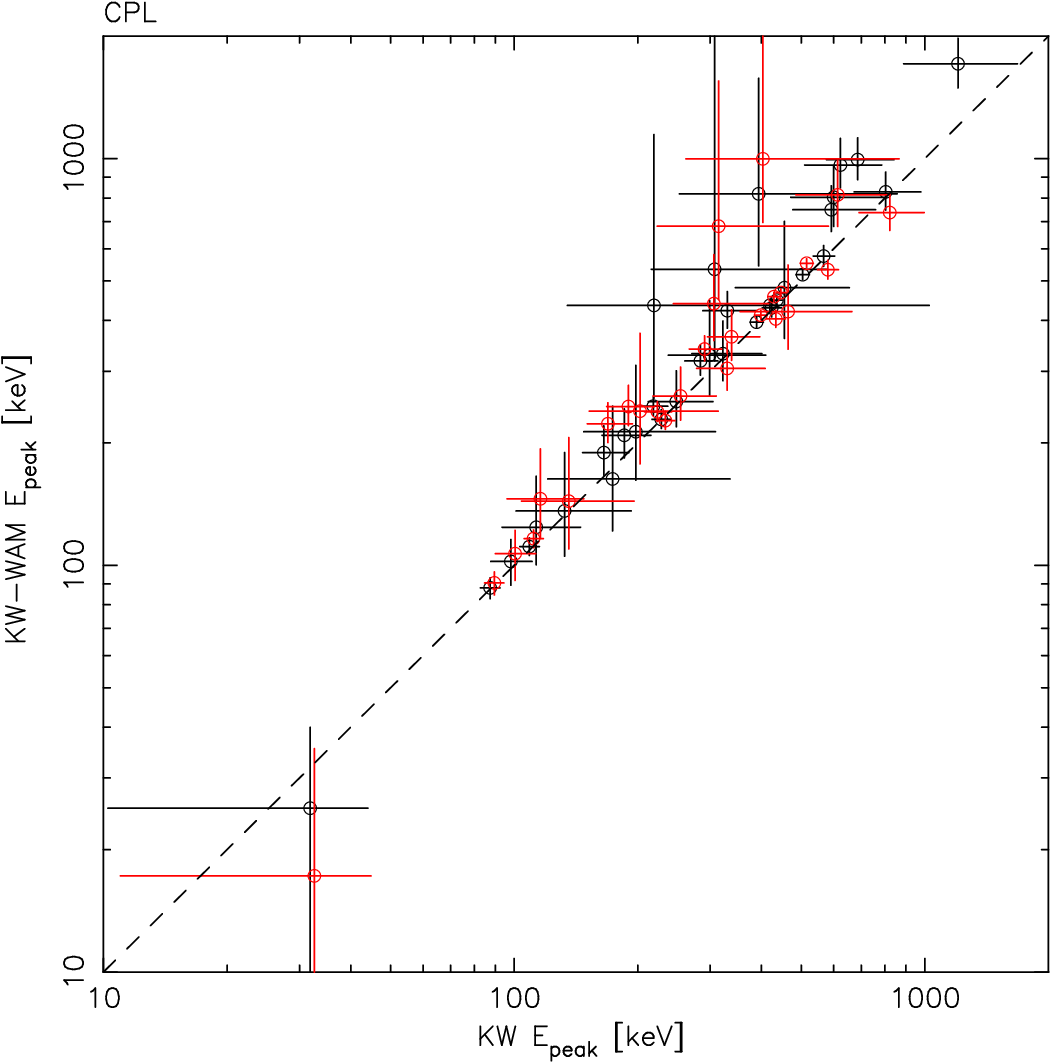}
\vspace{5mm}
\includegraphics[scale=0.3,angle=-90]{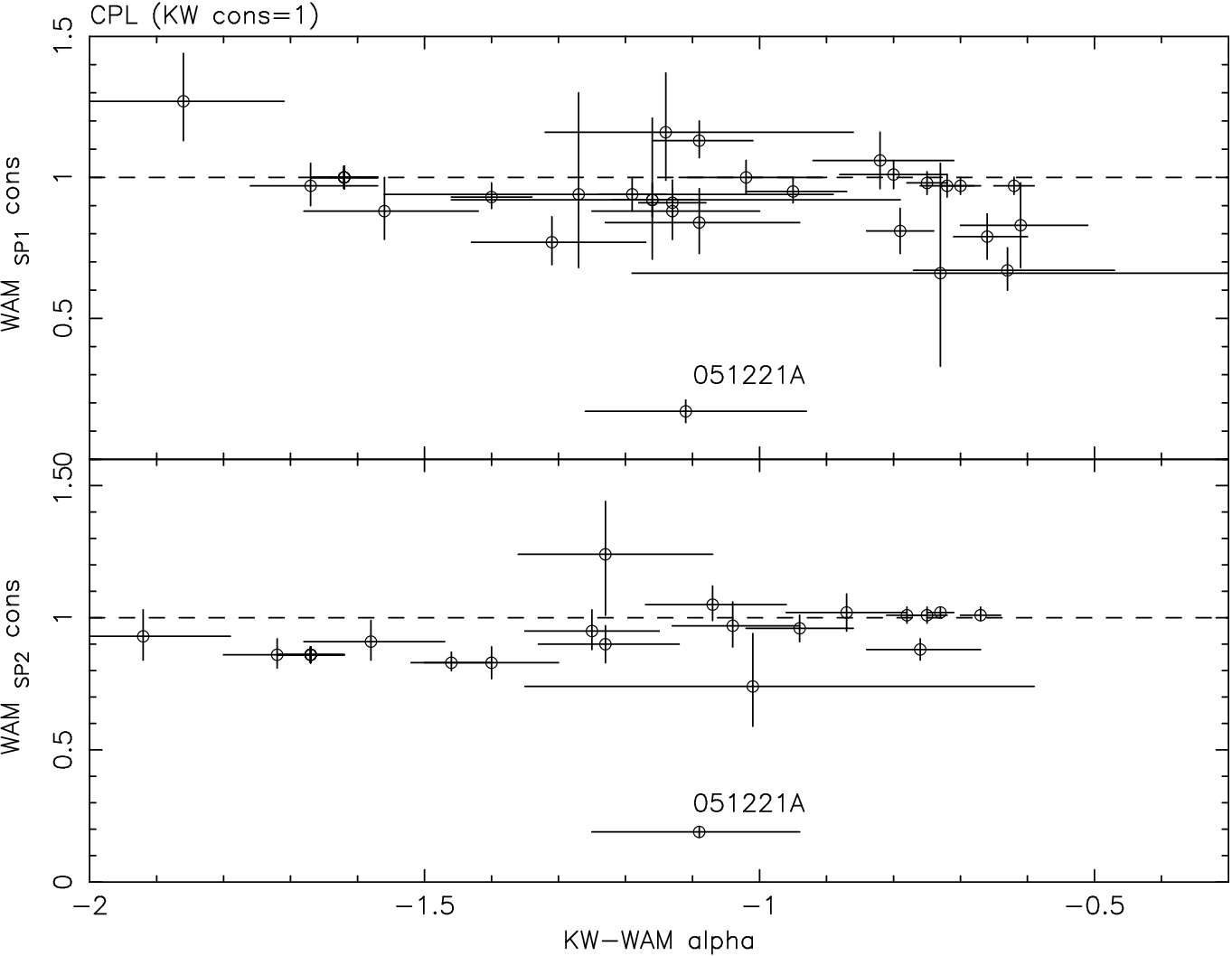}
\hspace{2mm}
\includegraphics[scale=0.3,angle=-90]{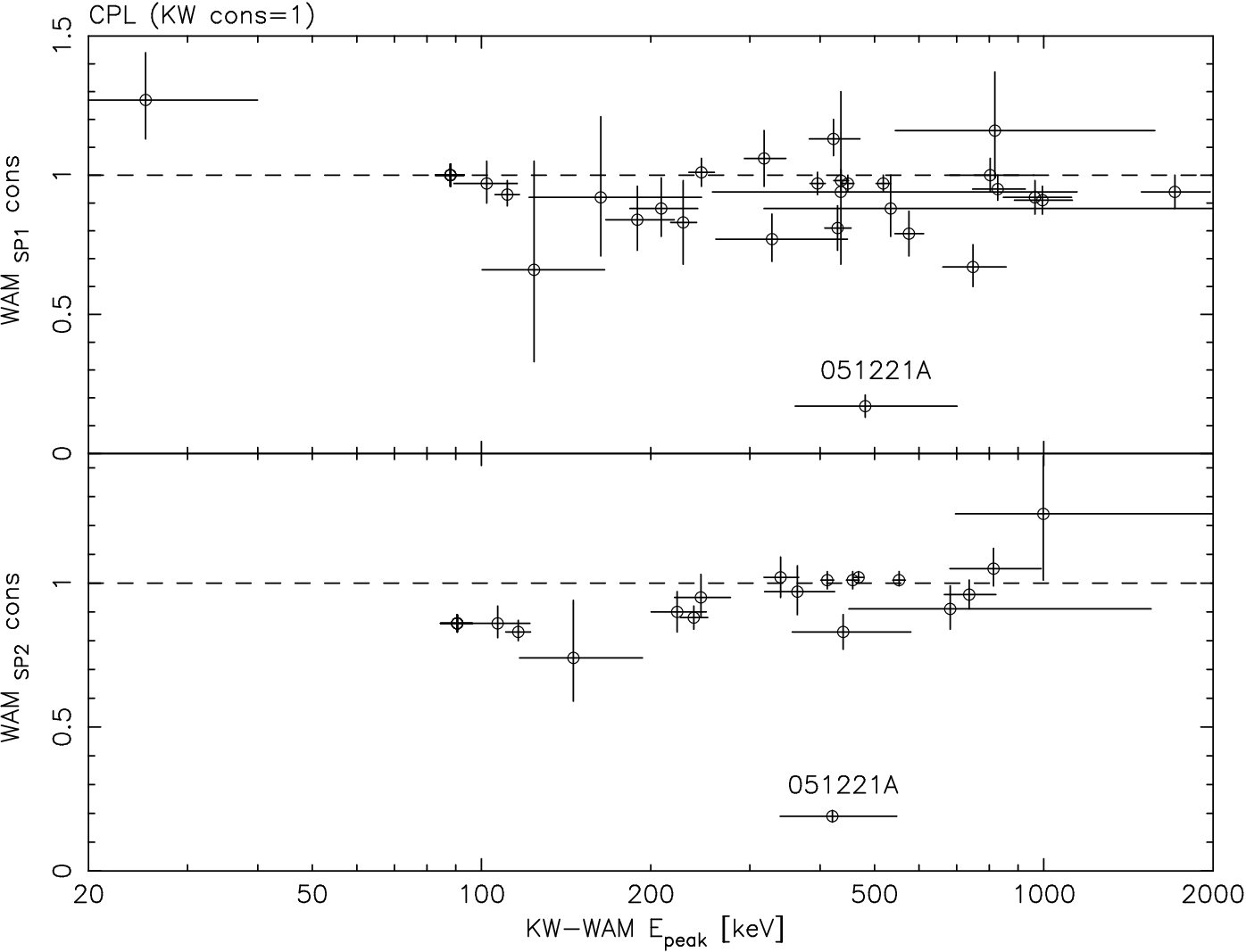}
\end{center}
\caption{Correlation between the KW and WAM joint fit spectral parameters, 
the low-energy photon index $\alpha$ (left-top) and $\ep$ (right-top), and those 
derived from the KW fit based on a CPL model.  The WAM$_{SP1}$ data are shown in 
black and the WAM$_{SP2}$ data are shown in red.  Just for display purposes, the 
WAM$_{SP2}$ data points are shifted by 0.01 in $\alpha$ and by 2\% of 
their $\ep$ value from the KW fits (x-axis).  
The WAM$_{SP1}$ and WAM$_{SP2}$ constant factors as a function of $\alpha$ (left-bottom) and $\ep$ 
(right-bottom) based on a CPL model.  The KW constant factor is fixed to 1.}
\label{fig:kw_wam_cpl}
\end{figure}

\begin{figure}
\begin{center}
\includegraphics[scale=0.3,angle=-90]{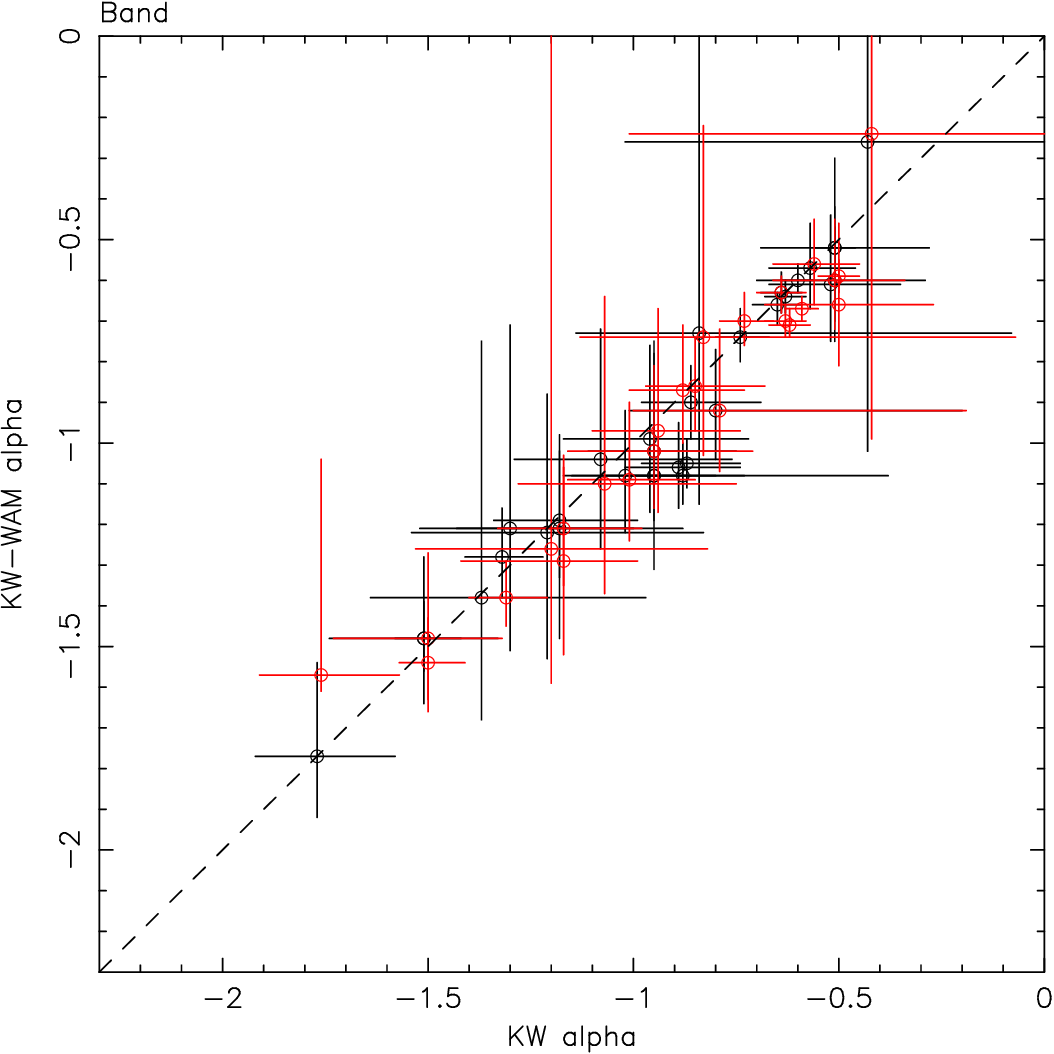}
\includegraphics[scale=0.3,angle=-90]{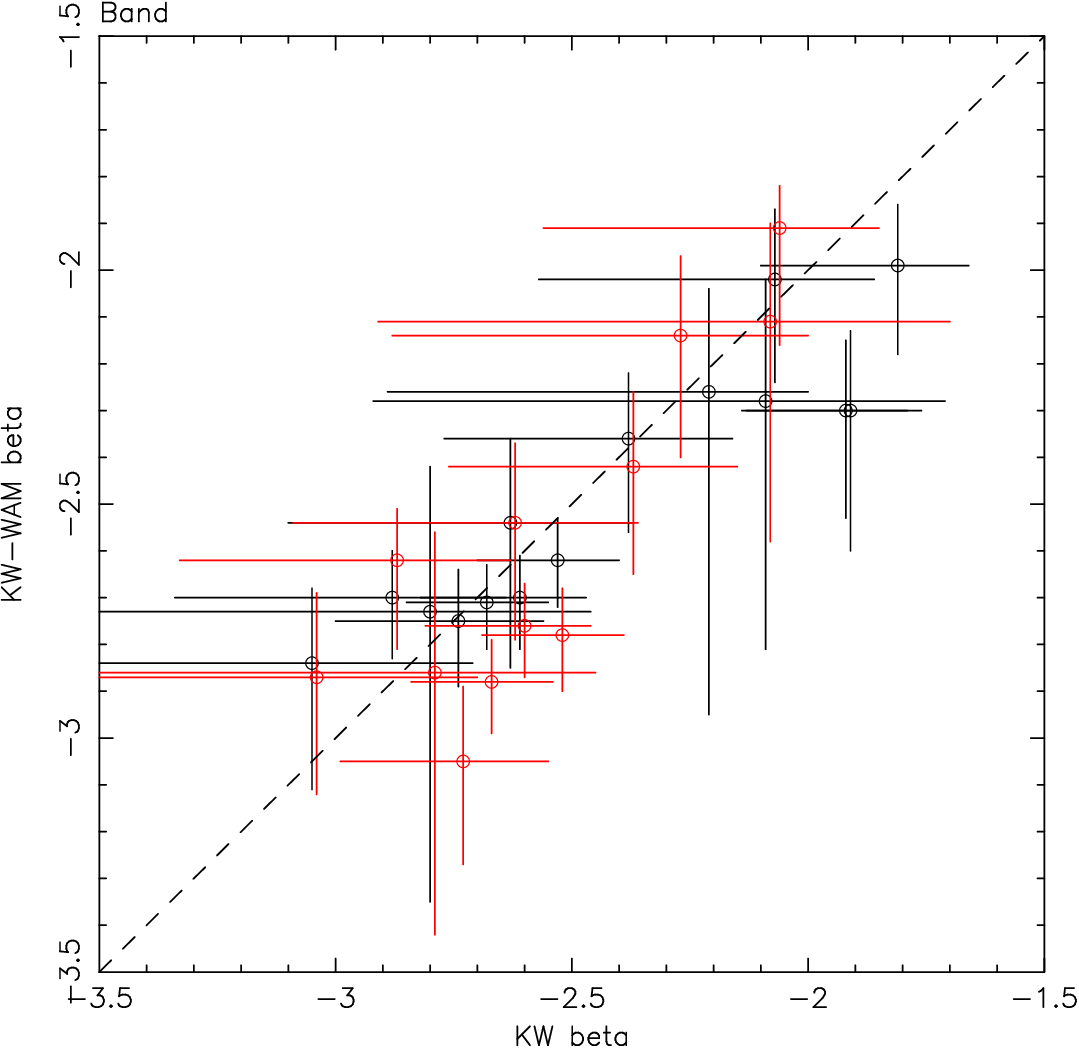}
\includegraphics[scale=0.3,angle=-90]{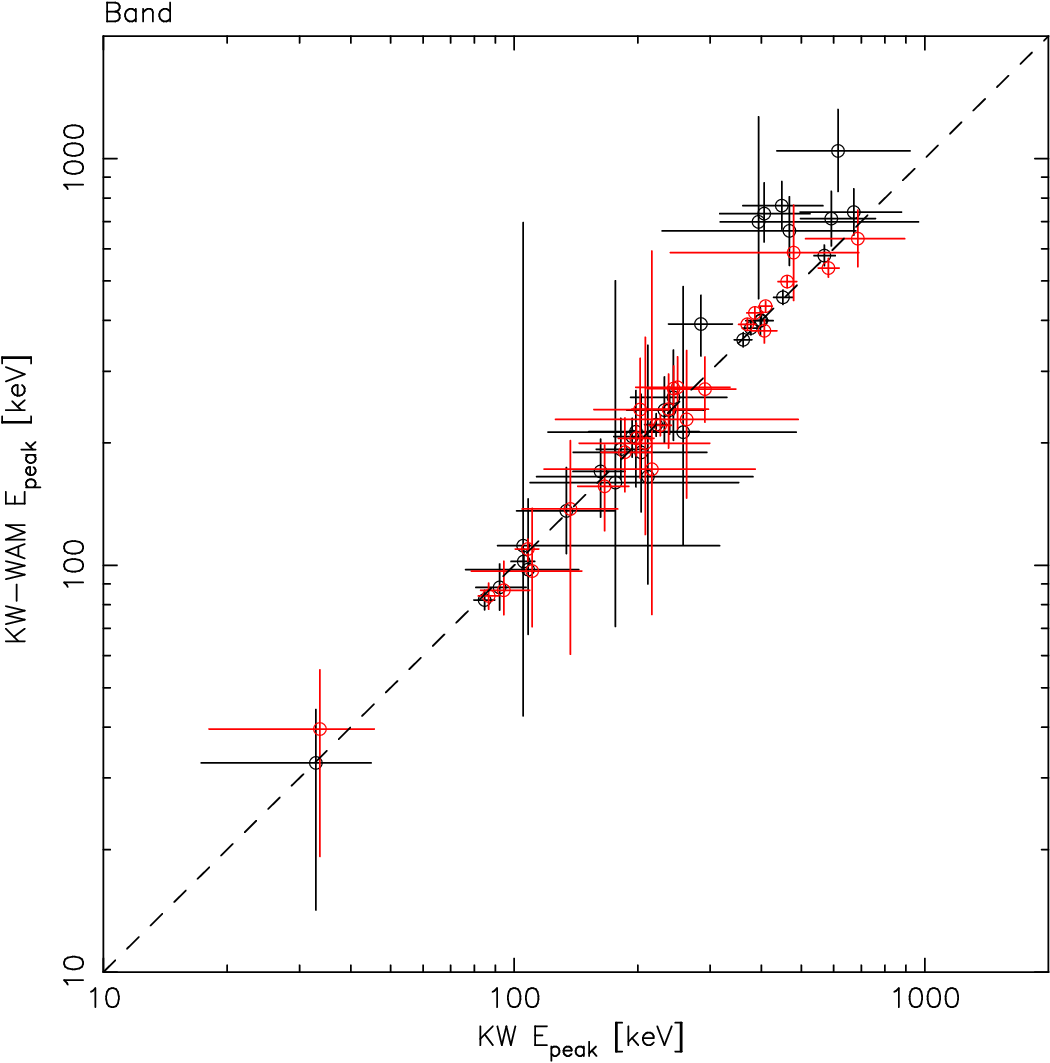}
\end{center}
\begin{center}
\includegraphics[scale=0.23,angle=-90]{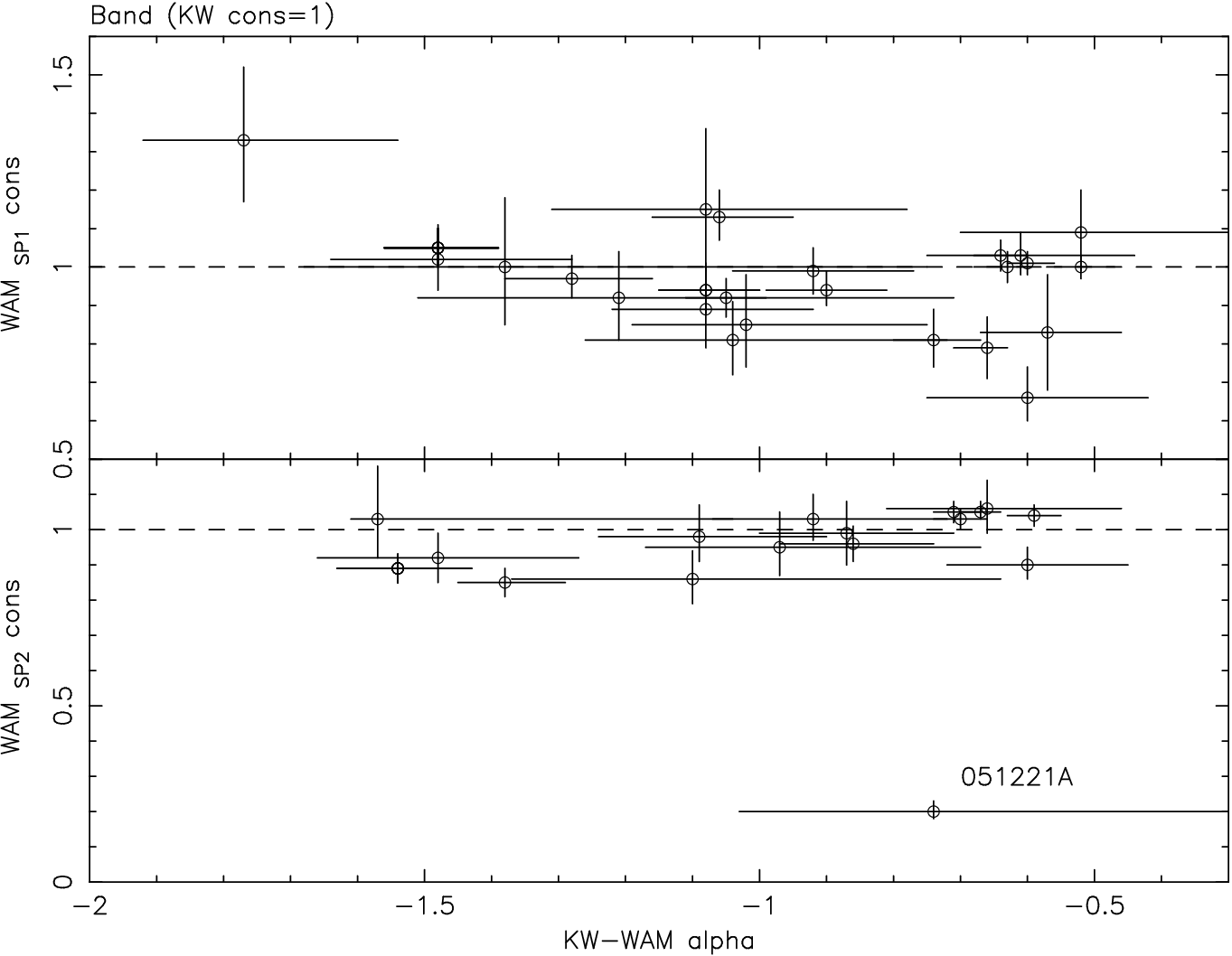}
\includegraphics[scale=0.23,angle=-90]{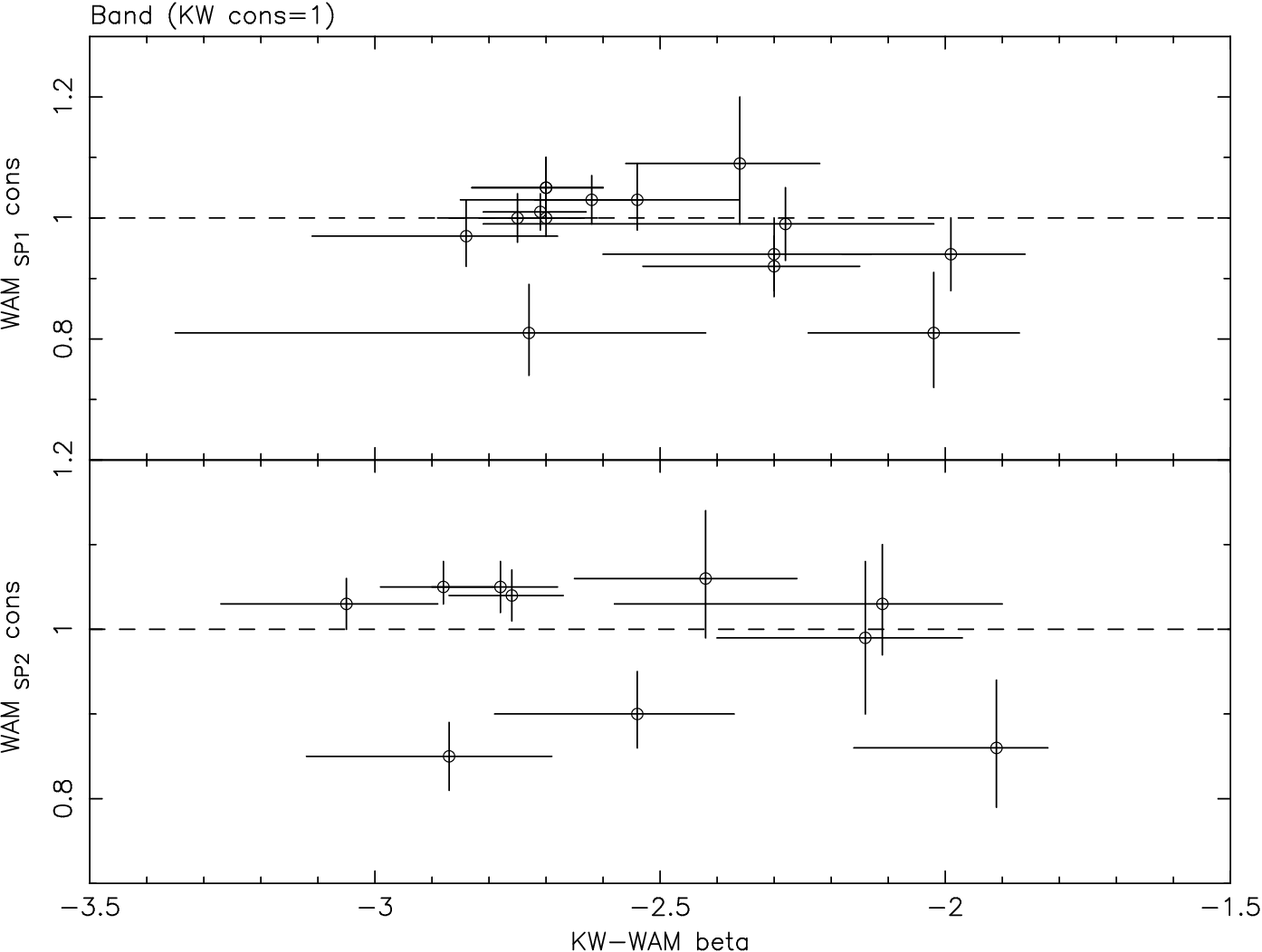}
\includegraphics[scale=0.23,angle=-90]{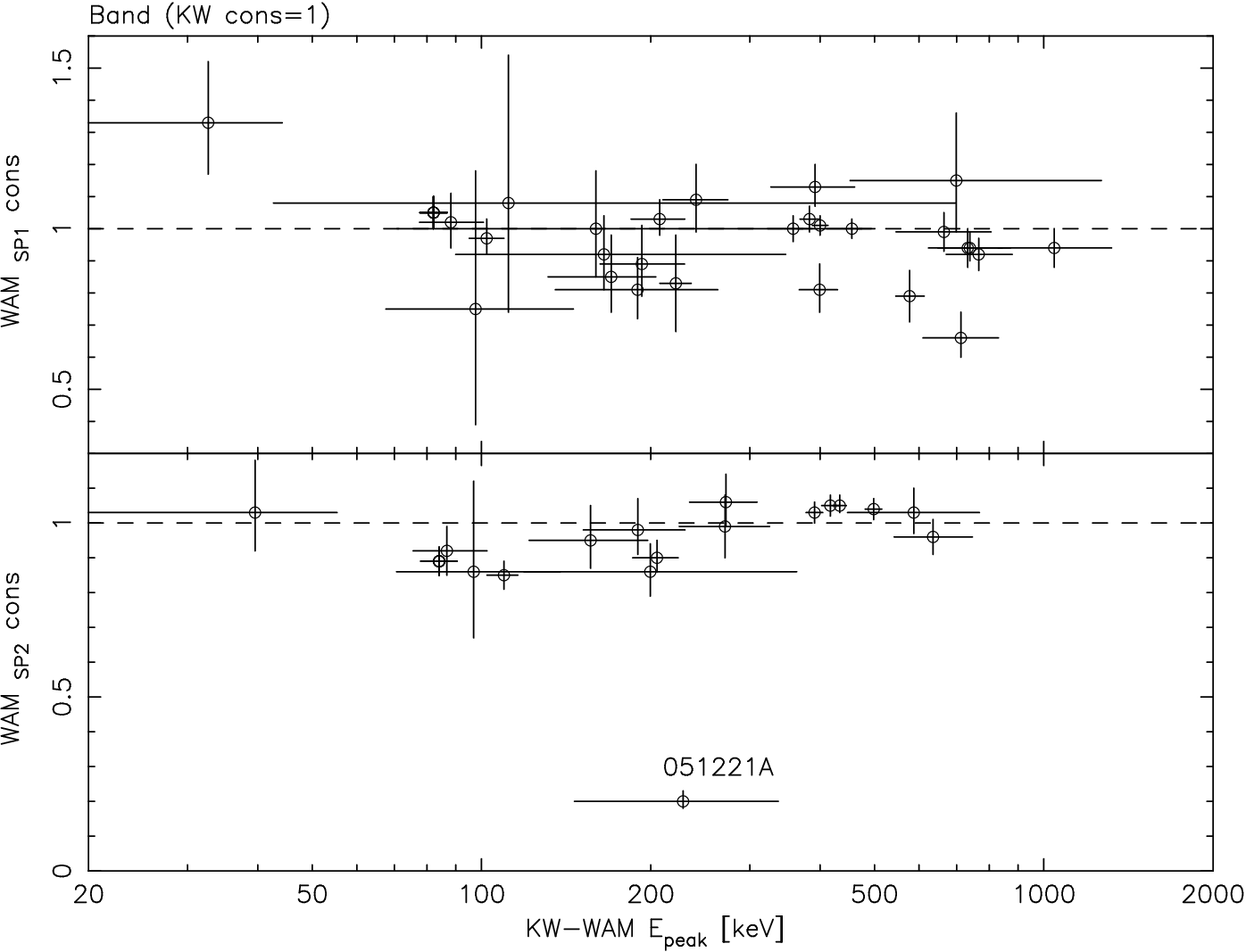}
\end{center}
\caption{Correlation between the KW and the WAM joint fit spectral parameters, 
the low-energy photon index $\alpha$ (left-top), the high-energy photon index $\beta$ 
(middle-top), and $\ep$ (right-top), and those derived from the KW fit based 
on the Band function.  The WAM$_{SP1}$ data are shown in black and the WAM$_{SP2}$ 
data are shown in red.  Just for display purposes, the 
WAM$_{SP2}$ data points are shifted by 0.01 in $\alpha$ and by 2\% of 
their $\ep$ value from the KW fits (x-axis).  
The WAM$_{SP1}$ and WAM$_{SP2}$ constant factors as a function of $\alpha$ (left-bottom), 
$\beta$ (middle-bottom) and $\ep$ (right-bottom) based on the Band function.  
The KW constant factor is fixed to 1.}
\label{fig:kw_wam_band}
\end{figure}

\begin{figure}
\begin{center}
\includegraphics[scale=0.4,angle=-90]{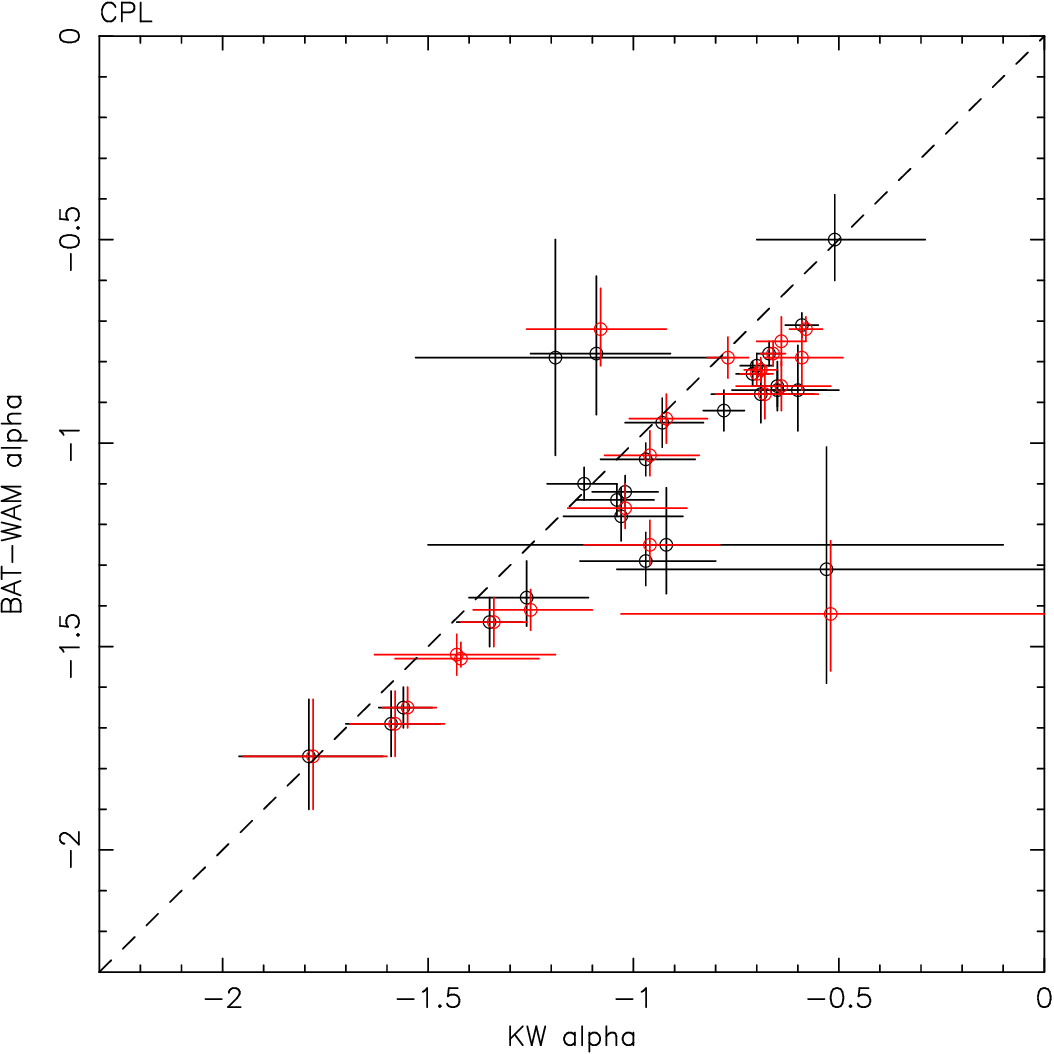}
\hspace{2mm}
\includegraphics[scale=0.4,angle=-90]{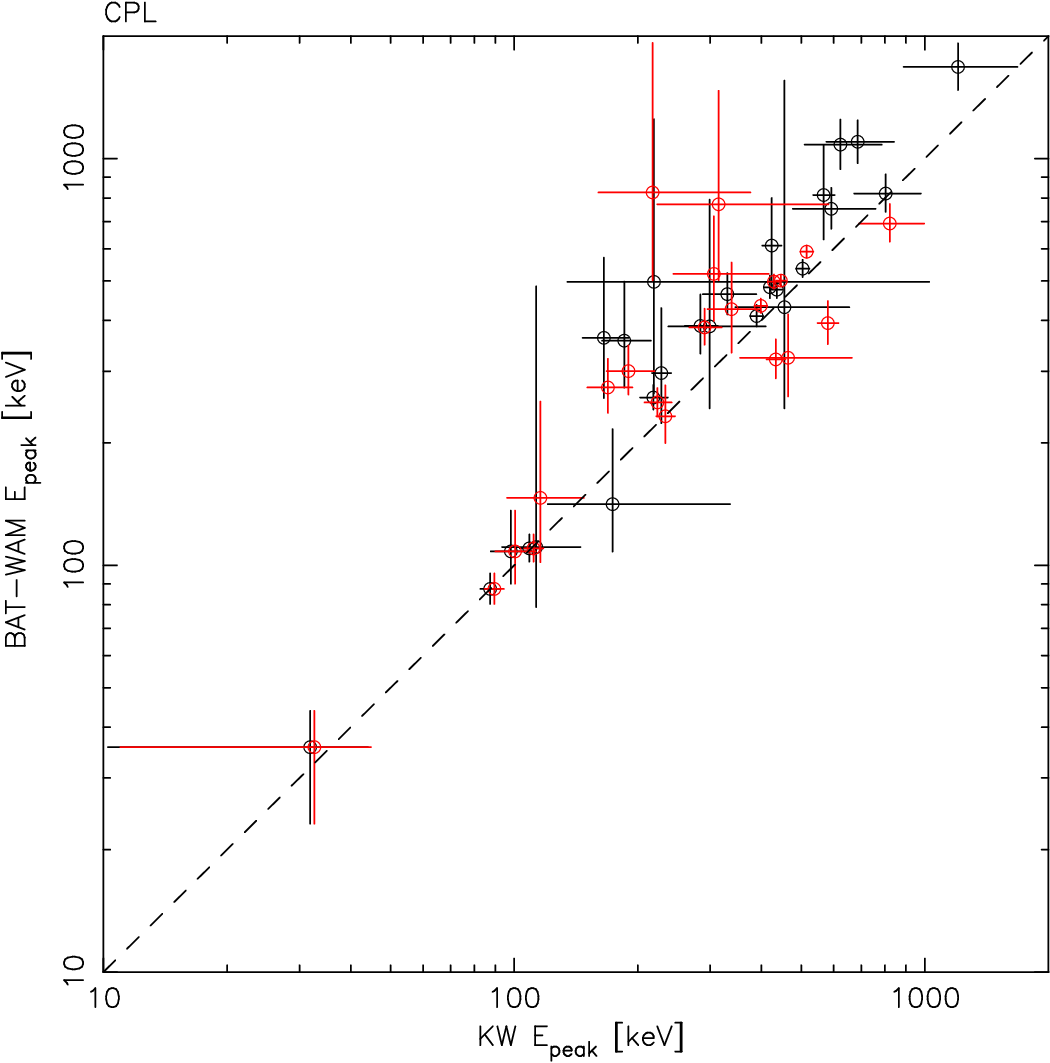}
\vspace{5mm}
\includegraphics[scale=0.3,angle=-90]{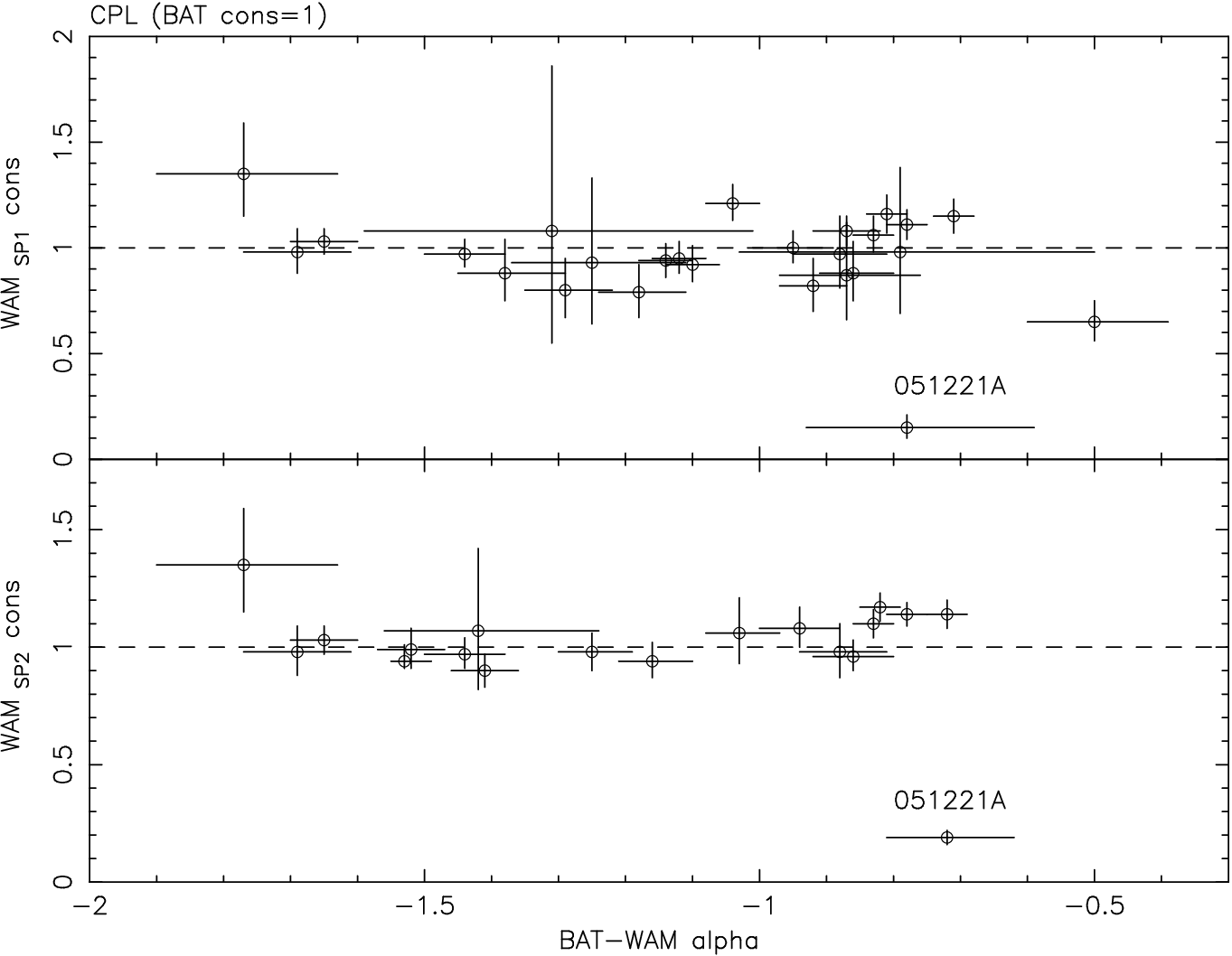}
\hspace{2mm}
\includegraphics[scale=0.3,angle=-90]{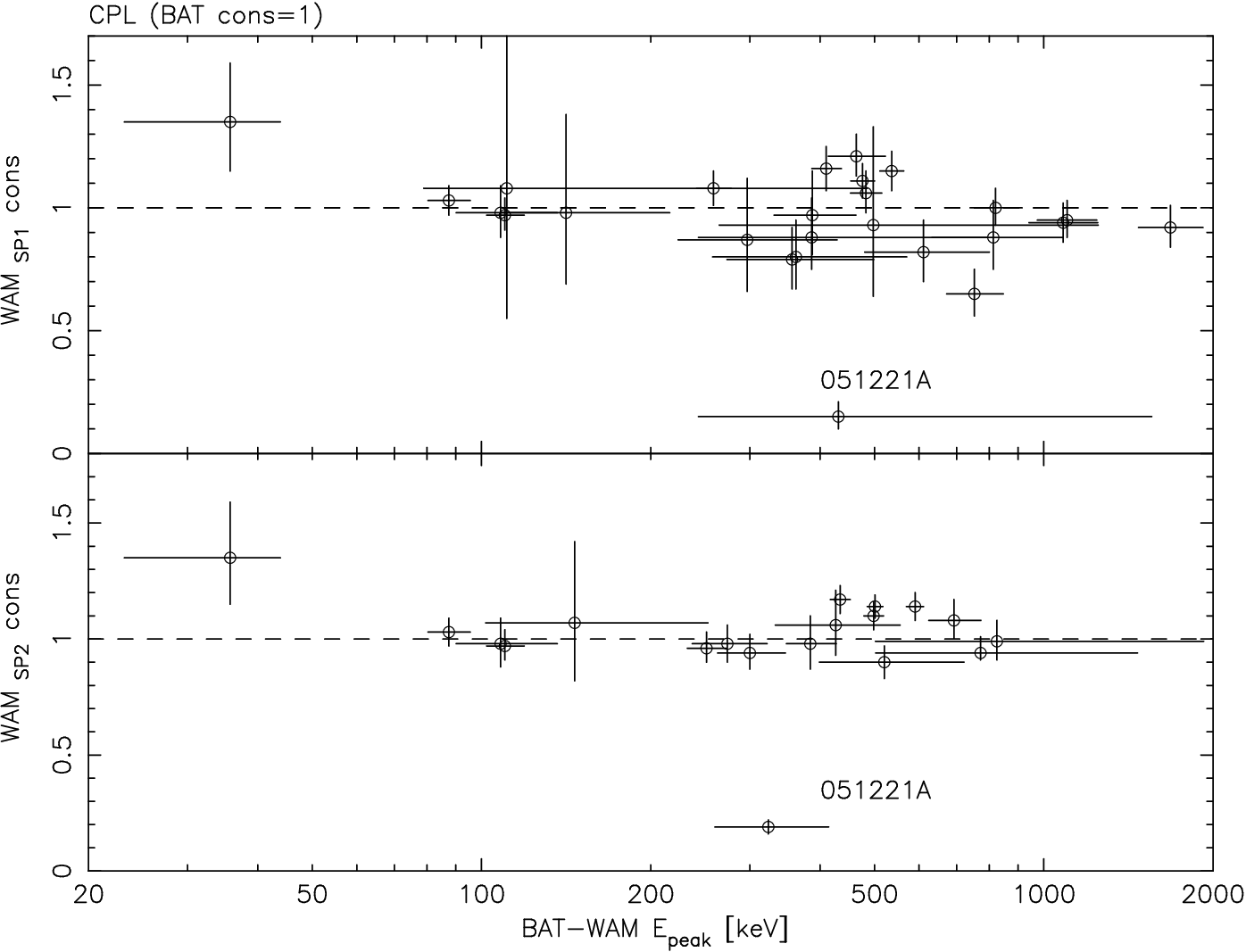}
\end{center}
\caption{Correlation between the BAT and WAM joint fit spectral parameters, 
the low-energy photon index $\alpha$ (left-top) and $\ep$ (right-top), and those 
derived from the KW fit based on a CPL model.  The WAM$_{SP1}$ data are shown in 
black and the WAM$_{SP2}$ data are shown in red.  Just for display purposes, the 
WAM$_{SP2}$ data points are shifted by 0.01 in $\alpha$ and by 2\% of 
their $\ep$ value from the KW fits (x-axis).  
The WAM$_{SP1}$ and WAM$_{SP2}$ constant factors as a function of $\alpha$ (left-bottom) and $\ep$ 
(right-bottom) based on a CPL model.  The BAT constant factor is fixed to 1.}
\label{fig:wam_bat_cpl}
\end{figure}

Third, the constant factor of the BAT data shows a systematically 
smaller value when we fix the constant factor of the KW value to 1.  
There is no correlation between the BAT smaller constant factor 
and the spectral parameters such as $\alpha$, $\ep$, and $\beta$ (see Figure 
\ref{fig:kw_bat_cpl}, \ref{fig:kw_bat_band}, \ref{fig:kw_wam_cpl}, 
\ref{fig:kw_wam_band}, \ref{fig:wam_bat_cpl}, \ref{fig:wam_bat_band}, 
\ref{fig:kw_wam_bat_cpl}, and \ref{fig:kw_wam_bat_band}).  
We investigate whether variations in the constant factor are correlated with the incident 
angle of each instrument or with the observed flux.  Figures \ref{fig:kw_wam_bat_kwtheta}, 
\ref{fig:kw_wam_bat_wamangle} and \ref{fig:kw_wam_bat_batangle} show the constant 
factors of the BAT and the WAM based on the three instrument joint fits plotted with 
respect to the incident angles of the KW, the WAM and the BAT respectively.  
Figure \ref{fig:kw_wam_bat_flux} shows the constant 
factors of the joint fits versus the observed flux in the 20-1000 keV band.  
Although there might be a hint of a correlation between the BAT 
constant factor and $\theta$, the correlation coefficient is 0.655 in 12 samples.  
The probability of such a correlation occurring by chance for this sample 
size is 0.131.
We find no systematic trend in the constant factor with either incident angles or the observed fluxes.  
Therefore, we conclude that the intrinsic effective area of the current 
energy response file of the BAT is 10-20\% smaller than that of the KW.  
Although there might be a similar trend in the WAM constant factor, the scatter 
in the WAM constant factor is too large for us to make a solid conclusion.  

Despite the systematic problems in the joint fits, we note that 
the uncertainties in the spectral parameters will improve significantly using 
the joint fits.  The joint 
KW-BAT fits provides smaller uncertainties in both $\alpha$ and $\ep$, compared 
to those of the KW fit.  The averaged uncertainties of $\alpha$ and $\ep$ in the 
joint KW-BAT fits are 0.06 and 63 keV, whereas the uncertainties in the KW 
fits are 0.1 and 71 keV.  This is because the BAT high statistics data will provide 
a better constraint on $\alpha$, which also improves the uncertainties in $\ep$.  
The KW-WAM joint fits provide better constraints on $\beta$ than does the KW fit.  
The averaged uncertainties of $\beta$ in the joint KW-WAM fits are 0.8, on the 
other hand, the averaged uncertainty of $\beta$ on the KW fit is 2.8.  The high 
statistics WAM data above 100 keV helps to constrain $\beta$.  The three instrument 
joint fits provide the smallest uncertainties in both $\alpha$ and $\ep$.  The 
averaged uncertainties in $\alpha$ and $\ep$ in the three instrument fits are 
0.04 and 45 keV.  The joint fit effort is important not only to enable us to 
understand the 
systematic problems among the instruments, but also to reduce the uncertainties 
in the spectral parameters such as $\alpha$ and $\ep$, allowing better constraints on 
the empirical relationships and ultimately better physical models of the prompt emission.  

Finally, we want to emphasize that it is very difficult to understand the instrumental 
effects buried in the scientific results such as GRB empirical relations 
without these kind of cross-calibration efforts.  
We believe that the systematic problems in the spectral parameters which we derived in this 
work by combining different instrument data would be added as a systematic error to the 
statistical uncertainties of the spectral parameters.  This is especially important when we display 
the spectral parameters derived from different instruments such as the empirical relationships.
We hope that this kind of joint analysis effort will expand to the data of the previous 
missions and also to the current/future missions to understand the instrumental 
problems and the instrumental effects on the interpretation of GRB prompt emission data.

We would like to thank the anonymous referee for comments and suggestions 
that materially improved the paper.  The Konus-Wind experiment is supported by a Russian Space Agency
contract and RFBR grant 09-02-00166a.  This research has made use of data obtained from the Suzaku 
satellite, a collaborative mission between the space agencies of Japan (JAXA) and the USA (NASA).  
It also has been supported in part by a Grant-in-Aid for Scientific Research (19047001 KY, 21740214 MO) 
of the Ministry of Education, Culture, Sports, Science and Technology (MEXT).


\newpage

\newpage


\newpage


\begin{figure}
\begin{center}
\includegraphics[scale=0.3,angle=-90]{figure15a.eps}
\includegraphics[scale=0.3,angle=-90]{figure15b.eps}
\includegraphics[scale=0.3,angle=-90]{figure15c.eps}
\end{center}
\begin{center}
\includegraphics[scale=0.23,angle=-90]{figure15d.eps}
\includegraphics[scale=0.23,angle=-90]{figure15e.eps}
\includegraphics[scale=0.23,angle=-90]{figure15f.eps}
\end{center}
\caption{Correlation between the BAT and the WAM joint fit spectral parameters, 
the low-energy photon index $\alpha$ (left-top), the high-energy photon index $\beta$ 
(middle-top), and $\ep$ (right-top), and those derived from the KW fit based 
on the Band function.  The WAM$_{SP1}$ data are shown in black and the WAM$_{SP2}$ 
data are shown in red.  Just for display purposes, the 
WAM$_{SP2}$ data points are shifted by 0.01 in $\alpha$ and by 2\% of 
their $\ep$ value from the KW fits (x-axis).  
The WAM$_{SP1}$ and WAM$_{SP2}$ constant factors as a function of $\alpha$ (left-bottom), 
$\beta$ (middle-bottom) and $\ep$ (right-bottom) based on the Band function.  
The BAT constant factor is fixed to 1.}
\label{fig:wam_bat_band}
\end{figure}

\begin{figure}
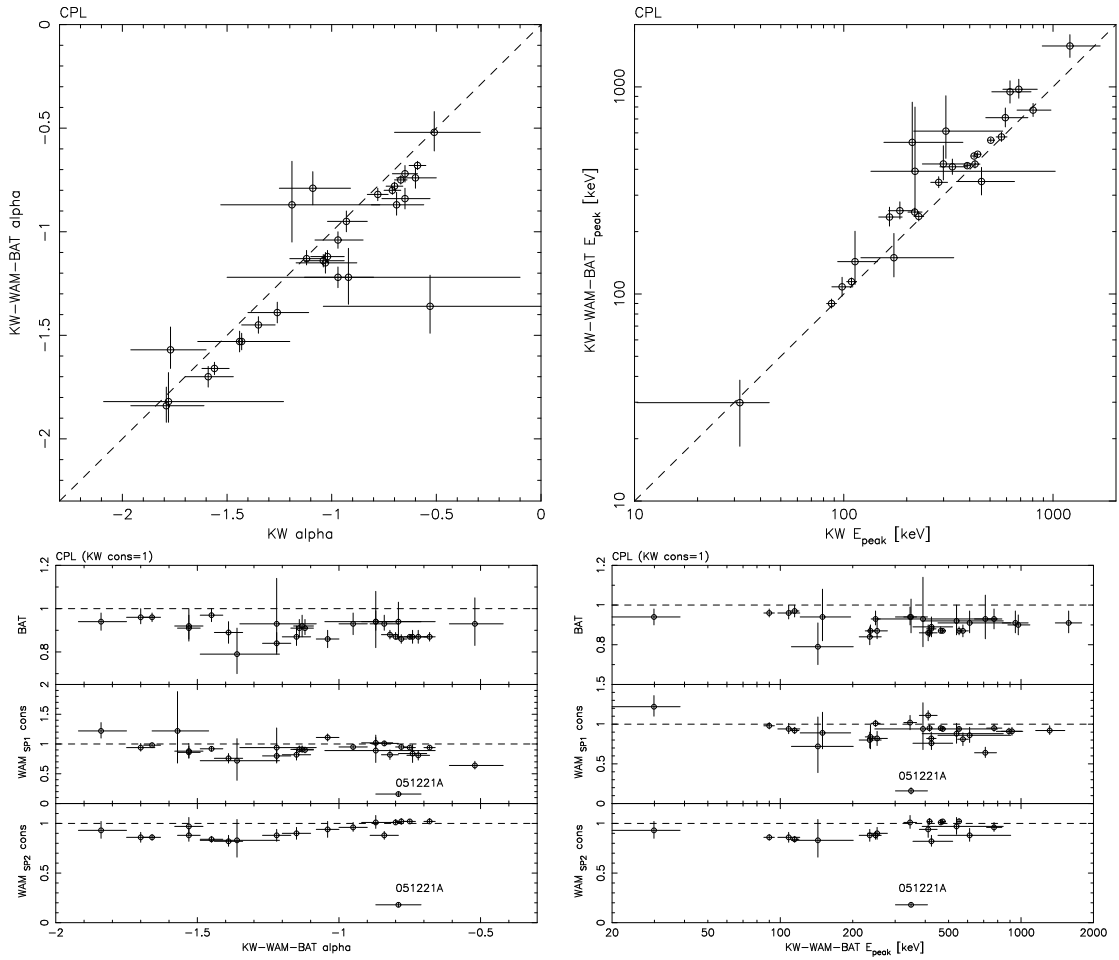

\begin{center}
\includegraphics[scale=0.4,angle=-90]{figure16a.eps}
\hspace{2mm}
\includegraphics[scale=0.4,angle=-90]{figure16b.eps}
\vspace{5mm}
\includegraphics[scale=0.3,angle=-90]{figure16c.eps}
\hspace{2mm}
\includegraphics[scale=0.3,angle=-90]{figure16d.eps}
\end{center}
\caption{Correlation between the KW, WAM and BAT joint fit spectral parameters, 
the low-energy photon index $\alpha$ (left-top) and $\ep$ (right-top), and those 
derived from the KW fit based on a CPL model.  
The BAT, WAM1 and WAM2 constant factors as a function of $\alpha$ (left-bottom) and $\ep$ 
(right-bottom) based on a CPL model.  The KW constant factor is fixed to 1.}
\label{fig:kw_wam_bat_cpl}
\end{figure}
\begin{figure}
\begin{center}
\includegraphics[scale=0.3,angle=-90]{figure17a.eps}
\includegraphics[scale=0.3,angle=-90]{figure17b.eps}
\includegraphics[scale=0.3,angle=-90]{figure17c.eps}
\end{center}
\begin{center}
\includegraphics[scale=0.23,angle=-90]{figure17d.eps}
\includegraphics[scale=0.23,angle=-90]{figure17e.eps}
\includegraphics[scale=0.23,angle=-90]{figure17f.eps}
\end{center}
\caption{Correlation between the KW, WAM and BAT joint fit spectral parameters, 
the low-energy photon index $\alpha$ (left-top), the high-energy photon index $\beta$ 
(middle-top), and $\ep$ (right-top), and those derived from the KW fit based 
on the Band function.  
The BAT, WAM1 and WAM2 constant factors as a function of $\alpha$ (left-bottom), 
$\beta$ (middle-bottom) and $\ep$ (right-bottom) based on the Band function.  
The KW constant factor is fixed to 1.}
\label{fig:kw_wam_bat_band}
\end{figure}

\newpage
\begin{figure}
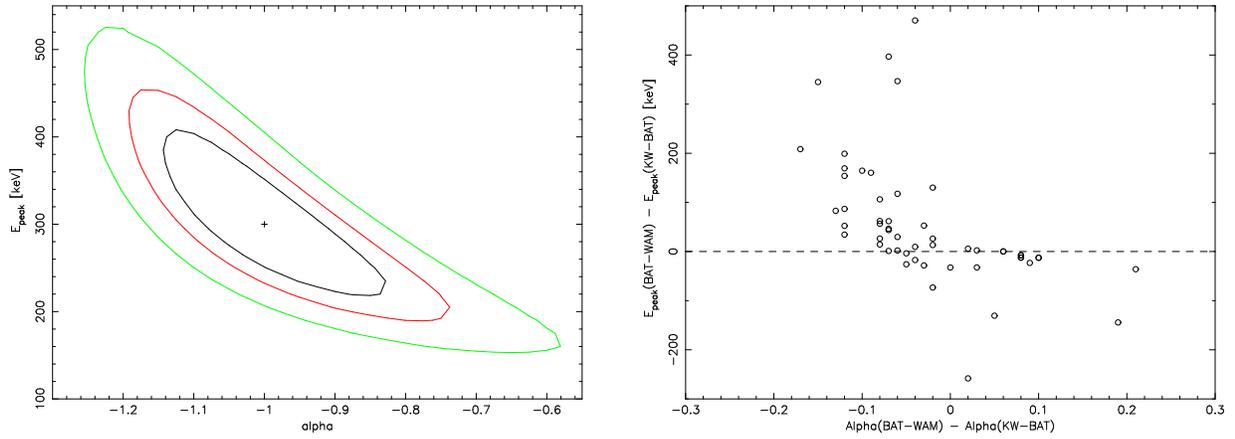

\begin{center}
\includegraphics[scale=0.33,angle=-90]{figure18a.eps}
\hspace{0.5cm}
\includegraphics[scale=0.33,angle=-90]{figure18b.eps}
\end{center}
\caption{Left: Confidence contours between $\ep$ and $\alpha$ derived from the KW simulated 
spectrum with the input Band parameters of $\alpha = -1$, $\beta=-2.5$, and $\ep = 300$ keV.  
Right: the difference in $\ep$ between the BAT-WAM joint fit and the KW-BAT joint fit vs. 
the difference in $\alpha$ between the BAT-WAM joint fit and the KW-BAT joint fit.}
\label{fig:alpha_ep}
\end{figure}

\newpage
\begin{figure}
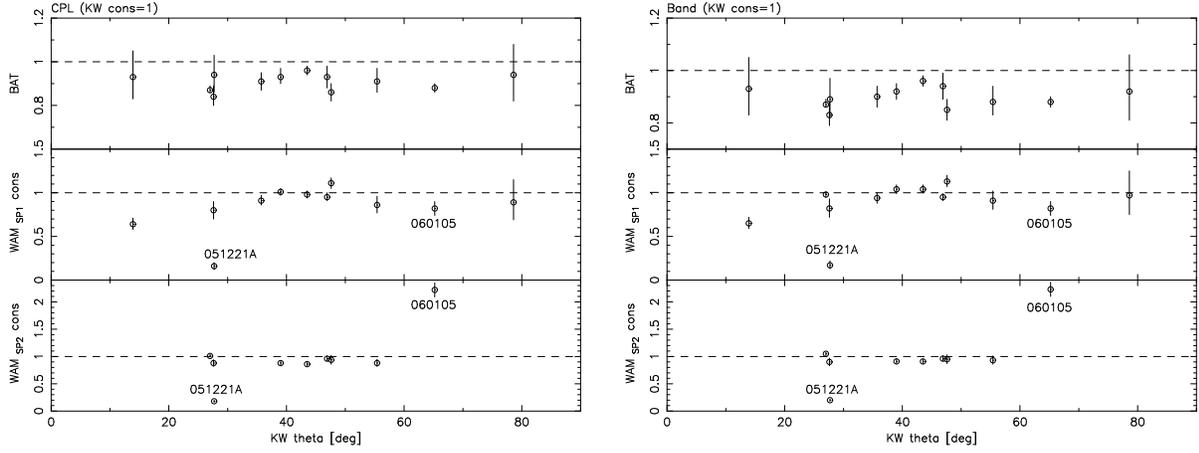

\begin{center}
\includegraphics[scale=0.33,angle=-90]{figure19a.eps}
\hspace{3mm}
\includegraphics[scale=0.33,angle=-90]{figure19b.eps}
\end{center}
\caption{The BAT, WAM$_{SP1}$ and WAM$_{SP2}$ constant factor based on a CPL fit (left) and 
the Band fit (right) as a function of the incident angle $\theta$ of KW.}
\label{fig:kw_wam_bat_kwtheta}
\end{figure}

\begin{figure}
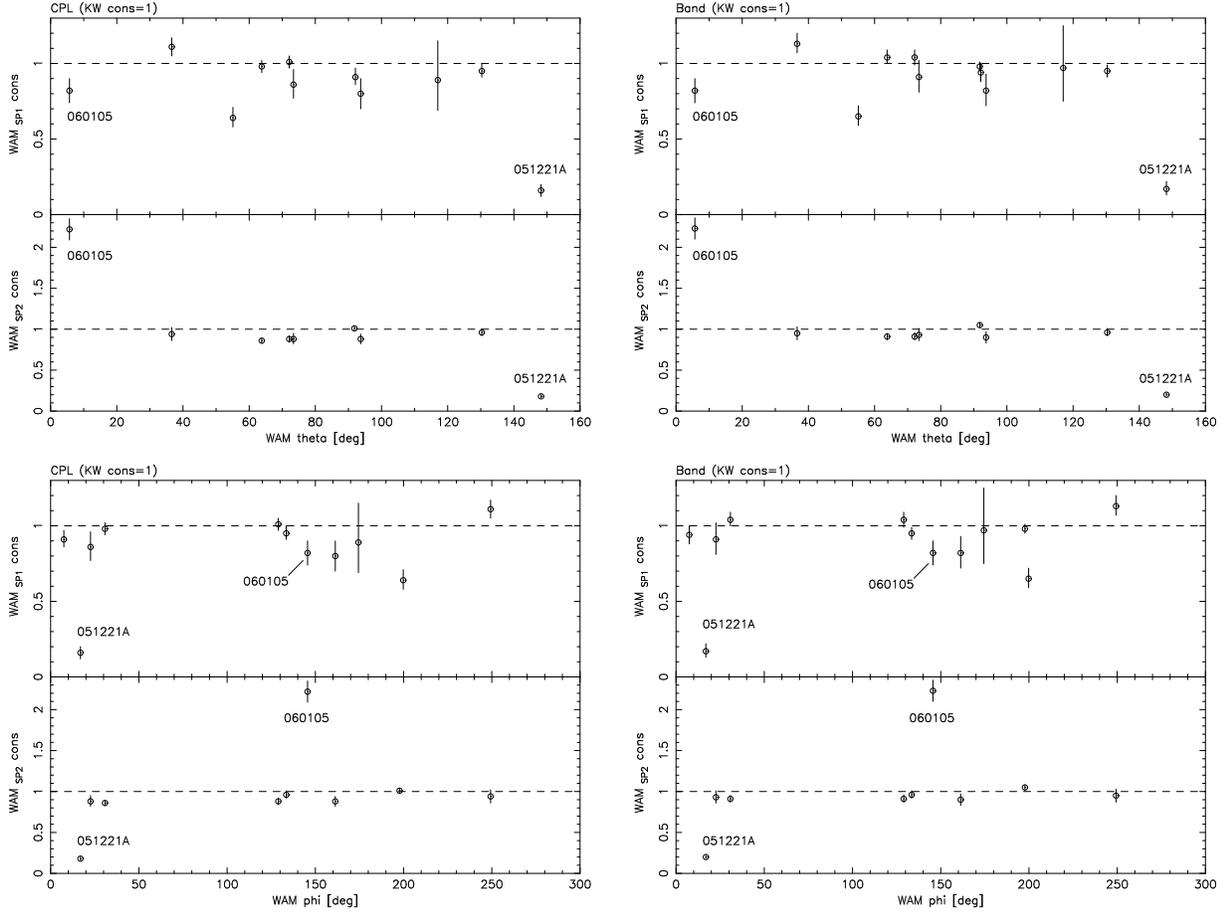

\begin{center}
\includegraphics[scale=0.33,angle=-90]{figure20a.eps}
\hspace{3mm}
\includegraphics[scale=0.33,angle=-90]{figure20b.eps}
\end{center}
\begin{center}
\includegraphics[scale=0.33,angle=-90]{figure20c.eps}
\hspace{3mm}
\includegraphics[scale=0.33,angle=-90]{figure20d.eps}
\end{center}
\caption{The WAM$_{SP1}$ and WAM$_{SP2}$ constant factor based on a CPL fit (left) and 
the Band fit (right) as a function of the incident angle $\theta$ (top) and 
$\phi$ (bottom) of WAM.}
\label{fig:kw_wam_bat_wamangle}
\end{figure}

\begin{figure}
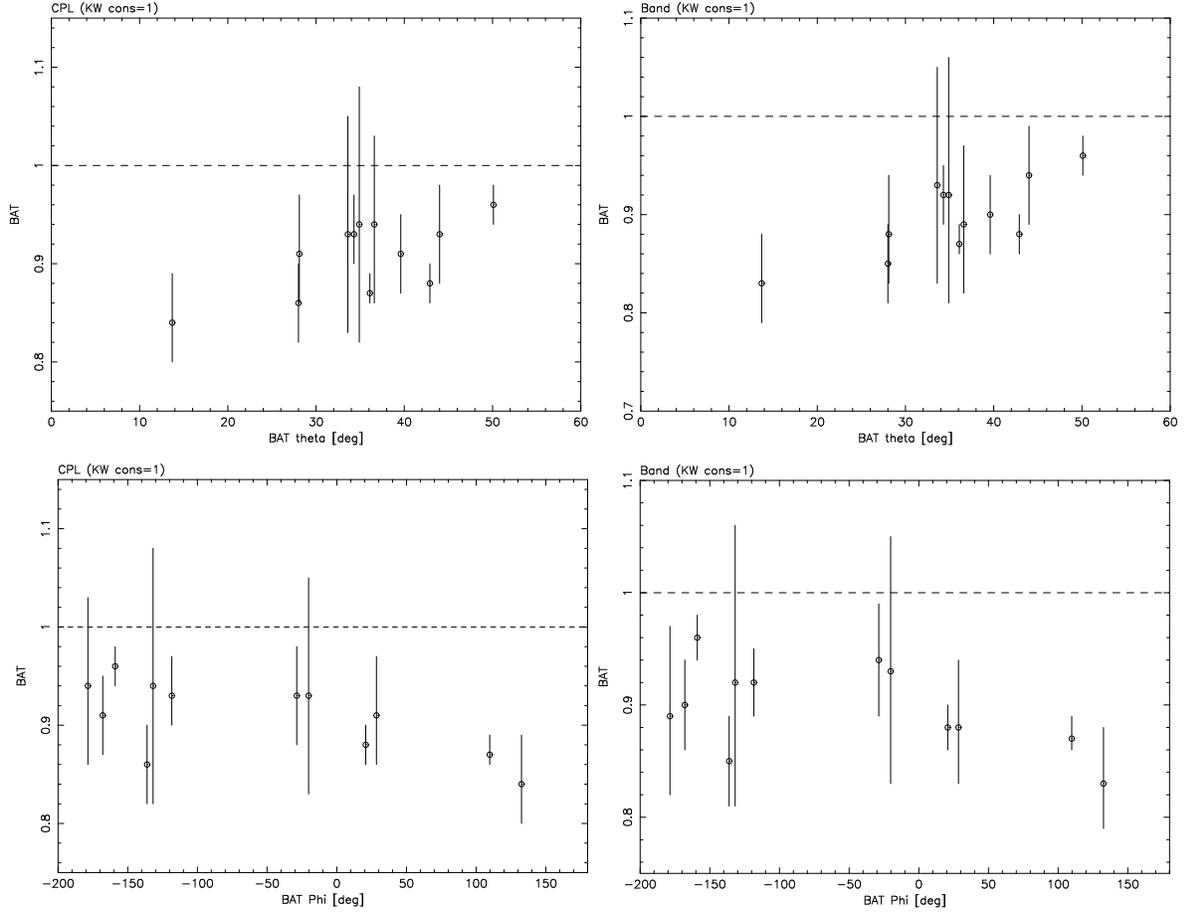

\begin{center}
\includegraphics[scale=0.33,angle=-90]{figure21a.eps}
\includegraphics[scale=0.33,angle=-90]{figure21b.eps}
\end{center}
\begin{center}
\includegraphics[scale=0.33,angle=-90]{figure21c.eps}
\includegraphics[scale=0.33,angle=-90]{figure21d.eps}
\end{center}
\caption{The BAT constant factor based on a CPL fit (left) and 
the Band fit (right) as a function of the incident angle $\theta$ (top) and 
$\phi$ (bottom) of BAT.}
\label{fig:kw_wam_bat_batangle}
\end{figure}

\begin{figure}
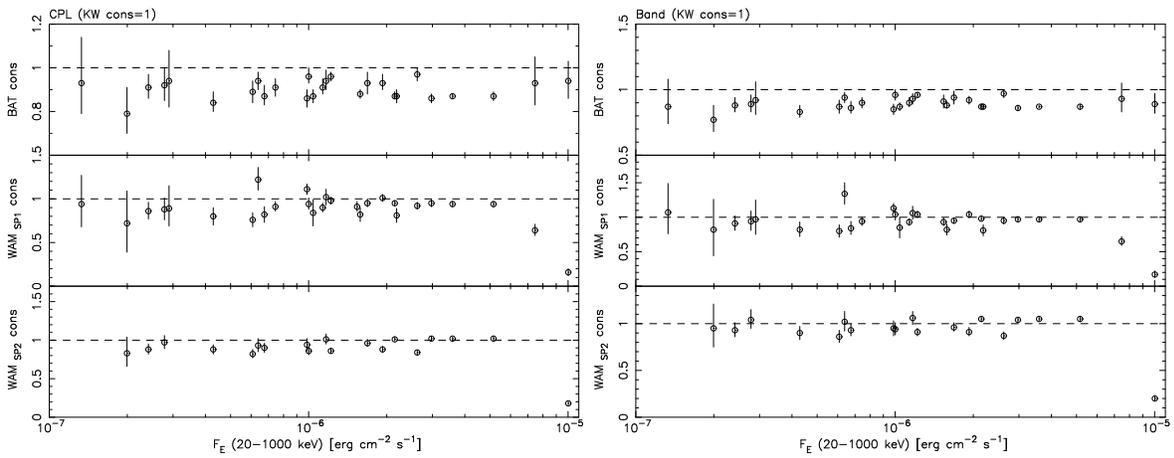

\begin{center}
\includegraphics[scale=0.33,angle=-90]{figure22a.eps}
\includegraphics[scale=0.33,angle=-90]{figure22b.eps}
\end{center}
\caption{The BAT, WAM$_{SP1}$ and WAM$_{SP2}$ constant factor based on a CPL fit (left) and 
the Band fit (right) as a function of the observed energy flux in the 20-1000 keV band.}
\label{fig:kw_wam_bat_flux}
\end{figure}







\newpage
\begin{table}
\caption{The GRB trigger time of each instrument.  Times are all in UT.}\label{tab:grb_trigtime}
\begin{center}
\begin{tabular}{lccc}\hline\hline
GRB     & T$_{0}$ (KW)           & T$_{0}$ (BAT) & T$_{0}$ (WAM)\\\hline
051008  & 2005/10/08 16:33:20.762 & 16:33:21.316  & 16:33:17.470\\
051221A & 2005/12/21 01:51:12.976 & 01:51:15.868  & 01:51:15.888\\
060105  & 2006/01/05 06:49:04.371 & 06:49:27.446  & 06:49:25.696\\
060117  & 2006/01/17 06:49:57.852 & 06:50:01.599  & 06:50:00.572\\
060124\footnotemark[$*$] & 2006/01/24 16:04:13.894 & 15:54:51.825  & 16:04:10.646\\
060502A & 2006/05/02 03:03:33.119 & 03:03:32.144  & 03:03:31.501\\
060813  & 2006/08/13 22:50:21.576 & 22:50:22.685  & 22:50:24.295\\
060814  & 2006/08/14 23:02:34.447 & 23:02:19.035  & 23:02:17.783\\
060904A & 2006/09/04 01:04:13.821 & 01:03:21.201  & 01:04:09.040\\
060912  & 2006/09/12 13:55:57.788 & 13:55:54.144  & 13:55:55.435\\
061006  & 2006/10/06 16:45:26.896 & 16:45:50.510  & 16:45:27.638\\
061007  & 2006/10/07 10:08:09.344 & 10:08:08.812  & 10:08:05.628\\
061222A & 2006/12/22 03:30:14.682 & 03:28:52.110  & 03:30:13.420\\
070328  & 2007/03/28 03:53:49.993 & 03:53:53.155  & 03:53:46.348\\
\hline\hline
\multicolumn{4}{@{}l@{}}{\hbox to 0pt{\parbox{100mm}{\footnotesize
\par\noindent
\footnotemark[$*$] The BAT triggered on the weak precursor, whereas the 
KW and WAM triggered on the main part of the burst which is $\sim$560 
sec later. }\hss}}
\end{tabular}
\end{center}
\end{table}

\begin{table}
  \caption{The GRB sky position information from the BAT instrument, the incident 
angle of GRBs, and the WAM detector ID number used in the analysis for WAM$_{SP1}$ and 
WAM$_{SP2}$ pair.  The KW detector which triggered on the GRB is shown in square 
brackets in the incidnet angle column of the KW.}\label{tab:list_grb}
  \begin{center}
    \begin{tabular}{lcccccc}
      \hline
      GRB & Position (deg) & \multicolumn{3}{c}{Incident Angle (deg)} & \multicolumn{2}{c}{WAM Det ID} \\
          & (R.A.$_{\rm J2000}$, Dec.$_{\rm J2000}$) & KW ($\theta$) & BAT ($\theta$, $\phi$) & WAM ($\theta$, $\phi$) & WAM$_{SP1}$ & WAM$_{SP2}$\\\hline
051008    & $(202.865, +42.103)$ & 46.9 [S2] & $(44.0, -28.7)$  & (130.3, 133.5) & 0 & 3\\
051221A   & $(328.715, +16.888)$ & 27.7 [S2] & $(36.6, -178.6)$ & (148.2, 16.9)  & 0 & 1\\
060105    & $(297.485, +46.356)$ & 65.2 [S2] & $(42.9, +20.8)$  & (5.7, 145.6)   & 0 & 3\\
060117    & $(327.912, -59.982)$ & 43.5 [S1] & $(50.1, -159.0)$ & (63.8, 30.8)   & 0 & 1\\
060124    & $(77.097, +69.727)$  & 46.6 [S2] & $(37.8, -158.3)$ & (134.5, 215.3) & 2 & 3\\
060502A   & $(240.937, +66.604)$ & 78.6 [S2] & $(34.9, -131.8)$ & (117.0,174.4)  & 3 & -\\
060813    & $(111.890, -29.844)$ & 39.0 [S1] & $(34.3, -118.4)$ & (72.1, 129.0)  & 0 & 3\\
060814    & $(221.338, +20.591)$ & 55.4 [S2] & $(28.1, +28.5)$  & (73.4, 22.6)   & 0 & 1\\
060904A   & $(237.731, +44.984)$ & 27.6 [S2] & $(13.7, +132.6)$ & (93.7, 161.3)  & 0 & 3\\
060912    & $(5.285, +20.971)$   & 72.9 [S2] & $(33.2, +125.7)$ & (116.4, 274.1) & 2 & -\\
061006    & $(110.998, -79.195)$ & 13.9 [S1] & $(33.6, -20.2)$  & (55.1, 199.8)  & 3 & -\\
061007    & $(46.299, -50.496)$  & 27.0 [S1] & $(36.1, +109.8)$ & (91.8, 197.7)  & 2 & 3\\
061222A   & $(358.254, 46.524)$  & 47.6 [S2] & $(28.0, -136.1)$ & (36.6, 249.3)  & 2 & 3\\
070328    & $(65.113, -34.079)$  & 35.7 [S1] & $(39.6, -167.9)$ & (92.1, 7.5)    & 1 & -\\
      \hline
    \end{tabular}
  \end{center}
\end{table}

\begin{table}
\caption{The energy fluence measured with KW, WAM and BAT in their full energy band.  
The KW, BAT and WAM fluence is measured using the time-averaged spectrum 
(T$_{100}$ interval).  For the KW, in case the burst emission started before the trigger 
time (e.g. GRB 051008 and GRB 060502A), the fluence for the pre-trigger part is estimated 
using a product of the count fluence for this part and the conversion factor from a count 
fluence to an energy fluence using the spectral information of the time-averaged spectrum.  
Then, the fluence of pre-trigger part is added to the fluence of the main part to obtain the 
total fluence.
\label{tbl:fluence}}
\begin{center}
\begin{tabular}{c|c|c|c}\hline
GRB     & S$^{\rm KW}$(20-10000 keV) & S$^{\rm WAM}$(100-1000 keV) & S$^{\rm BAT}$(15-150 keV)\\
        & [erg cm$^{-2}$] & [erg cm$^{-2}$] & [erg cm$^{-2}$]\\\hline
051008  & $5.2_{-2.1}^{+3.0} \times 10^{-5}$ & $(2.85 \pm 0.08) \times 10^{-5}$ & -     \\
051221A & $3.4_{-0.5}^{+0.8} \times 10^{-6}$ & $(1.5 \pm 0.1) \times 10^{-6}$ & $(1.16 \pm 0.04) \times 10^{-6}$\\
060105  & $(7.9 \pm 0.3) \times 10^{-5}$ & $(5.0 \pm 0.5) \times 10^{-5}$ & $(1.82 \pm 0.03) \times 10^{-5}$\\ 
060117  & $(3.1 \pm 0.2) \times 10^{-5}$ & $(1.54 \pm 0.07) \times 10^{-5}$ & $(2.05 \pm 0.03) \times 10^{-5}$\\
060124  & $3.1_{-0.7}^{+0.8} \times 10^{-5}$ & $(7.8 \pm 0.9) \times 10^{-6}$ & - \\
060502A & $3.8_{-0.8}^{+1.3} \times 10^{-6}$ & $(2.7 \pm 0.5) \times 10^{-6}$ & $(2.3 \pm 0.1) \times 10^{-6}$\\
060813  & $1.8_{-0.2}^{+0.3} \times 10^{-5}$ & $(1.00 \pm 0.05) \times 10^{-5}$ & $(5.5 \pm 0.1) \times 10^{-6}$\\ 
060814  & $3.7_{-0.6}^{+1.1} \times 10^{-5}$ & $(2.5 \pm 0.1) \times 10^{-5}$ & $(1.48 \pm 0.02) \times 10^{-5}$\\
060904A & $1.3_{-0.2}^{+0.4} \times 10^{-5}$ & $(6.3 \pm 0.7) \times 10^{-6}$ & $(7.8 \pm 0.2) \times 10^{-6}$\\ 
060912  & $5.7_{-1.6}^{+2.2} \times 10^{-6}$ & $(1.3 \pm 0.3) \times 10^{-6}$ & $(1.36 \pm 0.06) \times 10^{-6}$\\ 
061006  & $3.7_{-0.6}^{+0.8} \times 10^{-6}$ & $(2.6 \pm 0.1) \times 10^{-6}$ & $(1.4 \pm 0.1) \times 10^{-6}$\\
061007  & $(2.6 \pm 0.2) \times 10^{-4}$ & $(1.70 \pm 0.07) \times 10^{-4}$ & $(4.50 \pm 0.05) \times 10^{-5}$\\ 
061222A & $3.4_{-0.7}^{+0.9} \times 10^{-5}$ & $(1.75 \pm 0.08) \times 10^{-5}$ & $(8.1 \pm 0.2) \times 10^{-6}$\\
070328  & $7.9_{-1.4}^{+1.5} \times 10^{-5}$ & $(3.2 \pm 0.2) \times 10^{-5}$ & $(9.2 \pm 0.2) \times 10^{-6}$\\\hline
    \end{tabular}
  \end{center}
\end{table}

\begin{table}
\caption{The 1-s peak energy flux measured with KW, WAM and BAT in their full energy band.  
The KW 1-s peak energy flux is calculated as a product of 1-s peak count-rate and 
the conversion factor from a count rate to an energy flux using the spectrum accumulated over the 
time interval which comprises the peak.  The BAT and WAM 1-s peak 
energy flux is measured using the 1-s duration spectrum including the brightest part of the burst emission.  
\label{tbl:peakflux}}
\begin{center}
\begin{tabular}{c|c|c|c}\hline
GRB     & F$_{1s}^{\rm KW}$(20-10000 keV) & F$_{1s}^{\rm WAM}$(100-1000 keV) & F$_{1s}^{\rm BAT}$(15-150 keV)\\
        & [erg cm$^{-2}$ s$^{-1}$] & [erg cm$^{-2}$ s$^{-1}$] & [erg cm$^{-2}$ s$^{-1}$]\\\hline
051008  & $(5.1 \pm 1.1) \times 10^{-6}$ & $(2.0 \pm 0.1) \times 10^{-6}$ & -     \\
051221A & $3.4_{-0.5}^{+0.8} \times 10^{-6}$ & $(1.5 \pm 0.1) \times 10^{-6}$ & $(1.02 \pm 0.03) \times 10^{-6}$\\
060105  & $(5.5 \pm 0.6) \times 10^{-6}$ & $(3.3 \pm 0.6) \times 10^{-6}$ & $(7.1 \pm 0.4) \times 10^{-7}$\\ 
060117  & $6.5_{-0.5}^{+0.6} \times 10^{-6}$ & $(2.9 \pm 0.2) \times 10^{-6}$ & $(3.70 \pm 0.08) \times 10^{-6}$\\
060124  & $(3.4 \pm 0.9) \times 10^{-6}$ & $(1.0 \pm 0.2) \times 10^{-6}$ & - \\
060502A & $5.5_{-1.6}^{+2.2} \times 10^{-7}$ & $(3.8 \pm 1.1) \times 10^{-7}$ & $(1.7 \pm 0.2) \times 10^{-7}$\\
060813  & $3.6_{-0.5}^{+0.6} \times 10^{-6}$ & $(2.0 \pm 0.2) \times 10^{-6}$ & $(8.0 \pm 0.3) \times 10^{-7}$\\ 
060814  & $(2.1 \pm 0.3) \times 10^{-6}$ & $(1.4 \pm 0.1) \times 10^{-6}$ & $(6.1 \pm 0.2) \times 10^{-7}$\\
060904A & $1.3_{-0.2}^{+0.4} \times 10^{-6}$ & $(5.4 \pm 1.4) \times 10^{-7}$ & $(4.4 \pm 0.2) \times 10^{-7}$\\ 
060912  & $2.5_{-0.8}^{+1.0} \times 10^{-6}$ & $(5.2 \pm 1.5) \times 10^{-7}$ & $(5.8 \pm 0.3) \times 10^{-7}$\\ 
061006  & $3.7_{-0.6}^{+0.8} \times 10^{-6}$ & $(2.6 \pm 0.1) \times 10^{-6}$ & $(5.3 \pm 0.2) \times 10^{-7}$\\
061007  & $(1.2 \pm 0.1) \times 10^{-5}$ & $(9.2 \pm 0.3) \times 10^{-6}$ & $(1.56 \pm 0.04) \times 10^{-6}$\\ 
061222A & $4.8_{-1.2}^{+1.4} \times 10^{-6}$ & $(3.2 \pm 0.2) \times 10^{-6}$ & $(5.8 \pm 0.2) \times 10^{-7}$\\
070328  & $(5.9 \pm 1.2) \times 10^{-6}$ & $(1.9 \pm 0.1) \times 10^{-6}$ & $(3.9 \pm 0.2) \times 10^{-7}$\\\hline
    \end{tabular}
  \end{center}
\end{table}

\begin{longtable}{ll|cc|cc|cc|ccc}
\caption{Spectral intervals used in the analysis and time of flight from the spacecraft to the Earth. } \label{tab:spec_interval} 
\hline
GRB & Region & \multicolumn{2}{c|}{K-W} & \multicolumn{2}{c|}{BAT} & \multicolumn{2}{c|}{WAM} & \multicolumn{3}{c}{Time of flight}\\
    &        & t$^{\rm K-W}_{\rm start}$  & t$^{\rm K-W}_{\rm stop}$  & 
t$^{\rm BAT}_{\rm start}$ & t$^{\rm BAT}_{\rm stop}$ & 
t$^{\rm WAM}_{\rm start}$ & t$^{\rm WAM}_{\rm stop}$ & K-W$\rightarrow$$\oplus$ & BAT$\rightarrow$$\oplus$
& WAM$\rightarrow$$\oplus$ \\\hline
\endfirsthead
\hline
\endhead
\hline
\endfoot
\endlastfoot

051008  & Reg1  & 0.128  & 8.192  & 2.152     & 10.216    & 6.0  & 14.0  &  2.576   & $-0.001$   &  0.021   \\
        & Reg2  & 16.384 & 49.152 & NA        & NA        & 22.0  & 55.0  &          &            &          \\
        & Reg12 & 0.128  & 49.152 & NA        & NA        & 6.0  & 55.0  &          &            &          \\\hline
051221A & Reg1  & 0.0    & 0.256  & 0.064     & 0.320     & 0.0  & 1.0  &  2.962   &  0.006     & 0.015    \\\hline
060105  & Reg1  & 16.64  & 37.376 & $-4.094$  & 16.642    & $-2.0$  & 19.0  &  2.362   &  0.021     & $-0.000$ \\
        & Reg2$^{\rm S}$  & 37.376 & 58.88  & 16.642    & 38.146    & 19.0  & 40.0  &          &            &          \\
        & Reg12$^{\rm S}$ & 16.64  & 58.88  & $-4.094$  & 16.642    & $-2.0$  & 40.0  &          &            &          \\\hline
060117  & Reg1  & 0.192  & 8.448  & $-0.055$  & 8.200     & 1.0  & 9.0  &  3.514   &  0.015     & $-0.000$ \\
        & Reg2  & 8.448  & 13.312 & 8.200     & 13.064    & 9.0  & 14.0 &          &            &          \\
        & Reg3  & 13.312 & 21.504 & 13.064    & 21.256    & 14.0 & 22.0 &          &            &          \\
	& Reg13 & 0.192  & 21.504 & $-0.055$  & 21.256    & 1.0  & 22.0 &          &            &          \\\hline
060124  & Reg1  & 0.0    & 8.448  &  -        & -         & 0.0  & 9.0  & $-3.024$ &  0.011     &  0.004   \\
        & Reg2  & 8.448  & 16.640 &  -        & -         & 9.0  & 17.0 &          &            &          \\
        & Reg3  & 16.640 & 33.024 &  -        & -         & 17.0 & 33.0 &          &            &          \\
	& Reg13 & 0.0    & 33.024 &  -        & -         & 0.0  & 33.0 &          &            &          \\\hline
060502A & Reg1  & 0.0    & 8.448  & 0.638     & 9.086     & 2.0 & 10.5 & $-0.326$ & 0.010      &  0.006   \\\hline
060813  & Reg1  & 0.0    & 6.656  & 2.168     & 8.824     & 0.5 &  7.0 & 3.293    & 0.016      &  0.012   \\\hline
060814  & Reg1$^{\rm S}$ & 0.0    & 16.64     & 15.247    & 31.887    & 16.0  & 33.0  & $-0.152$ & 0.014      &  0.015   \\ 
        & Reg2  & 49.408 & 73.984 & 64.655    & 89.231    & 65.5 & 90.5 &          &            &          \\
	& Reg12$^{\rm S}$ & 0.0    & 73.984 & 15.247    & 89.231 & 16.5 & 90.5 &          &            &          \\\hline
060904A & Reg1  & 0.0    & 8.448  & 52.942    & 61.390    & 5.0  & 13.5  & 0.3267   & 0.003      &  0.003   \\
        & Reg2  & 8.448  & 16.64  & 61.390    & 69.582    & 13.5 & 21.5  &          &            &          \\
        & Reg12 & 0.0    & 16.64  & 52.942    & 69.582    & 5.0  & 21.5  &          &            &          \\\hline
060912  & Reg1  & 0.0    & 8.448  & 0.319     & 8.767     & -0.5 & 8.0  & $-3.306$ & 0.020      & 0.016    \\\hline
061006  & Reg1  & 0.0    & 0.256  & $-22.703$ & $-22.447$ & 0.0  & 0.5  & 0.921    & 0.010      & $-0.003$ \\\hline
061007  & Reg1  & 0.0    & 15.872 & $-1.057$  & 14.815    & 2.0  & 18.0  & $-1.577$ & 0.011      & 0.013    \\
        & Reg2$^{\rm S}$  & 24.064 & 40.704 & 23.007    & 39.647    & 26.0  & 43.0 &          &            &          \\
        & Reg3$^{\rm S}$  & 39.680 & 70.912 & 38.623    & 69.855    & 42.0  & 73.0 &          &            &          \\
	& Reg4$^{\rm S}$  & 24.064 & 70.912 & 23.007    & 69.855    & 26.0  & 73.0  &          &            &          \\
	& Reg5  & 70.912 & 87.296 & 69.855    & 86.239    & 73.0  & 89.0  &          &            &          \\
	& Reg15$^{\rm S}$ & 0.0    & 87.296 & $-1.057$  & 86.239    & 2.0  & 89.0  &          &            &          \\\hline
061222A & Reg1$^{\rm S}$  & 0.0    & 15.104 & 81.808    & 96.912    & 0.5  & 15.5  & $-0.745$ & 0.018      & -0.000   \\\hline
070328  & Reg1  & 0.0    & 8.448  & $-1.107$  & 7.341     & 5.5  & 14.0  & 2.071    & 0.016      & 0.012    \\
        & Reg2$^{\rm S}$  & 0.0    & 24.832 & $-1.107$  & 23.725    & 5.5  & 30.5  &          &            &          \\
	& Reg3$^{\rm S}$  & 24.832 & 41.216 & 23.725    & 40.109    & 30.5 & 47.0  &          &            &          \\
	& Reg13$^{\rm S}$ & 0.0    & 41.216 & $-1.107$  & 40.109    & 5.5  & 47.0  &          &            &          \\\hline
\hline\hline
\multicolumn{4}{@{}l@{}}{\hbox to 0pt{\parbox{180mm}{\footnotesize
\par\noindent
\footnotemark[S] The spectral interval which contains the $Swift$ spacecraft slew in the BAT data.}\hss}}
\end{longtable}

\newpage
\begin{table}
\caption{The best fit parameters by a linear function between parameters in a CPL model 
derived by the joint fit and the KW fit.  
For example, the linear function for $\alpha$ in the KW-BAT fits vs. the KW fit is 
$\alpha_{\rm (KW-BAT)} = {\rm a} \times \alpha_{\rm (KW)} + {\rm b}$.  The ``Cons'' row is the constant factor 
of the instrument shown in the parenthesis (B: BAT, W: WAM).  
}\label{tab:fit_param_cpl}
\begin{center}
\footnotesize
\begin{tabular}{c|cc|cc|cc|cc}\hline
    &  \multicolumn{2}{c|}{KW-BAT} & \multicolumn{2}{c|}{KW-WAM} & \multicolumn{2}{c|}{WAM-BAT} & \multicolumn{2}{c}{KW-WAM-BAT} \\\hline
    & a & b & a & b & a & b & a & b\\
    $\alpha$ & $0.92\pm0.05$ & $-(0.11 \pm 0.05)$ & $1.06\pm0.04$ & $-(0.01\pm0.04)$ & $0.93\pm0.04$ & $-(0.17\pm0.04)$ & $1.00\pm0.04$ & $-(0.09\pm0.03)$\\
    $\ep$    & $1.04\pm0.03$ & $-(2\pm6)$     & $1.06\pm0.02$ & $1\pm3$      & $1.11\pm0.03$ & $-(4\pm7)$ & $1.09\pm0.03$ & $-(2\pm6)$\\\hline
    Cons & \multicolumn{2}{c|}{$0.886\pm0.009$(B)} & \multicolumn{2}{c|}{$0.95\pm0.01$(W)} & \multicolumn{2}{c|}{$1.02\pm0.02$(W)} & $0.900\pm0.009$(B) & $0.948\pm0.011$(W)\\\hline
    \end{tabular}
  \end{center}
\end{table}

\newpage
\begin{table}
\caption{The best fit parameters by a linear function between parameters in the Band function derived by the joint 
fit and the KW fit.  
For example, the linear function for $\alpha$ in the KW-BAT fits vs. the KW fit is 
$\alpha_{\rm (KW-BAT)} = {\rm a} \times \alpha_{\rm (KW)} + {\rm b}$.  The ``Cons'' row is the constant factor 
of the instrument shown in the parenthesis (B: BAT, W: WAM).}\label{tab:fit_param_band}
\begin{center}
\footnotesize
\begin{tabular}{c|cc|cc|cc|cc}\hline
    &  \multicolumn{2}{c|}{KW-BAT} & \multicolumn{2}{c|}{KW-WAM} & \multicolumn{2}{c|}{WAM-BAT} & \multicolumn{2}{c}{KW-WAM-BAT} \\\hline
    & a & b & a & b & a & b & a & b \\
    $\alpha$ & $0.87\pm0.07$ & $-(0.20 \pm 0.06)$ & $0.97\pm0.07$ & $-(0.06\pm0.05)$ & $0.87\pm0.08$ & $-(0.26\pm0.05)$ & $0.90\pm0.07$ & $-(0.20\pm0.05)$\\
    $\ep$    & $1.09\pm0.04$ & $-(7\pm7)$     & $1.06\pm0.02$ & $-(6\pm6)$      & $1.14\pm0.04$ & $-(12\pm6)$ & $1.12\pm0.03$ & $-(9\pm6)$\\
    $\beta$  & $0.7\pm0.2$ & $-(0.9\pm0.4)$         & $1.15\pm0.19$ & $0.3\pm0.5$ & $0.4\pm0.3$ & $-(1.7\pm0.7)$ & $0.5\pm0.2$ & $-(1.3\pm0.4)$\\\hline
    Cons & \multicolumn{2}{c|}{$0.881\pm0.009$(B)} & \multicolumn{2}{c|}{$0.97\pm0.01$(W)} & \multicolumn{2}{c|}{$1.08\pm0.02$(W)} & $0.887\pm0.008$(B) & $0.981\pm0.012$(W)\\\hline
    \end{tabular}
  \end{center}
\end{table}

\begin{table}
\caption{Spectral parameters: 051008 Reg1 \label{tbl:051008_reg1_spec}}
\begin{center}
\footnotesize
\begin{tabular}{c|c|c|c|c|c|c|c|c}\hline
Instrument & Model & $\alpha$ & $\beta$ & $\ep$ [keV] & C(WAM0) & C(WAM3) & C(BAT) & $\chi^{2}$/dof\\\hline
KW & PL & $-1.40_{-0.03}^{+0.03}$ & - & - & - & - & - & 215/73 (2.95)\\
KW & CPL & $-0.93_{-0.09}^{+0.10}$ & - & $803_{-130}^{+175}$ & - & - & - & 77/72 (1.07)\\
KW & Band & $-0.86_{-0.12}^{+0.17}$ & $-2.3_{-1.3}^{+0.4}$ & $671_{-174}^{+206}$ & - & - & - & 73/71 (1.04)\\\hline
WAM0 & PL & $-1.66_{-0.03}^{+0.03}$ & - & - & - & - & - & 144/25 (5.79)\\
WAM0 & CPL & $-1.02_{-0.13}^{+0.15}$ & - & $871_{-117}^{+155}$ & - & - & - & 27/24 (1.13)\\
WAM0 & Band & $-0.91_{-0.18}^{+0.21}$ & $-2.6_{-1.5}^{+0.3}$ & $757_{-128}^{+170}$ & - & - & - & 24/23 (1.05)\\\hline
WAM3 & PL & $-1.72_{-0.04}^{+0.04}$ & - & - & - & - & - & 111/25 (4.44)\\
\hline
\end{tabular}
\end{center}
{\footnotesize
\footnotemark[$\dagger$] The joint fit of two WAM spectra.
}
\end{table}
\begin{table}
\caption{Spectral parameters: 051008 Reg2}
\begin{center}
\footnotesize
\begin{tabular}{c|c|c|c|c|c|c|c}\hline
Instrument & Model & $\alpha$ & $\beta$ & $\ep$ [keV] & C(WAM0) & C(WAM3) & $\chi^{2}$/dof\\\hline
KW & PL & $-1.45_{-0.11}^{+0.11}$ & - & - & - & - & 81/60 (1.36)\\
KW & CPL & $-0.95_{-0.33}^{+0.44}$ & - & $394_{-142}^{+462}$ & - & - & 73/59 (1.24)\\
KW & Band & $-0.96_{-0.20}^{+0.46}$ & -* & $406_{-115}^{+628}$ & - & - & 73/58 (1.26)\\\hline
WAM0 & PL & $-1.54_{-0.12}^{+0.00}$ & - & - & - & - & 43/23 (1.87)\\
WAM0 & CPL & $-0.61_{-0.53}^{+0.66}$ & - & $604_{-138}^{+350}$ & - & - & 31/22 (1.45)\\
WAM0 & Band & $-0.62_{-0.58}^{+1.78}$ & $<-1.78$ & $607_{-283}^{+286}$ & - & - & 31/21 (1.51)\\\hline
WAM3 & PL & $-1.51_{-0.13}^{+0.00}$ & - & - & - & - & 19/23 (0.86)\\
\hline
\end{tabular}
\end{center}
{\footnotesize
\footnotemark[*] The parameter is not constrained.\\
\footnotemark[$\dagger$] The joint fit of two WAM spectra.
}
\end{table}
\begin{table}
\caption{Spectral parameters: 051008 Reg12}
\begin{center}
\footnotesize
\begin{tabular}{c|c|c|c|c|c|c|c}\hline
Instrument & Model & $\alpha$ & $\beta$ & $\ep$ [keV] & C(WAM0) & C(WAM3) & $\chi^{2}$/dof\\\hline
KW & PL & $-1.43_{-0.04}^{+0.04}$ & - & - & - & - & 129/73 (1.77)\\
KW & CPL & $-0.91_{-0.15}^{+0.17}$ & - & $597_{-126}^{+211}$ & - & - & 68/72 (0.96)\\
KW & Band & $-0.81_{-0.20}^{+0.61}$ & $-2.12_{-1.15}^{+0.40}$ & $477_{-248}^{+202}$ & - & - & 65/71 (0.92)\\\hline
WAM0 & PL & $-1.63_{-0.04}^{+0.04}$ & - & - & - & - & 101/25 (4.06)\\
WAM0 & CPL & $-1.05_{-0.24}^{+0.21}$ & - & $885_{-175}^{+262}$ & - & - & 41/24 (1.73)\\
WAM0 & Band & $-0.83_{-0.25}^{+0.55}$ & $-2.30_{-0.71}^{+0.33}$ & $671_{-207}^{+218}$ & - & - & 37/23 (1.62)\\\hline
WAM3 & PL & $-1.66_{-0.05}^{+0.05}$ & - & - & - & - & 51/25 (2.06)\\
\hline
\end{tabular}
\end{center}
{\footnotesize
\footnotemark[$\dagger$] The joint fit of two WAM spectra.
}
\end{table}
\begin{table}
\caption{Spectral parameters: 051221A Reg1}
\begin{center}
\footnotesize
\begin{tabular}{c|c|c|c|c|c|c|c|c}\hline
Instrument & Model & $\alpha$ & $\beta$ & $\ep$ [keV] & C(WAM0) & C(WAM1) & C(BAT) & $\chi^{2}$/dof\\\hline
KW & PL & $-1.49_{-0.05}^{+0.05}$ & - & - & - & - & - & 54/37 (1.48)\\
KW & CPL & $-1.09_{-0.16}^{+0.18}$ & - & $455_{-109}^{+200}$ & - & - & - & 30/36 (0.83)\\
KW & Band & $-0.84_{-0.30}^{+0.76}$ & $-1.91_{-2.37}^{+0.24}$ & $258_{-137}^{+228}$ & - & - & - & 27/35 (0.78)\\\hline
WAM0 & PL & $-1.56_{-0.48}^{+0.00}$ & - & - & - & - & - & 18/14 (1.35)\\
WAM0 & CPL & $-1.54_{-0.51}^{+0.42}$ & - & -* & - & - & - & 18/13 (1.45)\\
WAM0 & Band & - & - & - & - & - & - & -\\\hline
WAM1 & PL & $-1.98_{-0.18}^{+0.00}$ & - & - & - & - & - & 7/14 (0.51)\\
\hline
\end{tabular}
\end{center}
{\footnotesize
\footnotemark[*] The parameter is not constrained. \\
\footnotemark[$\dagger$] The joint fit of two WAM spectra.
}
\end{table}
\begin{table}
\caption{Spectral parameters: 060105 Reg1}
\begin{center}
\footnotesize
\begin{tabular}{c|c|c|c|c|c|c|c|c}\hline
Instrument & Model & $\alpha$ & $\beta$ & $\ep$ [keV] & C(WAM0) & C(WAM3) & C(BAT) & $\chi^{2}$/dof\\\hline
KW & PL & $-1.35$ & - & - & - & - & - & 894/62 (14.44)\\
KW & CPL & $-0.65_{-0.06}^{+0.06}$ & - & $567_{-32}^{+37}$ & - & - & - & 69/61 (1.13)\\
KW & Band & $-0.65_{-0.06}^{+0.06}$ & -* & $569_{-32}^{+36}$ & - & - & - & 69/60 (1.15)\\\hline
WAM0 & PL & $-1.64_{-0.15}^{+0.00}$ & - & - & - & - & - & 15/14 (1.12)\\
WAM0 & CPL & $-0.69_{-0.67}^{+0.82}$ & - & $724_{-165}^{+583}$ & - & - & - & 9/13 (0.76)\\
WAM0 & Band & $-0.64_{-0.69}^{+1.08}$ & $<-1.81$ & $692_{-251}^{+305}$ & - & - & - & 9/12 (0.81)\\\hline
WAM3 & PL & $-2.10_{-0.09}^{+0.09}$ & - & - & - & - & - & 19/14 (1.39)\\
\hline
\end{tabular}
\end{center}
{\footnotesize
\footnotemark[*] The parameter is not constrained.\\
\footnotemark[$\dagger$] The joint fit of two WAM spectra.
}
\end{table}
\begin{table}
\caption{Spectral parameters: 060105 Reg2}
\begin{center}
\footnotesize
\begin{tabular}{c|c|c|c|c|c|c|c|c}\hline
Instrument & Model & $\alpha$ & $\beta$ & $\ep$ [keV] & C(WAM0) & C(WAM3) & C(BAT) & $\chi^{2}$/dof\\\hline
KW & PL & $-1.61$ & - & - & - & - & - & 626/62 (10.11)\\
KW & CPL & $-0.60_{-0.09}^{+0.10}$ & - & $228_{-12}^{+13}$ & - & - & - & 50/61 (0.83)\\
KW & Band & $-0.57_{-0.10}^{+0.11}$ & $<-2.79$ & $222_{-14}^{+15}$ & - & - & - & 47/60 (0.80)\\\hline
WAM0 & PL & $-2.27_{-0.43}^{+0.34}$ & - & - & - & - & - & 18/14 (1.32)\\
WAM0 & CPL & $-1.70_{-0.54}^{+4.23}$ & - & -* & - & - & - & 18/13 (1.40)\\
WAM0 & Band & -* & $-2.55_{-1.15}^{+0.54}$ & $185_{-164}^{+52}$ & - & - & - & 17/12 (1.44)\\\hline
WAM3 & PL & $-2.52_{-0.21}^{+0.19}$ & - & - & - & - & - & 20/14 (1.49)\\
\hline
\end{tabular}
\end{center}
{\footnotesize
\footnotemark[*] The parameter is not constrained.\\
\footnotemark[$\dagger$] The joint fit of two WAM spectra.
}
\end{table}
\begin{table}
\caption{Spectral parameters: 060105 Reg12}
\begin{center}
\footnotesize
\begin{tabular}{c|c|c|c|c|c|c|c|c}\hline
Instrument & Model & $\alpha$ & $\beta$ & $\ep$ [keV] & C(WAM0) & C(WAM3) & C(BAT) & $\chi^{2}$/dof\\\hline
KW & PL & $-1.45$ & - & - & - & - & - & 1043/62 (16.84)\\
KW & CPL & $-0.78_{-0.05}^{+0.05}$ & - & $424_{-22}^{+24}$ & - & - & - & 61/61 (1.00)\\
KW & Band & $-0.74_{-0.06}^{+0.07}$ & $-2.80_{-0.80}^{+0.34}$ & $398_{-31}^{+29}$ & - & - & - & 57/60 (0.95)\\\hline
WAM0 & PL & $-1.81_{-0.15}^{+0.00}$ & - & - & - & - & - & 12/14 (0.88)\\
WAM0 & CPL & $-1.30_{-0.58}^{+0.70}$ & - & $>495$ & - & - & - & 10/13 (0.78)\\
WAM0 & Band & $-1.27_{-8.73}^{+6.27}$ & $<-1.75$ & -* & - & - & - & 10/12 (0.84)\\\hline
WAM3 & PL & $-2.24_{-0.09}^{+0.09}$ & - & - & - & - & - & 27/14 (1.98)\\
\hline
\end{tabular}
\end{center}
{\footnotesize
\footnotemark[*] The parameter is not constrained.\\
\footnotemark[$\dagger$] The joint fit of two WAM spectra.
}
\end{table}
\begin{table}
\caption{Spectral parameters: 060117 Reg1}
\begin{center}
\footnotesize
\begin{tabular}{c|c|c|c|c|c|c|c|c}\hline
Instrument & Model & $\alpha$ & $\beta$ & $\ep$ [keV] & C(WAM0) & C(WAM1) & C(BAT) & $\chi^{2}$/dof\\\hline
KW & PL & $-2.02_{-0.03}^{+0.03}$ & - & - & - & - & - & 135/61 (2.22)\\
KW & CPL & $-1.59_{-0.11}^{+0.12}$ & - & $98_{-10}^{+12}$ & - & - & - & 73/60 (1.22)\\
KW & Band & $-1.51_{-0.17}^{+0.18}$ & $<-2.41$ & $92_{-11}^{+15}$ & - & - & - & 72/59 (1.23)\\\hline
WAM0 & PL & $-2.41_{-0.16}^{+0.14}$ & - & - & - & - & - & 29/22 (1.33)\\
WAM0 & CPL & - & - & - & - & - & - & -\\
WAM0 & Band & - & - & - & - & - & - & -\\\hline
WAM1 & PL & $-2.32_{-0.12}^{+0.11}$ & - & - & - & - & - & 26/22 (1.19)\\
\hline
\end{tabular}
\end{center}
{\footnotesize
\footnotemark[$\dagger$] The joint fit of two WAM spectra.
}
\end{table}
\begin{table}
\caption{Spectral parameters: 060117 Reg2}
\begin{center}
\footnotesize
\begin{tabular}{c|c|c|c|c|c|c|c|c}\hline
Instrument & Model & $\alpha$ & $\beta$ & $\ep$ [keV] & C(WAM0) & C(WAM1) & C(BAT) & $\chi^{2}$/dof\\\hline
KW & PL & $-1.98_{-0.02}^{+0.02}$ & - & - & - & - & - & 335/60 (5.60)\\
KW & CPL & $-1.35_{-0.08}^{+0.08}$ & - & $109_{-6}^{+6}$ & - & - & - & 58/59 (0.99)\\
KW & Band & $-1.32_{-0.09}^{+0.10}$ & $-3.05_{-0.71}^{+0.34}$ & $106_{-7}^{+7}$ & - & - & - & 53/58 (0.92)\\\hline
WAM0 & PL & $-2.70_{-0.10}^{+0.10}$ & - & - & - & - & - & 20/22 (0.95)\\
WAM0 & CPL & - & - & - & - & - & - & - \\
WAM0 & Band & - & - & - & - & - & - & - \\\hline
WAM1 & PL & $-2.53_{-0.07}^{+0.07}$ & - & - & - & - & - & 31/22 (1.43)\\
\hline
\end{tabular}
\end{center}
{\footnotesize
\footnotemark[$\dagger$] The joint fit of two WAM spectra.
}
\end{table}
\begin{table}
\caption{Spectral parameters: 060117 Reg3}
\begin{center}
\footnotesize
\begin{tabular}{c|c|c|c|c|c|c|c|c}\hline
Instrument & Model & $\alpha$ & $\beta$ & $\ep$ [keV] & C(WAM0) & C(WAM1) & C(BAT) & $\chi^{2}$/dof\\\hline
KW & PL & $-2.31_{-0.05}^{+0.05}$ & - & - & - & - & - & 107/60 (1.79)\\
KW & CPL & $-1.79_{-0.17}^{+0.18}$ & - & $32_{-22}^{+12}$ & - & - & - & 67/59 (1.14)\\
KW & Band & $-1.77_{-0.15}^{+0.19}$ & $-3.62_{-0.83}^{+6.28}$ & $33_{-16}^{+12}$ & - & - & - & 67/58 (1.16)\\\hline
WAM0 & PL & $-2.95_{-0.32}^{+0.28}$ & - & - & - & - & - & 29/22 (1.33)\\
WAM0 & CPL & - & - & - & - & - & - & -\\
WAM0 & Band & - & - & - & - & - & - & -\\\hline
WAM1 & PL & $-2.73_{-0.25}^{+0.22}$ & - & - & - & - & - & 35/22 (1.62)\\
\hline
\end{tabular}
\end{center}
{\footnotesize
\footnotemark[$\dagger$] The joint fit of two WAM spectra.
}
\end{table}
\begin{table}
\caption{Spectral parameters: 060117 Reg13}
\begin{center}
\footnotesize
\begin{tabular}{c|c|c|c|c|c|c|c|c}\hline
Instrument & Model & $\alpha$ & $\beta$ & $\ep$ [keV] & C(WAM0) & C(WAM1) & C(BAT) & $\chi^{2}$/dof\\\hline
KW & PL & $-2.05_{-0.02}^{+0.02}$ & - & - & - & - & - & 302/60 (5.04)\\
KW & CPL & $-1.56_{-0.06}^{+0.07}$ & - & $87_{-5}^{+5}$ & - & - & - & 50/59 (0.86)\\
KW & Band & $-1.51_{-0.07}^{+0.09}$ & $-2.88_{-0.46}^{+0.24}$ & $85_{-5}^{+5}$ & - & - & - & 45/58 (0.78)\\\hline
WAM0 & PL & $-2.63_{-0.09}^{+0.09}$ & - & - & - & - & - & 20/22 (0.92)\\
WAM0 & CPL & - & - & - & - & - & - & -\\
WAM0 & Band & - & - & - & - & - & - & -\\\hline
WAM1 & PL & $-2.48_{-0.07}^{+0.06}$ & - & - & - & - & - & 33/22 (1.53)\\
\hline
\end{tabular}
\end{center}
{\footnotesize
\footnotemark[$\dagger$] The joint fit of two WAM spectra.
}
\end{table}
\begin{table}
\caption{Spectral parameters: 060124 Reg1}
\begin{center}
\footnotesize
\begin{tabular}{c|c|c|c|c|c|c|c}\hline
Instrument & Model & $\alpha$ & $\beta$ & $\ep$ [keV] & C(WAM2) & C(WAM3) & $\chi^{2}$/dof\\\hline
KW & PL & $-1.71_{-0.09}^{+0.09}$ & - & - & - & - & 83/60 (1.39)\\
KW & CPL & $-1.18_{-0.28}^{+0.33}$ & - & $198_{-50}^{+112}$ & - & - & 70/59 (1.19)\\
KW & Band & $-1.17_{-0.27}^{+0.33}$ & -* & $199_{-64}^{+110}$ & - & - & 70/58 (1.21)\\\hline
WAM2 & PL & $-2.17_{-0.62}^{+0.45}$ & - & - & - & - & 10/14 (0.72)\\
WAM2 & CPL & - & - & - & - & - & -\\
WAM2 & Band & - & - & - & - & - & -\\\hline
WAM3 & PL & $-2.00_{-0.39}^{+0.31}$ & - & - & - & - & 14/14 (1.02)\\
WAM3 & CPL & - & - & - & - & - & -\\
WAM3 & Band & - & - & - & - & - & -\\\hline
KW+WAM2 & PL & $-1.73_{-0.09}^{+0.09}$ & - & - & $0.77_{-0.22}^{+0.27}$ & - & 96/75 (1.29)\\
KW+WAM2 & CPL & $-1.21_{-0.25}^{+0.29}$ & - & $213_{-51}^{+97}$ & $0.81_{-0.22}^{+0.26}$ & - & 78/74 (1.06)\\
KW+WAM2 & Band & $-1.30_{-0.16}^{+0.28}$ & -* & $228_{-51}^{+97}$ & $0.82_{-0.22}^{+0.26}$ & - & 79/73 (1.08)\\\hline
KW+WAM3 & PL & $-1.73_{-0.09}^{+0.09}$ & - & - & - & $0.78_{-0.16}^{+0.20}$ & 100/75 (1.33)\\
KW+WAM3 & CPL & $-1.29_{-0.23}^{+0.27}$ & - & $239_{-61}^{+133}$ & - & $0.74_{-0.15}^{+0.17}$ & 85/74 (1.15)\\
KW+WAM3 & Band & $-1.29_{-0.13}^{+0.21}$ & -* & $240_{-54}^{+132}$ & - & $0.74_{-0.13}^{+0.15}$ & 85/73 (1.17)\\\hline
KW+2WAM$^{\dagger}$ & PL & $-1.82_{-0.11}^{+0.10}$ & - & - & $0.83_{-0.24}^{+0.29}$ & $0.82_{-0.17}^{+0.21}$ & 105/88 (1.20)\\
KW+2WAM$^{\dagger}$ & CPL & $-1.33_{-0.26}^{+0.30}$ & - & $244_{-60}^{+119}$ & $0.80_{-0.22}^{+0.25}$ & $0.74_{-0.15}^{+0.18}$ & 91/87 (1.05)\\
KW+2WAM$^{\dagger}$ & Band & $-1.44_{-0.16}^{+0.43}$ & -* & $268_{-60}^{+119}$ & $0.82_{-0.22}^{+0.25}$  & $0.76_{-0.15}^{+0.18}$ & 91/86 (1.07)\\
\hline
\end{tabular}
\end{center}
{\footnotesize
\footnotemark[*] The parameter is not constrained.\\
\footnotemark[$\dagger$] The joint fit of two WAM spectra.
}
\end{table}
\begin{table}
\caption{Spectral parameters: 060124 Reg2}
\begin{center}
\footnotesize
\begin{tabular}{c|c|c|c|c|c|c|c}\hline
Instrument & Model & $\alpha$ & $\beta$ & $\ep$ [keV] & C(WAM2) & C(WAM3) & $\chi^{2}$/dof\\\hline
KW & PL & $-1.63_{-0.04}^{+0.04}$ & - & - & - & - & 133/69 (1.94)\\
KW & CPL & $-1.12_{-0.13}^{+0.15}$ & - & $323_{-52}^{+78}$ & - & - & 60/68 (0.90)\\
KW & Band & $-0.96_{-0.21}^{+0.24}$ & $-2.21_{-0.68}^{+0.21}$ & $244_{-52}^{+85}$ & - & - & 55/67 (0.83)\\\hline
WAM2 & PL & $-2.02_{-0.20}^{+0.18}$ & - & - & - & - & 17/14 (1.23)\\
WAM2 & CPL & $-1.21_{-0.78}^{+1.20}$ & - & $<785$ & - & - & 14/13 (1.11)\\
WAM2 & Band & - & - & - & - & - & - \\\hline
WAM3 & PL & $-2.11_{-0.14}^{+0.13}$ & - & - & - & - & 18/14 (1.30)\\
WAM3 & CPL & $-1.23_{-0.65}^{+0.83}$ & - & $281_{-132}^{+74}$ & - & - & 12/13 (0.94)\\
WAM3 & Band & $-1.19_{-0.67}^{+1.20}$ & $<-2.27$ & $279_{-65}^{+72}$ & - & - & 11/12 (1.00)\\\hline
KW+WAM2 & PL & $-1.65_{-0.04}^{+0.03}$ & - & - & $0.70_{-0.08}^{+0.08}$ & - & 164/84 (1.95)\\
KW+WAM2 & CPL & $-1.14_{-0.12}^{+0.13}$ & - & $331_{-47}^{+67}$ & $0.66_{-0.07}^{+0.07}$ & - & 75/83 (0.91)\\
KW+WAM2 & Band & $-0.99_{-0.18}^{+0.23}$ & $-2.26_{-0.69}^{+0.22}$ & $259_{-56}^{+79}$ & $0.66_{-0.07}^{+0.08}$ & - & 70/82 (0.86)\\\hline
KW+WAM3 & PL & $-1.67_{-0.04}^{+0.03}$ & - & - & - & $0.70_{-0.06}^{+0.07}$ & 189/84 (2.26)\\
KW+WAM3 & CPL & $-1.09_{-0.12}^{+0.13}$ & - & $305_{-35}^{+47}$ & - & $0.64_{-0.05}^{+0.05}$ & 73/83 (0.89)\\
KW+WAM3 & Band & $-1.02_{-0.14}^{+0.21}$ & $-2.51_{-1.06}^{+0.36}$ & $274_{-57}^{+51}$ & - & $0.64_{-0.05}^{+0.06}$ & 69/82 (0.85)\\\hline
KW+2WAM$^{\dagger}$ & PL & $-1.69_{-0.04}^{+0.03}$ & - & - & $0.73_{-0.08}^{+0.09}$ & $0.72_{-0.08}^{+0.07}$ & 217/99 (2.20)\\
KW+2WAM$^{\dagger}$ & CPL & $-1.11_{-0.11}^{+0.12}$ & - & $313_{-34}^{+45}$ & $0.66_{-0.07}^{+0.07}$ & $0.64_{-0.05}^{+0.05}$ & 88/98 (0.91)\\
KW+2WAM$^{\dagger}$ & Band & $-1.03_{-0.14}^{+0.19}$ & $-2.50_{-0.93}^{+0.32}$ & $279_{-54}^{+51}$ & $0.66_{-0.07}^{+0.08}$  & $0.64_{-0.05}^{+0.05}$ & 84/97 (0.87)\\
\hline
\end{tabular}
\end{center}
{\footnotesize
\footnotemark[$\dagger$] The joint fit of two WAM spectra.
}
\end{table}
\begin{table}
\caption{Spectral parameters: 060124 Reg3}
\begin{center}
\footnotesize
\begin{tabular}{c|c|c|c|c|c|c|c}\hline
Instrument & Model & $\alpha$ & $\beta$ & $\ep$ [keV] & C(WAM2) & C(WAM3) & $\chi^{2}$/dof\\\hline
KW & PL & $-1.85_{-0.11}^{+0.11}$ & - & - & - & - & 80/60 (1.34)\\
KW & CPL & $-1.21_{-0.34}^{+0.41}$ & - & $133_{-32}^{+60}$ & - & - & 66/59 (1.13)\\
KW & Band & $-1.20_{-0.35}^{+0.21}$ & $<-2.32$ & $132_{-31}^{+40}$ & - & - & 66/58 (1.15)\\\hline
WAM2 & PL & $-2.42_{-1.02}^{+0.67}$ & - & - & - & - & 11/14 (0.85)\\
WAM2 & CPL & $<2.5$ & - & $231_{-62}^{+41}$ & - & - & 9/13 (0.71)\\
WAM2 & Band & - & - & - & - & - & -\\\hline
WAM3 & PL & $-2.29_{-0.62}^{+0.46}$ & - & - & - & - & 20/14 (1.49)\\
WAM3 & CPL & - & - & - & - & - & -\\
WAM3 & Band & - & - & - & - & - & -\\\hline
KW+WAM2 & PL & $-1.86_{-0.11}^{+0.11}$ & - & - & $0.88_{-0.32}^{+0.39}$ & - & 94/75 (1.26)\\
KW+WAM2 & CPL & $-1.22_{-0.32}^{+0.37}$ & - & $136_{-31}^{+53}$ & $0.97_{-0.33}^{+0.41}$ & - & 77/74 (1.05)\\
KW+WAM2 & Band & $-1.22_{-0.30}^{+0.34}$ & $<-2.44$ & $137_{-30}^{+39}$ & $0.97_{-0.16}^{+0.40}$ & - & 77/73 (1.07)\\\hline
KW+WAM3 & PL & $-1.87_{-0.11}^{+0.10}$ & - & - & - & $0.98_{-0.24}^{+0.30}$ & 103/75 (1.39)\\
KW+WAM3 & CPL & $-1.29_{-0.30}^{+0.36}$ & - & $144_{-34}^{+62}$ & - & $0.94_{-0.22}^{+0.28}$ & 89/74 (1.21)\\
KW+WAM3 & Band & $-1.24_{-0.35}^{+1.57}$ & $<-2.11$ & $136_{-85}^{+66}$ & - & $0.96_{-0.24}^{+0.32}$ & 89/73 (1.22)\\\hline
KW+2WAM$^{\dagger}$ & PL & $-2.00_{-0.14}^{+0.12}$ & - & - & $0.98_{-0.35}^{+0.44}$ & $1.05_{-0.26}^{+0.33}$ & 108/88 (1.24)\\
KW+2WAM$^{\dagger}$ & CPL & $-1.44_{-0.35}^{+0.40}$ & - & $139_{-43}^{+61}$ & $0.97_{-0.33}^{+0.41}$ & $0.96_{-0.23}^{+0.30}$ & 99/87 (1.14)\\
KW+2WAM$^{\dagger}$ & Band & $-1.42_{-0.36}^{+0.46}$ & $<-2.24$ & $137_{-42}^{+60}$ & $0.97_{-0.33}^{+0.41}$  & $0.97_{-0.23}^{+0.30}$ & 99/86 (1.15)\\
\hline
\end{tabular}
\end{center}
{\footnotesize
\footnotemark[$\dagger$] The joint fit of two WAM spectra.
}
\end{table}
\begin{table}
\caption{Spectral parameters: 060124 Reg13}
\begin{center}
\footnotesize
\begin{tabular}{c|c|c|c|c|c|c|c}\hline
Instrument & Model & $\alpha$ & $\beta$ & $\ep$ [keV] & C(WAM2) & C(WAM3) & $\chi^{2}$/dof\\\hline
KW & PL & $-1.72_{-0.04}^{+0.04}$ & - & - & - & - & 111/69 (1.61)\\
KW & CPL & $-1.22_{-0.14}^{+0.15}$ & - & $248_{-36}^{+57}$ & - & - & 55/68 (0.81)\\
KW & Band & $-1.18_{-0.16}^{+0.19}$ & $<-2.17$ & $232_{-44}^{+60}$ & - & - & 54/67 (0.81)\\\hline
WAM2 & PL & $-2.14_{-0.25}^{+0.22}$ & - & - & - & - & 17/14 (1.24)\\
WAM2 & CPL & $0.15_{-1.77}^{+2.85}$ & - & $284_{-73}^{+76}$ & - & - & 11/13 (0.88)\\
WAM2 & Band & $3.32_{-4.45}^{+1.77}$ & $<-2.26$ & $242_{-51}^{+85}$ & - & - & 9/12 (0.82)\\\hline
WAM3 & PL & $-2.10_{-0.17}^{+0.15}$ & - & - & - & - & 19/14 (1.39)\\
WAM3 & CPL & $-1.60_{-0.59}^{+0.80}$ & - & $<414$ & - & - & 17/13 (1.33)\\
WAM3 & Band & - & - & - & - & - & -\\\hline
KW+WAM2 & PL & $-1.74_{-0.04}^{+0.04}$ & - & - & $0.74_{-0.09}^{+0.10}$ & - & 138/84 (1.65)\\
KW+WAM2 & CPL & $-1.22_{-0.13}^{+0.14}$ & - & $253_{-34}^{+48}$ & $0.73_{-0.09}^{+0.09}$ & - & 68/83 (0.82)\\
KW+WAM2 & Band & $-1.19_{-0.14}^{+0.17}$ & $-2.76_{-7.24}^{+0.50}$ & $241_{-41}^{+50}$ & $0.73_{-0.09}^{+0.09}$ & - & 67/82 (0.82)\\\hline
KW+WAM3 & PL & $-1.75_{-0.04}^{+0.04}$ & - & - & $0.77_{-0.07}^{+0.07}$ & - & 148/84 (1.77)\\
KW+WAM3 & CPL & $-1.25_{-0.12}^{+0.13}$ & - & $261_{-33}^{+46}$ & $0.71_{-0.06}^{+0.06}$ & - & 73/83 (0.89)\\
KW+WAM3 & Band & $-1.21_{-0.14}^{+0.18}$ & $-2.62_{-7.38}^{+0.42}$ & $242_{-48}^{+53}$ & $0.71_{-0.06}^{+0.07}$ & - & 72/82 (0.88)\\\hline
KW+2WAM$^{\dagger}$ & PL & $-1.77_{-0.04}^{+0.04}$ & - & - & $0.77_{-0.10}^{+0.10}$ & $0.78_{-0.07}^{+0.08}$ & 174/99 (1.76)\\
KW+2WAM$^{\dagger}$ & CPL & $-1.25_{-0.12}^{+0.13}$ & - & $261_{-31}^{+41}$ & $0.73_{-0.09}^{+0.09}$ & $0.71_{-0.06}^{+0.06}$ & 86/98 (0.88)\\
KW+2WAM$^{\dagger}$ & Band & $-1.22_{-0.13}^{+0.16}$ & $-2.74_{-7.26}^{+0.46}$ & $248_{-42}^{+45}$ & $0.73_{-0.09}^{+0.09}$  & $0.71_{-0.06}^{+0.07}$ & 84/97 (0.88)\\
\hline
\end{tabular}
\end{center}
{\footnotesize
\footnotemark[$\dagger$] The joint fit of two WAM spectra.
}
\end{table}
\begin{table}
\caption{Spectral parameters: 060502A Reg1}
\begin{center}
\footnotesize
\begin{tabular}{c|c|c|c|c|c|c|c}\hline
Instrument & Model & $\alpha$ & $\beta$ & $\ep$ [keV] & C(WAM3) & C(BAT) & $\chi^{2}$/dof\\\hline
KW & PL & $-1.71_{-0.11}^{+0.10}$ & - & - & - & - & 73/60 (1.22)\\
KW & CPL & $-1.19_{-0.34}^{+0.42}$ & - & $174_{-53}^{+162}$ & - & - & 64/59 (1.10)\\
KW & Band & $-0.90_{-0.60}^{+4.75}$ & $<-1.79$ & $117_{-80}^{+103}$ & - & - & 64/58 (1.11)\\\hline
WAM3 & PL & $-2.75_{-0.81}^{+0.58}$ & - & - & - & - & 11/14 (0.81)\\
WAM3 & CPL & - & - & - & - & - & -\\
WAM3 & Band & - & - & - & - & - & -\\\hline
BAT & PL & $-1.26_{-0.08}^{+0.08}$ & - & - & - & - & 54/57 (0.96)\\
\hline
\end{tabular}
\end{center}
\end{table}
\begin{table}
\caption{Spectral parameters: 060813 Reg1}
\begin{center}
\footnotesize
\begin{tabular}{c|c|c|c|c|c|c|c|c}\hline
Instrument & Model & $\alpha$ & $\beta$ & $\ep$ [keV] & C(WAM0) & C(WAM3) & C(BAT) & $\chi^{2}$/dof\\\hline
KW & PL & $-1.57_{-0.02}^{+0.02}$ & - & - & - & - & - & 431/60 (7.18)\\
KW & CPL & $-0.65_{-0.11}^{+0.12}$ & - & $218_{-16}^{+18}$ & - & - & - & 82/59 (1.40)\\
KW & Band & $-0.52_{-0.15}^{+0.17}$ & $-2.63_{-0.47}^{+0.26}$ & $194_{-19}^{+21}$ & - & - & - & 74/58 (1.28)\\\hline
WAM0 & PL & $-2.23_{-0.07}^{+0.07}$ & - & - & - & - & - & 42/22 (1.94)\\
WAM0 & CPL & $-1.48_{-0.30}^{+0.34}$ & - & $223_{-71}^{+38}$ & - & - & - & 20/21 (0.98)\\
WAM0 & Band & - & - & - & - & - & - & -\\\hline
WAM3 & PL & $-2.19_{-0.08}^{+0.07}$ & - & - & - & - & - & 38/22 (1.76)\\
\hline
\end{tabular}
\end{center}
{\footnotesize
\footnotemark[$\dagger$] The joint fit of two WAM spectra.
}
\end{table}
\begin{table}
\caption{Spectral parameters: 060814 Reg1}
\begin{center}
\footnotesize
\begin{tabular}{c|c|c|c|c|c|c|c|c}\hline
Instrument & Model & $\alpha$ & $\beta$ & $\ep$ [keV] & C(WAM0) & C(WAM1) & C(BAT) & $\chi^{2}$/dof\\\hline
KW & PL & $-1.64_{-0.04}^{+0.04}$ & - & - & - & - & - & 76/60 (1.28)\\
KW & CPL & $-1.26_{-0.14}^{+0.15}$ & - & $299_{-62}^{+111}$ & - & - & - & 45/59 (0.78)\\
KW & Band & $-1.08_{-0.21}^{+0.32}$ & $-2.07_{-0.50}^{+0.21}$ & $204_{-65}^{+91}$ & - & - & - & 41/58 (0.72)\\\hline
WAM0 & PL & $-2.01_{-0.27}^{+0.23}$ & - & - & - & - & - & 16/22 (0.73)\\
WAM0 & CPL & - & - & - & - & - & - & -\\
WAM0 & Band & - & - & - & - & - & - & -\\\hline
WAM1 & PL & $-1.84_{-0.09}^{+0.09}$ & - & - & - & - & - & 27/22 (1.24)\\
\hline
\end{tabular}
\end{center}
{\footnotesize
\footnotemark[$\dagger$] The joint fit of two WAM spectra.
}
\end{table}
\begin{table}
\caption{Spectral parameters: 060814 Reg2}
\begin{center}
\footnotesize
\begin{tabular}{c|c|c|c|c|c|c|c|c}\hline
Instrument & Model & $\alpha$ & $\beta$ & $\ep$ [keV] & C(WAM0) & C(WAM1) & C(BAT) & $\chi^{2}$/dof\\\hline
KW & PL & $-1.78_{-0.07}^{+0.07}$ & - & - & - & - & - & 63/60 (1.07)\\
KW & CPL & $-1.44_{-0.20}^{+0.24}$ & - & $212_{-57}^{+159}$ & - & - & - & 53/59 (0.91)\\
KW & Band & $-1.37_{-0.27}^{+0.40}$ & $<-1.92$ & $177_{-67}^{+176}$ & - & - & - & 53/58 (0.92)\\\hline
WAM0 & PL & $-1.61_{-0.27}^{-0.22}$ & - & - & - & - & - & 16/22 (0.77)\\
WAM0 & CPL & - & - & - & - & - & - & -\\
WAM0 & Band & - & - & - & - & - & - & -\\\hline
WAM1 & PL & $-1.77_{-0.12}^{+0.11}$ & - & - & - & - & - & 22/22 (1.04)\\
\hline
\end{tabular}
\end{center}
{\footnotesize
\footnotemark[$\dagger$] The joint fit of two WAM spectra.
}
\end{table}
\begin{table}
\caption{Spectral parameters: 060814 Reg12}
\begin{center}
\footnotesize
\begin{tabular}{c|c|c|c|c|c|c|c|c}\hline
Instrument & Model & $\alpha$ & $\beta$ & $\ep$ [keV] & C(WAM0) & C(WAM1) & C(BAT) & $\chi^{2}$/dof\\\hline
KW & PL & $-1.70_{-0.06}^{+0.06}$ & - & - & - & - & - & 72/60 (1.20)\\
KW & CPL & $-1.43_{-0.16}^{+0.19}$ & - & $308_{-92}^{+268}$ & - & - & - & 61/59 (1.04)\\
KW & Band & $-1.30_{-0.22}^{+0.42}$ & $<-1.81$ & $212_{-98}^{+170}$ & - & - & - & 59/58 (1.02)\\\hline
WAM0 & PL & $-1.76_{-0.25}^{+0.21}$ & - & - & - & - & - & 15/22 (0.71)\\
WAM0 & CPL & - & - & - & - & - & - & -\\
WAM0 & Band & - & - & - & - & - & - & -\\\hline
WAM1 & PL & $-1.77_{-0.10}^{+0.09}$ & - & - & - & - & - & 19/22 (0.88)\\
\hline
\end{tabular}
\end{center}
{\footnotesize
\footnotemark[$\dagger$] The joint fit of two WAM spectra.
}
\end{table}
\begin{table}
\caption{Spectral parameters: 060904A Reg1}
\begin{center}
\footnotesize
\begin{tabular}{c|c|c|c|c|c|c|c|c}\hline
Instrument & Model & $\alpha$ & $\beta$ & $\ep$ [keV] & C(WAM0) & C(WAM1) & C(BAT) & $\chi^{2}$/dof\\\hline
KW & PL & $-1.67_{-0.04}^{+0.04}$ & - & - & - & - & - & 128/60 (2.15)\\
KW & CPL & $-1.03_{-0.14}^{+0.15}$ & - & $186_{-22}^{+30}$ & - & - & - & 45/59 (0.77)\\
KW & Band & $-1.02_{-0.15}^{+0.16}$ & $<-2.44$ & $182_{-23}^{+30}$ & - & - & - & 44/58 (0.77)\\\hline
WAM0 & PL & $-2.16_{-0.25}^{+0.21}$ & - & - & - & - & - & 14/14 (1.03)\\
WAM0 & CPL & $-1.07_{-0.66}^{+1.36}$ & - & $<430$ & - & - & - & 10/13 (0.84)\\
WAM0 & Band & $-1.14_{-0.99}^{+6.02}$ & $<-2.25$ & $309_{-145}^{+122}$ & - & - & - & 10/12 (0.91)\\\hline
WAM3 & PL & $-2.10_{-0.10}^{+0.09}$ & - & - & - & - & - & 10/14 (0.73)\\
\hline
\end{tabular}
\end{center}
{\footnotesize
\footnotemark[$\dagger$] The joint fit of two WAM spectra.
}
\end{table}
\begin{table}
\caption{Spectral parameters: 060904A Reg2}
\begin{center}
\footnotesize
\begin{tabular}{c|c|c|c|c|c|c|c|c}\hline
Instrument & Model & $\alpha$ & $\beta$ & $\ep$ [keV] & C(WAM0) & C(WAM1) & C(BAT) & $\chi^{2}$/dof\\\hline
KW & PL & $-1.76_{-0.11}^{+0.11}$ & - & - & - & - & - & 60/60 (1.01)\\
KW & CPL & $-0.53_{-0.51}^{+0.65}$ & - & $113_{-20}^{+32}$ & - & - & - & 36/59 (0.62)\\
KW & Band & $-0.43_{-0.59}^{+1.17}$ & $<-2.26$ & $108_{-32}^{+35}$ & - & - & - & 36/58 (0.62)\\\hline
WAM0 & PL & $-1.76_{-1.03}^{+0.65}$ & - & - & - & - & - & 14/14 (1.06)\\
WAM0 & CPL & - & - & - & - & - & - & -\\
WAM0 & Band & - & - & - & - & - & - & -\\\hline
WAM3 & PL & $-2.38_{-0.48}^{+0.38}$ & - & - & - & - & - & 32/14 (2.31)\\
\hline
\end{tabular}
\end{center}
{\footnotesize
\footnotemark[$\dagger$] The joint fit of two WAM spectra.
}
\end{table}
\begin{table}
\caption{Spectral parameters: 060904A Reg12}
\begin{center}
\footnotesize
\begin{tabular}{c|c|c|c|c|c|c|c|c}\hline
Instrument & Model & $\alpha$ & $\beta$ & $\ep$ [keV] & C(WAM0) & C(WAM1) & C(BAT) & $\chi^{2}$/dof\\\hline
KW & PL & $-1.69_{-0.04}^{+0.04}$ & - & - & - & - & - & 120/60 (2.02)\\
KW & CPL & $-0.97_{-0.16}^{+0.17}$ & - & $165_{-19}^{+25}$ & - & - & - & 32/59 (0.56)\\
KW & Band & $-0.95_{-0.16}^{+0.20}$ & $<-2.42$ & $162_{-23}^{+24}$ & - & - & - & 31/58 (0.55)\\\hline
WAM0 & PL & $-2.06_{-0.26}^{+0.22}$ & - & - & - & - & - & 14/14 (1.06)\\
WAM0 & CPL & $-0.57_{-1.33}^{+2.12}$ & - & $333_{-169}^{+163}$ & - & - & - & 10/13 (0.84)\\
WAM0 & Band & $0.80_{-2.63}^{+4.20}$ & $<-2.08$ & $285_{-96}^{+187}$ & - & - & - & 10/12 (0.89)\\\hline
WAM3 & PL & $-2.14_{-0.11}^{+0.10}$ & - & - & - & - & - & 15/14 (1.13)\\
\hline
\end{tabular}
\end{center}
{\footnotesize
\footnotemark[$\dagger$] The joint fit of two WAM spectra.
}
\end{table}
\begin{table}
\caption{Spectral parameters: 060912 Reg1}
\begin{center}
\footnotesize
\begin{tabular}{c|c|c|c|c|c|c|c}\hline
Instrument & Model & $\alpha$ & $\beta$ & $\ep$ [keV] & C(WAM2) & C(BAT) & $\chi^{2}$/dof\\\hline
KW & PL & $-1.92_{-0.16}^{+0.15}$ & - & - & - & - & 63/60 (1.06)\\
KW & CPL & - & - & - & - & - & -\\
KW & Band & - & - & - & - & - & -\\\hline
WAM2 & PL & $-1.71_{-0.78}^{+0.51}$ & - & - & - & - & 8/14 (0.58)\\
WAM2 & CPL & - & - & - & - & - & -\\
WAM2 & Band & - & - & - & - & - & -\\\hline
BAT & PL & $-1.82_{-0.10}^{+0.10}$ & - & - & - & - & 57/57 (1.01)\\
BAT & CPL & - & - & - & - & - & -\\
BAT & Band & - & - & - & - & - & -\\\hline
BAT+WAM2 & PL & $-1.81_{-0.10}^{+0.10}$ & - & - & $0.91_{-0.28}^{+0.34}$ & - & 65/72 (0.91)\\
BAT+WAM2 & CPL & - & - & - & - & - & -\\
BAT+WAM2 & Band & - & - & - & - & - & -\\\hline
KW+BAT & PL & $-1.85_{-0.08}^{+0.08}$ & - & - & - & $0.79_{-0.11}^{+0.15}$ & 121/118 (1.03)\\
KW+BAT & CPL & - & - & - & - & - & -\\
KW+BAT & Band & - & - & - & - & - & -\\\hline
KW+WAM2 & PL & $-1.91_{-0.16}^{+0.15}$ & - & - & $0.82_{-0.25}^{+0.33}$ & - & 72/75 (0.96)\\
KW+WAM2 & CPL & - & - & - & - & - & -\\
KW+WAM2 & Band & - & - & - & - & - & -\\\hline
KW+WAM2+BAT & PL & $-1.85_{-0.08}^{+0.08}$ & - & - & $0.77_{-0.22}^{+0.26}$ & $0.79_{-0.11}^{+0.15}$ & 130/133 (0.98)\\
KW+WAM2+BAT & CPL & - & - & - & - & - & -\\
KW+WAM2+BAT & Band & - & - & - & -  & - & -\\
\hline
\end{tabular}
\end{center}
\end{table}
\begin{table}
\caption{Spectral parameters: 061006 Reg1}
\begin{center}
\footnotesize
\begin{tabular}{c|c|c|c|c|c|c|c}\hline
Instrument & Model & $\alpha$ & $\beta$ & $\ep$ [keV] & C(WAM3) & C(BAT) & $\chi^{2}$/dof\\\hline
KW & PL & $-1.20_{-0.06}^{+0.06}$ & - & - & - & - & 87/31 (2.81)\\
KW & CPL & $-0.51_{-0.19}^{+0.22}$ & - & $592_{-115}^{+166}$ & - & - & 37/30 (1.25)\\
KW & Band & $-0.51_{-0.19}^{+0.22}$ & -* & $592_{-94}^{+166}$ & - & - & 37/29 (1.29)\\\hline
WAM3 & PL & $-1.59_{-0.07}^{+0.06}$ & - & - & - & - & 65/24 (2.72)\\
WAM3 & CPL & $-0.64_{-0.29}^{+0.33}$ & - & $795_{-116}^{+170}$ & - & - & 22/23 (0.96)\\
WAM3 & Band & $-0.44_{-0.41}^{+0.57}$ & $<-2.07$ & $693_{-150}^{+209}$ & - & - & 20/22 (0.93)\\\hline
BAT & PL & $-0.61_{-0.09}^{+0.09}$ & - & - & - & - & 60/57 (1.07)\\
\hline
\end{tabular}
\end{center}
{\footnotesize
\footnotemark[*] The parameter is not constrained.
}
\end{table}
\begin{table}
\caption{Spectral parameters: 061007 Reg1}
\begin{center}
\footnotesize
\begin{tabular}{c|c|c|c|c|c|c|c|c}\hline
Instrument & Model & $\alpha$ & $\beta$ & $\ep$ [keV] & C(WAM2) & C(WAM3) & C(BAT) & $\chi^{2}$/dof\\\hline
KW & PL & $-1.56_{-0.03}^{+0.03}$ & - & - & - & - & - & 328/66 (4.98)\\
KW & CPL & $-0.69_{-0.12}^{+0.13}$ & - & $284_{-24}^{+29}$ & - & - & - & 71/65 (1.10)\\
KW & Band & $-0.51_{-0.18}^{+0.23}$ & $-2.38_{-0.39}^{+0.22}$ & $239_{-34}^{+35}$ & - & - & - & 58/64 (0.92)\\\hline
WAM2 & PL & $-2.23_{-0.12}^{+0.00}$ & - & - & - & - & - & 19/25 (0.77)\\
WAM2 & CPL & $-1.98_{-0.01}^{+0.36}$ & - & $<421$ & - & - & - & 16/24 (0.68)\\
WAM2 & Band & $-1.47_{-1.01}^{+6.47}$ & $<-2.20$ & $<415$ & - & - & - & 16/23 (0.71)\\\hline
WAM3 & PL & $-2.15_{-0.07}^{+0.07}$ & - & - & - & - & - & 43/25 (1.76)\\
\hline
\end{tabular}
\end{center}
{\footnotesize
\footnotemark[$\dagger$] The joint fit of two WAM spectra.
}
\end{table}
\begin{table}
\caption{Spectral parameters: 061007 Reg2}
\begin{center}
\footnotesize
\begin{tabular}{c|c|c|c|c|c|c|c|c}\hline
Instrument & Model & $\alpha$ & $\beta$ & $\ep$ [keV] & C(WAM2) & C(WAM3) & C(BAT) & $\chi^{2}$/dof\\\hline
KW & PL & $-1.47$ & - & - & - & - & - & 3121/86 (36.30)\\
KW & CPL & $-0.59_{-0.04}^{+0.04}$ & - & $504_{-18}^{+19}$ & - & - & - & 117/85 (1.38)\\
KW & Band & $-0.51_{-0.05}^{+0.05}$ & $-2.61_{-0.21}^{+0.14}$ & $452_{-23}^{+25}$ & - & - & - & 82/84 (0.99)\\\hline
WAM2 & PL & $-2.20_{-0.03}^{+0.03}$ & - & - & - & - & - & 231/25 (9.25)\\
WAM2 & CPL & $-1.34_{-0.16}^{+0.18}$ & - & $508_{-39}^{+38}$ & - & - & - & 50/24 (2.11)\\
WAM2 & Band & $-0.76_{-0.30}^{+0.41}$ & $-2.76_{-0.17}^{+0.13}$ & $463_{-33}^{+33}$ & - & - & - & 16/23 (0.70)\\\hline
WAM3 & PL & $-2.11$ & - & - & - & - & - & 738/25 (29.52)\\
\hline
\end{tabular}
\end{center}
{\footnotesize
\footnotemark[$\dagger$] The joint fit of two WAM spectra.
}
\end{table}
\begin{table}
\caption{Spectral parameters: 061007 Reg3}
\begin{center}
\footnotesize
\begin{tabular}{c|c|c|c|c|c|c|c|c}\hline
Instrument & Model & $\alpha$ & $\beta$ & $\ep$ [keV] & C(WAM2) & C(WAM3) & C(BAT) & $\chi^{2}$/dof\\\hline
KW & PL & $-1.54$ & - & - & - & - & - & 2463/86 (28.64)\\
KW & CPL & $-0.70_{-0.04}^{+0.04}$ & - & $389_{-14}^{+15}$ & - & - & - & 108/85 (1.28)\\
KW & Band & $-0.64_{-0.05}^{+0.05}$ & $-2.74_{-0.26}^{+0.18}$ & $361_{-17}^{+18}$ & - & - & - & 87/84 (1.04)\\\hline
WAM2 & PL & $-2.31_{-0.04}^{+0.04}$ & - & - & - & - & - & 87/25 (3.51)\\
WAM2 & CPL & $-1.54_{-0.27}^{+0.29}$ & - & $321_{-112}^{+59}$ & - & - & - & 42/24 (1.77)\\
WAM2 & Band & $-0.83_{-0.51}^{+0.91}$ & $-2.75_{-0.21}^{+0.14}$ & $347_{-33}^{+30}$ & - & - & - & 17/23 (0.78)\\\hline
WAM3 & PL & $-2.24$ & - & - & - & - & - & 388/25 (15.56)\\
\hline
\end{tabular}
\end{center}
{\footnotesize
\footnotemark[$\dagger$] The joint fit of two WAM spectra.
}
\end{table}
\begin{table}
\caption{Spectral parameters: 061007 Reg4}
\begin{center}
\footnotesize
\begin{tabular}{c|c|c|c|c|c|c|c|c}\hline
Instrument & Model & $\alpha$ & $\beta$ & $\ep$ [keV] & C(WAM2) & C(WAM3) & C(BAT) & $\chi^{2}$/dof\\\hline
KW & PL & $-1.51$ & - & - & - & - & - & 4518/86 (52.54)\\
KW & CPL & $-0.67_{-0.03}^{+0.03}$ & - & $436_{-12}^{+13}$ & - & - & - & 152/85 (1.80)\\
KW & Band & $-0.60_{-0.04}^{+0.04}$ & $-2.68_{-0.17}^{+0.13}$ & $400_{-15}^{+16}$ & - & - & - & 111/84 (1.33)\\\hline
WAM2 & PL & $-2.29_{-0.03}^{+0.03}$ & - & - & - & - & - & 179/25 (7.19)\\
WAM2 & CPL & $-1.53_{-0.18}^{+0.20}$ & - & $386_{-67}^{+49}$ & - & - & - & 62/24 (2.59)\\
WAM2 & Band & $-0.80_{-0.36}^{+0.57}$ & $-2.74_{-0.14}^{+0.11}$ & $397_{-27}^{+28}$ & - & - & - & 15/23 (0.69)\\\hline
WAM3 & PL & $-2.23$ & - & - & - & - & - & 710/25 (28.41)\\
\hline
\end{tabular}
\end{center}
{\footnotesize
\footnotemark[$\dagger$] The joint fit of two WAM spectra.
}
\end{table}
\begin{table}
\caption{Spectral parameters: 061007 Reg5}
\begin{center}
\footnotesize
\begin{tabular}{c|c|c|c|c|c|c|c|c}\hline
Instrument & Model & $\alpha$ & $\beta$ & $\ep$ [keV] & C(WAM2) & C(WAM3) & C(BAT) & $\chi^{2}$/dof\\\hline
KW & PL & $-1.77_{-0.19}^{+0.16}$ & - & - & - & - & - & 73/86 (0.85)\\
KW & CPL & - & - & - & - & - & - & -\\
KW & Band & - & - & - & - & - & - & -\\\hline
WAM2 & PL & $-2.16_{-0.82}^{+0.51}$ & - & - & - & - & - & 10/15 (0.72)\\
WAM2 & CPL & - & - & - & - & - & - & -\\
WAM2 & Band & - & - & - & - & - & - & -\\\hline
WAM3 & PL & $-2.37_{-0.63}^{+0.43}$ & - & - & - & - & - & 8/15 (0.55)\\
\hline
\end{tabular}
\end{center}
{\footnotesize
\footnotemark[$\dagger$] The joint fit of two WAM spectra.
}
\end{table}
\begin{table}
\caption{Spectral parameters: 061007 Reg15}
\begin{center}
\footnotesize
\begin{tabular}{c|c|c|c|c|c|c|c|c}\hline
Instrument & Model & $\alpha$ & $\beta$ & $\ep$ [keV] & C(WAM2) & C(WAM3) & C(BAT) & $\chi^{2}$/dof\\\hline
KW & PL & $-1.50$ & - & - & - & - & - & 2968/86 (34.52)\\
KW & CPL & $-0.71_{-0.04}^{+0.04}$ & - & $419_{-14}^{+15}$ & - & - & - & 131/85 (1.55)\\
KW & Band & $-0.63_{-0.05}^{+0.05}$ & $-2.53_{-0.17}^{+0.13}$ & $377_{-17}^{+18}$ & - & - & - & 92/84 (1.10)\\\hline
WAM2 & PL & $-2.28_{-0.04}^{+0.04}$ & - & - & - & - & - & 108/25 (4.34)\\
WAM2 & CPL & $-1.62_{-0.19}^{+0.22}$ & - & $368_{-105}^{+70}$ & - & - & - & 38/24 (1.62)\\
WAM2 & Band & $-0.89_{-0.43}^{+0.74}$ & $-2.69_{-0.18}^{+0.13}$ & $392_{-34}^{+36}$ & - & - & - & 12/23 (0.54)\\\hline
WAM3 & PL & $-2.21$ & - & - & - & - & - & 444/25 (17.79)\\
\hline
\end{tabular}
\end{center}
{\footnotesize
\footnotemark[$\dagger$] The joint fit of two WAM spectra.
}
\end{table}
\begin{table}
\caption{Spectral parameters: 061222A Reg1}
\begin{center}
\footnotesize
\begin{tabular}{c|c|c|c|c|c|c|c|c}\hline
Instrument & Model & $\alpha$ & $\beta$ & $\ep$ [keV] & C(WAM2) & C(WAM3) & C(BAT) & $\chi^{2}$/dof\\\hline
KW & PL & $-1.50_{-0.03}^{+0.03}$ & - & - & - & - & - & 160/60 (2.68)\\
KW & CPL & $-0.97_{-0.11}^{+0.12}$ & - & $330_{-42}^{+58}$ & - & - & - & 53/59 (0.91)\\
KW & Band & $-0.89_{-0.13}^{+0.15}$ & $-2.28_{-0.61}^{+0.27}$ & $285_{-47}^{+55}$ & - & - & - & 48/58 (0.84)\\\hline
WAM2 & PL & $-1.92_{-0.07}^{+0.07}$ & - & - & - & - & - & 37/14 (2.71)\\
WAM2 & CPL & $-1.00_{-0.32}^{+0.37}$ & - & $467_{-50}^{+71}$ & - & - & - & 7/13 (0.57)\\
WAM2 & Band & $-0.96_{-0.35}^{+0.53}$ & $<-2.41$ & $459_{-73}^{+76}$ & - & - & - & 7/12 (0.60)\\\hline
WAM3 & PL & $-1.97_{-0.16}^{+0.14}$ & - & - & - & - & - & 12/14 (0.87)\\
\hline
\end{tabular}
\end{center}
{\footnotesize
\footnotemark[$\dagger$] The joint fit of two WAM spectra.
}
\end{table}
\begin{table}
\caption{Spectral parameters: 070328 Reg1}
\begin{center}
\footnotesize
\begin{tabular}{c|c|c|c|c|c|c|c}\hline
Instrument & Model & $\alpha$ & $\beta$ & $\ep$ [keV] & C(WAM1) & C(BAT) & $\chi^{2}$/dof\\\hline
KW & PL & $-1.38_{-0.03}^{+0.03}$ & - & - & - & - & 157/76 (2.07)\\
KW & CPL & $-1.12_{-0.08}^{+0.09}$ & - & $1205_{-317}^{+475}$ & - & - & 91/75 (1.23)\\
KW & Band & $-0.95_{-0.13}^{+0.15}$ & $-1.81_{-0.29}^{+0.15}$ & $615_{-178}^{+306}$ & - & - & 82/74 (1.12)\\\hline
WAM1 & PL & $-1.67_{-0.04}^{+0.04}$ & - & - & - & - & 84/25 (3.38)\\
WAM1 & CPL & $-1.25_{-0.12}^{+0.12}$ & - & $1905_{-296}^{+435}$ & - & - & 37/24 (1.54)\\
WAM1 & Band & $-0.90_{-0.28}^{+0.48}$ & $-2.02_{-0.24}^{+0.14}$ & $1001_{-311}^{+485}$ & - & - & 30/23 (1.34)\\\hline
BAT & PL & $-1.12_{-0.04}^{+0.04}$ & - & - & - & - & 62/57 (1.09)\\
\hline
\end{tabular}
\end{center}
\end{table}
\begin{table}
\caption{Spectral parameters: 070328 Reg2}
\begin{center}
\footnotesize
\begin{tabular}{c|c|c|c|c|c|c|c}\hline
Instrument & Model & $\alpha$ & $\beta$ & $\ep$ [keV] & C(WAM1) & C(BAT) & $\chi^{2}$/dof\\\hline
KW & PL & $-1.42_{-0.02}^{+0.02}$ & - & - & - & - & 248/76 (3.27)\\
KW & CPL & $-1.02_{-0.08}^{+0.08}$ & - & $686_{-110}^{+155}$ & - & - & 94/75 (1.27)\\
KW & Band & $-0.87_{-0.11}^{+0.13}$ & $-1.92_{-0.21}^{+0.13}$ & $449_{-88}^{+116}$ & - & - & 77/74 (1.05)\\\hline
WAM1 & PL & $-1.82_{-0.04}^{+0.04}$ & - & - & - & - & 130/25 (5.24)\\
WAM1 & CPL & $-1.19_{-0.14}^{+0.16}$ & - & $1143_{-153}^{+180}$ & - & - & 31/24 (1.33)\\
WAM1 & Band & $-0.74_{-0.31}^{+0.41}$ & $-2.36_{-0.30}^{+0.17}$ & $762_{-130}^{+185}$ & - & - & 23/23 (1.01)\\\hline
BAT & PL & $-1.17_{-0.03}^{+0.03}$ & - & - & - & - & 51/57 (0.91)\\
\hline
\end{tabular}
\end{center}
\end{table}
\begin{table}
\caption{Spectral parameters: 070328 Reg3}
\begin{center}
\footnotesize
\begin{tabular}{c|c|c|c|c|c|c|c}\hline
Instrument & Model & $\alpha$ & $\beta$ & $\ep$ [keV] & C(WAM1) & C(BAT) & $\chi^{2}$/dof\\\hline
KW & PL & $-1.56_{-0.16}^{+0.15}$ & - & - & - & - & 63/60 (1.06)\\
KW & CPL & $-0.92_{-0.58}^{+0.82}$ & - & $219_{-84}^{+806}$ & - & - & 59/59 (1.00)\\
KW & Band & $0.13_{-4.72}^{+1.52}$ & $-1.93_{-0.29}^{+0.31}$ & $105_{-14}^{+211}$ & - & - & 56/58 (0.98)\\\hline
WAM1 & PL & $-1.99_{-0.37}^{+0.28}$ & - & - & - & - & 21/15 (1.41)\\
WAM1 & CPL & - & - & - & - & - & -\\
WAM1 & Band & - & - & - & - & - & -\\\hline
BAT & PL & $-1.34_{-0.09}^{+0.09}$ & - & - & - & - & 37/57 (0.65)\\
\hline
\end{tabular}
\end{center}
\end{table}
\begin{table}
\caption{Spectral parameters: 070328 Reg13}
\begin{center}
\footnotesize
\begin{tabular}{c|c|c|c|c|c|c|c}\hline
Instrument & Model & $\alpha$ & $\beta$ & $\ep$ [keV] & C(WAM1) & C(BAT) & $\chi^{2}$/dof\\\hline
KW & PL & $-1.43_{-0.03}^{+0.03}$ & - & - & - & - & 190/76 (2.51)\\
KW & CPL & $-1.04_{-0.09}^{+0.10}$ & - & $622_{-112}^{+163}$ & - & - & 84/75 (1.13)\\
KW & Band & $-0.88_{-0.13}^{+0.15}$ & $-1.91_{-0.23}^{+0.15}$ & $407_{-90}^{+119}$ & - & - & 71/74 (0.96)\\\hline
WAM1 & PL & $-1.85_{-0.05}^{+0.04}$ & - & - & - & - & 105/25 (4.22)\\
WAM1 & CPL & $-1.21_{-0.16}^{+0.18}$ & - & $1115_{-160}^{+203}$ & - & - & 31/24 (1.31)\\
WAM1 & Band & $-0.65_{-0.44}^{+0.55}$ & $-2.33_{-0.37}^{+0.17}$ & $706_{-126}^{+242}$ & - & - & 25/23 (1.11)\\\hline
BAT & PL & $-1.19_{-0.03}^{+0.03}$ & - & - & - & - & 38/57 (0.68)\\
\hline
\end{tabular}
\end{center}
\end{table}

\clearpage
\begin{table}
\caption{Fluxes: 051008 Reg1 \label{tbl:051008_reg1_flux}}
\begin{center}
\footnotesize
\begin{tabular}{c|c|c|c|c|c}\hline
Instrument & Energy range & Flux (PL) & Flux (CPL) & Flux (Band) & Inst fixed to 1 \\
           & [keV]        & [10$^{-7}$ erg cm$^{-2}$ s$^{-1}$] & [10$^{-7}$ erg cm$^{-2}$ s$^{-1}$] & [10$^{-7}$ erg cm$^{-2}$ s$^{-1}$] & -\\\hline
KW & 20-1000 & $14.9_{-0.6}^{+0.6}$ & $17.3_{-0.4}^{+0.7}$ & $16.8_{-1.0}^{+0.9}$ & - \\
WAM0 & 120-1000 & $12.0_{-0.3}^{+0.3}$ & $13.7_{-0.4}^{+0.4}$ & $13.6_{-0.5}^{+0.5}$ & - \\
WAM3 & 120-1000 & $11.9_{-0.4}^{+0.4}$ & $13.4_{-0.5}^{+0.2}$ & $13.3_{-0.5}^{+0.5}$ & - \\
BAT & 15-150 & $ 3.4_{-0.1}^{+0.1}$ & $ 3.3_{-0.1}^{+0.1}$ & - & - \\
BAT+WAM0 & 15-1000 & $ 9.5_{-0.5}^{+0.5}$ & $16.3_{-1.1}^{+1.2}$ & $16.2_{-1.1}^{+1.2}$ & BAT \\
BAT+WAM3 & 15-1000 & $ 8.9_{-0.5}^{+0.5}$ & $15.1_{-1.0}^{+1.1}$ & $14.9_{-1.1}^{+1.1}$ & BAT \\
KW+BAT & 15-1000 & $15.4_{-0.6}^{+0.6}$ & $17.4_{-0.7}^{+0.7}$ & $16.9_{-0.9}^{+0.9}$ & KW \\
KW+WAM0 & 20-1000 & $13.8_{-0.6}^{+0.6}$ & $17.2_{-0.7}^{+0.7}$ & $17.2_{-0.7}^{+0.7}$ & KW \\
KW+WAM3 & 20-1000 & $13.6_{-0.6}^{+0.6}$ & $16.9_{-0.7}^{+0.7}$ & $16.7_{-0.7}^{+0.8}$ & KW \\
KW+2WAM+BAT & 15-1000 & $14.1_{-0.6}^{+0.6}$ & $17.2_{-0.6}^{+0.7}$ & $17.0_{-0.7}^{+0.7}$ & KW \\
\hline
\end{tabular}
\end{center}
\end{table}
\begin{table}
\caption{Fluxes: 051008 Reg2}
\begin{center}
\footnotesize
\begin{tabular}{c|c|c|c|c|c}\hline
Instrument & Energy range & Flux (PL) & Flux (CPL) & Flux (Band) & Inst fixed to 1 \\
           & [keV]        & [10$^{-7}$ erg cm$^{-2}$ s$^{-1}$] & [10$^{-7}$ erg cm$^{-2}$ s$^{-1}$] & [10$^{-7}$ erg cm$^{-2}$ s$^{-1}$] & -\\\hline
KW & 20-1000 & $ 1.9_{-0.3}^{+0.3}$ & $ 1.7_{-0.3}^{+0.3}$ & $ 1.7_{-0.3}^{+0.2}$ & - \\
WAM0 & 120-1000 & $ 1.8_{-0.2}^{+0.2}$ & $ 1.9_{-0.2}^{+0.2}$ & $ 1.9_{-0.2}^{+0.2}$ & - \\
WAM3 & 120-1000 & $ 1.9_{-0.2}^{+0.2}$ & $ 1.9_{-0.2}^{+0.2}$ & $ 1.9_{-0.2}^{+0.2}$ & - \\
KW-WAM0 & 20-1000 & $ 1.8_{-0.3}^{+0.3}$ & $ 1.9_{-0.3}^{+0.3}$ & $ 2.0_{-0.1}^{+0.2}$ & KW \\
KW-WAM3 & 20-1000 & $ 1.8_{-0.3}^{+0.3}$ & $ 1.9_{-0.3}^{+0.3}$ & $ 1.9_{-0.2}^{+0.3}$ & KW \\
KW-2WAM & 20-1000 & $ 1.8_{-0.2}^{+0.3}$ & $ 2.0_{-0.1}^{+0.3}$ & $ 1.9_{-0.3}^{+0.3}$ & KW \\
\hline
\end{tabular}
\end{center}
\end{table}
\begin{table}
\caption{Fluxes: 051008 Reg12}
\begin{center}
\footnotesize
\begin{tabular}{c|c|c|c|c|c}\hline
Instrument & Energy range & Flux (PL) & Flux (CPL) & Flux (Band) & Inst fixed to 1 \\
           & [keV]        & [10$^{-7}$ erg cm$^{-2}$ s$^{-1}$] & [10$^{-7}$ erg cm$^{-2}$ s$^{-1}$] & [10$^{-7}$ erg cm$^{-2}$ s$^{-1}$] & -\\\hline
KW & 20-1000 & $ 4.4_{-0.2}^{+0.2}$ & $ 4.6_{-0.3}^{+0.3}$ & $ 4.4_{-0.3}^{+0.4}$ & - \\
WAM0 & 120-1000 & $ 3.6_{-0.1}^{+0.1}$ & $ 4.0_{-0.2}^{+0.2}$ & $ 3.9_{-0.2}^{+0.2}$ & - \\
WAM3 & 120-1000 & $ 3.6_{-0.1}^{+0.1}$ & $ 4.0_{-0.2}^{+0.2}$ & $ 4.0_{-0.2}^{+0.2}$ & - \\
KW-WAM0 & 20-1000 & $ 4.1_{-0.2}^{+0.2}$ & $ 4.8_{-0.2}^{+0.3}$ & $ 4.7_{-0.3}^{+0.3}$ & KW \\
KW-WAM3 & 20-1000 & $ 4.1_{-0.2}^{+0.2}$ & $ 4.7_{-0.3}^{+0.3}$ & $ 4.5_{-0.3}^{+0.3}$ & KW \\
KW-2WAM & 20-1000 & $ 4.0_{-0.2}^{+0.2}$ & $ 4.7_{-0.2}^{+0.2}$ & $ 4.7_{-0.3}^{+0.3}$ & KW \\
\hline
\end{tabular}
\end{center}
\end{table}
\begin{table}
\caption{Fluxes: 051221A Reg1}
\begin{center}
\footnotesize
\begin{tabular}{c|c|c|c|c|c}\hline
Instrument & Energy range & Flux (PL) & Flux (CPL) & Flux (Band) & Inst fixed to 1 \\
           & [keV]        & [10$^{-7}$ erg cm$^{-2}$ s$^{-1}$] & [10$^{-7}$ erg cm$^{-2}$ s$^{-1}$] & [10$^{-7}$ erg cm$^{-2}$ s$^{-1}$] & -\\\hline
KW & 20-1000 & $100.2_{-7.6}^{+7.6}$ & $100.9_{-9.5}^{+9.2}$ & $97.7_{-8.4}^{+8.3}$ & - \\
WAM0 & 120-1000 & $14.1_{-3.6}^{+3.5}$ & $14.3_{-3.7}^{+3.6}$ & $14.2_{-3.9}^{+3.7}$ & - \\
WAM1 & 120-1000 & $13.9_{-1.5}^{+1.5}$ & $13.9_{-1.5}^{+1.5}$ & $13.9_{-1.1}^{+1.1}$ & - \\
BAT & 15-150 & $29.0_{-1.0}^{+1.0}$ & $28.8_{-1.0}^{+1.0}$ & - & - \\
BAT-WAM0 & 15-1000 & $224.0_{-23.7}^{+26.4}$ & $111.1_{-38.7}^{+74.9}$ & $117.6_{-32.5}^{+51.8}$ & BAT \\
BAT-WAM1 & 15-1000 & $174.4_{-18.0}^{+20.0}$ & $90.6_{-14.6}^{+17.7}$ & $86.1_{-11.8}^{+14.0}$ & BAT \\
KW-BAT & 15-1000 & $108.9_{-7.6}^{+7.6}$ & $98.2_{-10.2}^{+10.2}$ & $99.1_{-8.4}^{+8.3}$ & KW \\
KW-WAM0 & 20-1000 & $100.1_{-7.6}^{+7.6}$ & $101.8_{-4.6}^{+9.0}$ & $98.3_{-8.1}^{+8.1}$ & KW \\
KW-WAM1 & 20-1000 & $95.3_{-7.7}^{+7.7}$ & $98.1_{-8.5}^{+8.6}$ & $95.4_{-7.8}^{+7.9}$ & KW \\
KW-2WAM-BAT & 15-1000 & $109.8_{-7.4}^{+7.4}$ & $98.5_{-8.8}^{+9.0}$ & $97.9_{-7.6}^{+7.7}$ & KW \\
\hline
\end{tabular}
\end{center}
\end{table}
\begin{table}
\caption{Fluxes: 060105 Reg1}
\begin{center}
\footnotesize
\begin{tabular}{c|c|c|c|c|c}\hline
Instrument & Energy range & Flux (PL) & Flux (CPL) & Flux (Band) & Inst fixed to 1 \\
           & [keV]        & [10$^{-7}$ erg cm$^{-2}$ s$^{-1}$] & [10$^{-7}$ erg cm$^{-2}$ s$^{-1}$] & [10$^{-7}$ erg cm$^{-2}$ s$^{-1}$] & -\\\hline
KW & 20-1000 & $19.0_{-0.4}^{+0.4}$ & $21.8_{-0.5}^{+0.5}$ & $21.9_{-0.6}^{+0.4}$ & - \\
WAM0 & 120-1000 & $13.4_{-1.3}^{+1.3}$ & $15.3_{-1.7}^{+0.8}$ & $15.2_{-1.8}^{+1.6}$ & - \\
WAM3 & 120-1000 & $34.4_{-2.0}^{+2.0}$ & $36.4_{-1.1}^{+2.2}$ & $35.8_{-1.2}^{+2.6}$ & - \\
BAT & 15-150 & $ 3.8_{-0.1}^{+0.1}$ & - & - & - \\
BAT-WAM0 & 15-1000 & $28.0_{-2.2}^{+2.4}$ & $20.2_{-2.3}^{+2.4}$ & $20.4_{-0.4}^{+1.5}$ & BAT \\
BAT-WAM3 & 15-1000 & $23.5_{-1.9}^{+2.2}$ & $13.7_{-1.2}^{+1.3}$ & $14.2_{-1.2}^{-2.5}$ & BAT \\
KW-BAT & 15-1000 & $19.6_{-0.4}^{+0.4}$ & $21.9_{-0.5}^{+0.5}$ & $22.2_{-0.8}^{+0.3}$ & KW \\
KW-WAM0 & 20-1000 & $19.0_{-0.4}^{+0.4}$ & $21.9_{-0.5}^{+0.5}$ & $21.9_{-0.6}^{+0.4}$ & KW \\
KW-WAM3 & 20-1000 & $18.9_{-0.4}^{+0.4}$ & $21.5_{-0.5}^{+0.5}$ & $21.6_{-0.4}^{+0.4}$ & KW \\
KW-2WAM-BAT & 15-1000 & $19.5_{-0.4}^{+0.4}$ & $21.6_{-0.5}^{+0.5}$ & $21.8_{-0.6}^{+0.4}$ & KW \\
\hline
\end{tabular}
\end{center}
\end{table}
\begin{table}
\caption{Fluxes: 060105 Reg2}
\begin{center}
\footnotesize
\begin{tabular}{c|c|c|c|c|c}\hline
Instrument & Energy range & Flux (PL) & Flux (CPL) & Flux (Band) & Inst fixed to 1 \\
           & [keV]        & [10$^{-7}$ erg cm$^{-2}$ s$^{-1}$] & [10$^{-7}$ erg cm$^{-2}$ s$^{-1}$] & [10$^{-7}$ erg cm$^{-2}$ s$^{-1}$] & -\\\hline
KW & 20-1000 & $11.7_{-0.3}^{+0.3}$ & $10.2_{-0.4}^{+0.4}$ & $10.4_{-0.4}^{+0.4}$ & - \\
WAM0 & 120-1000 & $ 6.3_{-1.3}^{+1.3}$ & $ 6.4_{-1.8}^{+1.4}$ & $ 6.3_{-1.4}^{+1.1}$ & - \\
WAM3 & 120-1000 & $16.9_{-1.9}^{+1.9}$ & $16.8_{-2.2}^{+2.2}$ & $16.8_{-2.1}^{+2.0}$ & - \\
BAT & 15-150 & $ 3.82_{-0.07}^{+0.07}$ & $ 3.80_{-0.08}^{+0.08}$ & - & - \\
BAT-WAM0 & 15-1000 & $22.7_{-1.5}^{+1.6}$ & $10.4_{-1.9}^{+2.6}$ & $10.5_{-1.5}^{+2.4}$ & BAT \\
BAT-WAM3 & 15-1000 & $21.9_{-1.4}^{+1.5}$ & $ 8.9_{-1.0}^{+1.2}$ & $ 8.8_{-1.0}^{+1.2}$ & BAT \\
KW-BAT & 15-1000 & $12.73_{-0.3}^{+0.3}$ & $10.6_{-0.4}^{+0.4}$ & $10.6_{-0.4}^{+0.4}$ & KW \\
KW-WAM0 & 20-1000 & $11.7_{-0.3}^{+0.3}$ & $10.2_{-0.4}^{+0.4}$ & $10.4_{-0.4}^{+0.4}$ & KW \\
KW-WAM3 & 20-1000 & $11.6_{-0.3}^{+0.3}$ & $10.2_{-0.4}^{+0.4}$ & $10.3_{-0.4}^{+0.4}$ & KW \\
KW-2WAM-BAT & 15-1000 & $12.7_{-0.3}^{+0.3}$ & $10.5_{-0.4}^{+0.4}$ & $10.6_{-0.4}^{+0.4}$ & KW \\
\hline
\end{tabular}
\end{center}
\end{table}
\begin{table}
\caption{Fluxes: 060105 Reg12}
\begin{center}
\footnotesize
\begin{tabular}{c|c|c|c|c|c}\hline
Instrument & Energy range & Flux (PL) & Flux (CPL) & Flux (Band) & Inst fixed to 1 \\
           & [keV]        & [10$^{-7}$ erg cm$^{-2}$ s$^{-1}$] & [10$^{-7}$ erg cm$^{-2}$ s$^{-1}$] & [10$^{-7}$ erg cm$^{-2}$ s$^{-1}$] & -\\\hline
KW & 20-1000 & $15.7_{-0.3}^{+0.3}$ & $16.1_{-0.4}^{+0.4}$ & $15.8_{-0.4}^{+0.4}$ & - \\
WAM0 & 120-1000 & $10.1_{-1.0}^{+1.0}$ & $10.8_{-1.1}^{+1.2}$ & $10.6_{-1.2}^{+1.2}$ & - \\
WAM3 & 120-1000 & $25.7_{-1.4}^{+1.4}$ & $26.8_{-1.5}^{+1.5}$ & $26.3_{-1.5}^{+1.9}$ & - \\
BAT & 15-150 & $ 3.82_{-0.06}^{+0.06}$ & - & - & - \\
BAT-WAM0 & 15-1000 & $25.1_{-1.4}^{+1.5}$ & $16.6_{-1.9}^{+2.0}$ & $16.3_{-2.3}^{+2.0}$ & BAT \\
BAT-WAM3 & 15-1000 & $22.5_{-1.3}^{+1.4}$ & $11.5_{-0.9}^{+1.0}$ & $11.1_{-0.8}^{+0.9}$ & BAT \\
KW-BAT & 15-1000 & $16.4_{-0.3}^{+0.3}$ & $16.3_{-0.4}^{+0.4}$ & $16.1_{-0.4}^{+0.5}$ & KW \\
KW-WAM0 & 20-1000 & $15.7_{-0.3}^{+0.3}$ & $16.1_{-0.4}^{+0.4}$ & $15.8_{-0.4}^{+0.4}$ & KW \\
KW-WAM3 & 20-1000 & $15.6_{-0.3}^{+0.3}$ & $15.8_{-0.4}^{+0.4}$ & $15.6_{-0.3}^{+0.4}$ & KW \\
KW-2WAM-BAT & 15-1000 & $16.3_{-0.3}^{+0.3}$ & $16.1_{-0.3}^{+0.3}$ & $15.9_{-0.4}^{+0.4}$ & KW \\
\hline
\end{tabular}
\end{center}
\end{table}
\begin{table}
\caption{Fluxes: 060117 Reg1}
\begin{center}
\footnotesize
\begin{tabular}{c|c|c|c|c|c}\hline
Instrument & Energy range & Flux (PL) & Flux (CPL) & Flux (Band) & Inst fixed to 1 \\
           & [keV]        & [10$^{-7}$ erg cm$^{-2}$ s$^{-1}$] & [10$^{-7}$ erg cm$^{-2}$ s$^{-1}$] & [10$^{-7}$ erg cm$^{-2}$ s$^{-1}$] & -\\\hline
KW & 20-1000 & $12.0_{-0.5}^{+0.5}$ & $10.2_{-0.5}^{+0.6}$ & $10.2_{-0.6}^{+0.5}$ & - \\
WAM0 & 120-1000 & $ 5.2_{-0.4}^{+0.4}$ & $ 5.2_{-0.4}^{+0.4}$ & $ 5.1_{+0.2}^{+0.2}$ & - \\
WAM1 & 120-1000 & $ 4.9_{-0.3}^{+0.3}$ & $ 4.8_{-0.4}^{+0.3}$ & $ 5.6_{-0.4}^{-0.7}$ & - \\
BAT & 15-150 & $ 7.1_{-0.2}^{+0.2}$ & $ 7.0_{-0.2}^{+0.2}$ & $ 6.9_{-0.2}^{+0.2}$ & - \\
BAT-WAM0 & 15-1000 & $14.3_{-0.7}^{+0.7}$ & $11.1_{-0.8}^{+1.0}$ & $10.8_{-0.7}^{+0.8}$ & BAT \\
BAT-WAM1 & 15-1000 & $14.0_{-0.6}^{+0.7}$ & $11.6_{-0.7}^{+0.8}$ & $11.1_{-0.6}^{+0.7}$ & BAT \\
KW-BAT & 15-1000 & $13.5_{-0.5}^{+0.5}$ & $10.8_{-0.5}^{+0.6}$ & $11.2_{-0.6}^{+0.7}$ & KW \\
KW-WAM0 & 20-1000 & $11.8_{-0.5}^{+0.5}$ & $10.5_{-0.5}^{+0.5}$ & $10.8_{-0.5}^{+0.5}$ & KW \\
KW-WAM1 & 20-1000 & $11.7_{-0.5}^{+0.5}$ & $10.8_{-0.5}^{+0.5}$ & $11.0_{-0.5}^{+0.5}$ & KW \\
KW-2WAM-BAT & 15-1000 & $13.2_{-0.4}^{+0.4}$ & $11.6_{-0.5}^{+0.5}$ & $11.6_{-0.4}^{+0.6}$ & KW \\
\hline
\end{tabular}
\end{center}
\end{table}
\begin{table}
\caption{Fluxes: 060117 Reg2}
\begin{center}
\footnotesize
\begin{tabular}{c|c|c|c|c|c}\hline
Instrument & Energy range & Flux (PL) & Flux (CPL) & Flux (Band) & Inst fixed to 1 \\
           & [keV]        & [10$^{-7}$ erg cm$^{-2}$ s$^{-1}$] & [10$^{-7}$ erg cm$^{-2}$ s$^{-1}$] & [10$^{-7}$ erg cm$^{-2}$ s$^{-1}$] & -\\\hline
KW & 20-1000 & $31.9_{-0.8}^{+0.8}$ & $26.6_{-0.8}^{+0.9}$ & $27.5_{-1.0}^{+1.0}$ & - \\
WAM0 & 120-1000 & $12.2_{-0.6}^{+0.6}$ & $12.1_{-0.5}^{+0.5}$ & $12.6_{-0.2}^{-0.3}$ & - \\
WAM1 & 120-1000 & $11.8_{-0.4}^{+0.4}$ & $11.2_{-0.6}^{+0.6}$ & $11.3_{-0.5}^{+0.5}$ & - \\
BAT & 15-150 & $18.8_{-0.3}^{+0.3}$ & $18.5_{-0.4}^{+0.4}$ & $18.5_{-0.3}^{+0.3}$ & - \\
BAT-WAM0 & 15-1000 & $38.1_{-1.4}^{+1.4}$ & $27.8_{-1.1}^{+1.2}$ & $27.5_{-1.1}^{+1.2}$ & BAT \\
BAT-WAM1 & 15-1000 & $36.5_{-1.3}^{+1.3}$ & $29.3_{-1.0}^{+1.1}$ & $29.5_{-1.0}^{+1.1}$ & BAT \\
KW-BAT & 15-1000 & $35.8_{-0.8}^{+0.8}$ & $28.0_{-0.8}^{+0.9}$ & $29.0_{-1.0}^{+1.1}$ & KW \\
KW-WAM0 & 20-1000 & $30.5_{-0.8}^{+0.8}$ & $26.9_{-0.7}^{+0.7}$ & $27.8_{-0.8}^{+0.8}$ & KW \\
KW-WAM1 & 20-1000 & $30.1_{-0.8}^{+0.8}$ & $27.6_{-0.7}^{+0.7}$ & $28.3_{-0.8}^{+0.8}$ & KW \\
KW-2WAM-BAT & 15-1000 & $33.7_{-0.8}^{+0.8}$ & $29.1_{-0.7}^{+0.7}$ & $29.6_{-0.7}^{+0.7}$ & KW \\
\hline
\end{tabular}
\end{center}
\end{table}
\input{table53.tex}
\input{table54.tex}
\input{table55.tex}
\input{table56.tex}
\input{table57.tex}
\input{table58.tex}
\input{table59.tex}
\input{table60.tex}
\input{table61.tex}
\input{table62.tex}
\input{table63.tex}
\input{table64.tex}
\input{table65.tex}
\input{table66.tex}
\input{table67.tex}
\input{table68.tex}
\input{table69.tex}
\input{table70.tex}
\input{table71.tex}
\input{table72.tex}
\input{table73.tex}
\input{table74.tex}
\input{table75.tex}
\input{table76.tex}
\input{table77.tex}
\input{table78.tex}
\input{table79.tex}


%
%

\end{document}